%% file: rpithes.tex
\documentclass[chap]{thesis}

\usepackage[hang]{caption}      % to indent subsequent lines of captions
\renewcommand{\captionfont}{\bfseries} % bold caption (needed with caption 
\begin{document}

\include{rpititle-phd}   % titlepage material
\include{rpiack}

\include{rpiabs}

\include{rpichap1}

\include{rpichap2}

\include{rpichap3}
\include{rpichap4}

\include{rpichap5}
\include{rpibib}
\include{rpiapp}

\end{document}

%% file: rpititle-phd.tex
%%%%%%%%%%%%%%%%%%%%%%%%%%%%%%%%%%%%%%%%%%%%%%%%%%%%%%%%%%%%%%%%%%% 
%                                                                 %
%                            TITLE PAGE                           %
%                            PhD Thesis                           %
%                                                                 %
%%%%%%%%%%%%%%%%%%%%%%%%%%%%%%%%%%%%%%%%%%%%%%%%%%%%%%%%%%%%%%%%%%% 
    
% Supply information for use on title page:
%   
\thesistitle{\textbf{Synchronization Landscapes in Small-World-Connected Computer Networks}}        
\author{Hasan Guclu}        
\degree{Doctor of Philosophy}        
\department{Physics}        
\signaturelines{5}     %max number of signature lines is 7        
\thadviser{Gy\"orgy Korniss}

%\cothadviser{Second Adviser} % If you have 2 thesis advisers
\memberone{Saroj K. Nayak}        
\membertwo{Toh-Ming Lu}        
\memberthree{Wayne G. Roberge}
\memberfour{Boleslaw K. Szymanski}

\submitdate{July 2005\\(For Graduation August 2005)}        
\copyrightyear{2005}   % if omitted, current year is used.        

% Print titlepage and other prefatory material:
%    
\titlepage        
\abstitlepage          % required by Office of Graduate School (1 copy)
\copyrightpage         % optional           
\tableofcontents        
%\listoftables          % required if there are tables
\listoffigures         % required if there are figures

%% file: rpiack.tex
%%%%%%%%%%%%%%%%%%%%%%%%%%%%%%%%%%%%%%%%%%%%%%%%%%%%%%%%%%%%%%%%%%% 
%                                                                 %
%                         ACKNOWLEDGEMENT                         %
%                                                                 %
%%%%%%%%%%%%%%%%%%%%%%%%%%%%%%%%%%%%%%%%%%%%%%%%%%%%%%%%%%%%%%%%%%% 
 
\specialhead{ACKNOWLEDGMENT}
 
First and foremost I am deeply indebted to my academic and research 
adviser, Prof.~Gy\"orgy Korniss. 
I am particularly grateful for his guidance and support in the last four years
of my life in Rensselaer. I thank Zolt\'an Toroczkai from Los Alamos
National Laboratory for his mentorship when I was an intern in 
Center for Nonlinear Studies (Summer 2001) and Complex Systems group 
(Summer 2002). I also thank Mark Novotny and Zolt\'an R\'acz for collaboration
through fruitful discussions. Special thanks go to my thesis committee members, 
Prof.~Szymanski, Prof.~Lu, Prof.~Roberge and Prof.~Nayak.

I have benefited from time spent with group members and friends 
here at Rensselaer Polytechnic Institute. 
I would like to thank Balazs Kozma and Lauren O'Malley. I was
particularly fortunate to have had Dr.~Tansel Karabacak and Dr.~Ibrahim
Yilmaz as colleagues and close friends, with whom I have shared
many interesting discussions, both academic and otherwise, since
the day they accepted me as a guest in their house. Last, but not least, 
I also would like to thank dearest S.~Azra Konet for her support for the last two months.

I acknowledge the financial support of the National Science Foundation
(DMR-0113049) and Research Corporation (RI0761).

%% file: rpiabs.tex
%%%%%%%%%%%%%%%%%%%%%%%%%%%%%%%%%%%%%%%%%%%%%%%%%%%%%%%%%%%%%%%%%%% 
%                                                                 %
%                            ABSTRACT                             %
%                                                                 %
%%%%%%%%%%%%%%%%%%%%%%%%%%%%%%%%%%%%%%%%%%%%%%%%%%%%%%%%%%%%%%%%%%% 
 
\specialhead{ABSTRACT}
 
In this thesis we study synchronization phenomena in natural
and artificial coupled multi-component systems, applicable to the scalability 
of parallel discrete-event simulation for systems 
with asynchronous dynamics. We also study the role of various complex
communication topologies as synchronization networks.
We analyze the properties of the virtual time horizon 
or synchronization landscape (corresponding to the progress of the
processing elements) of these networks by using the framework of
non-equilibrium surface growth.

When the communication topology mimics that
of the short-range interacting underlying system, the virtual time
horizon exhibits Kardar-Parisi-Zhang-like kinetic roughening. 
Although the virtual times, on 
average, progress at a nonzero rate, their statistical spread diverges with
the number of processing elements, hindering efficient data
collection. We show that when the synchronization topology is extended
to include quenched random communication links (small-world links) 
between the processing
elements, they make a close-to-uniform progress with a nonzero rate,
without global synchronization. This leads to a fully scalable parallel 
simulation for underlying systems with asynchronous dynamics and short-range interactions.
We study both short-range and small-world synchronization topologies in one- and two-dimensional
systems. We also provide a coarse-grained
description for the small-world-synchronized virtual-time horizon and
compare the findings to those obtained by ``simulating the
simulations'' based on the exact algorithmic rules. We also present
numerical results for the evolution of the virtual-time
horizon on scale-free Barab\'asi-Albert networks serving as 
communication topology among the processing elements.

Finally, we investigate to what extent
small-world couplings (extending the original local relaxational dynamics
through the random links) lead to the suppression of extreme
fluctuations in the synchronization landscape. In the
absence of the random links, the steady-state landscape is
``rough'' (strongly de-synchronized state) and the average and the extreme
height fluctuations diverge in the same power-law fashion with the
system size (number of nodes). With small-world links present, the
average size of the fluctuations becomes finite (synchronized state).
For exponential-like noise the extreme heights diverge only
logarithmically with the number of nodes, while for power-law noise
they diverge in a power-law fashion. The statistics of the extreme heights are
governed by the Fisher--Tippett--Gumbel and the Fr\'echet distribution
for exponential and power-law noise, respectively.

%% file: rpichap1.tex
%%%%%%%%%%%%%%%%%%%%%%%%%%%%%%%%%%%%%%%%%%%%%%%%%%%%%%%%%%%%%%%%%%% 
%                                                                 %
%                            CHAPTER ONE                          %
%                                                                 %
%%%%%%%%%%%%%%%%%%%%%%%%%%%%%%%%%%%%%%%%%%%%%%%%%%%%%%%%%%%%%%%%%%% 

\chapter{INTRODUCTION}

\section{Complex Networks}

Cooperative behavior and collective phenomena have always been the center stage of statistical physics.
More recently, the study of complex systems has become widespread
across disciplines ranging from socio-economic systems, traffic models,
epidemic models, to the Internet, the World-Wide Web, and grid computer networks.  
With the tools and frameworks provided by modern statistical physics,
and with the availability of rapidly increasing computational
resources, there is a chance to gain deeper understanding of the
behavior of these systems.

One direction to study complexity is using minimal models where one
considers a {\em large} number of simple interacting entities (agents, individuals,
components, etc.) assuming a (typically simple) effective  interaction between
these entities. For example, in the Ising model for ferromagnets, the
entities are the two-state spins and the interaction energetically prefers
neighboring spins to be aligned. In simple models for social systems, the entities are
humans, and the interaction can be, e.g., mimicking (simple majority influence by their
social contacts).

While the interactions and the individual components may be simple, the
collective behavior of these interacting systems are often far from
trivial. For example, in the Ising model, in sufficiently high
dimension, spontaneous order (symmetry breaking) emerges below some critical
temperature. At the critical point the systems becomes strongly
correlated, even though the interaction between spins only extends to a few
neighbors. These are the kind of emergent behaviors we are interested in,
namely, how locally interacting entities can produce large-scale effects.
It needs to be emphasized that in these models, complexity emerges
through the ``outcome'' of the evolution of the system with a large number
of entities, not in the construction of the individual-level
(``microscopic'') dynamics or rules.

Despite the great complexity
and variety of systems, universal laws and phenomena are essential to our inquiry
and to our understanding \cite{BARYAM03}. One way of describing
complex systems is modeling them mathematically by using the
framework of networks which is essentially a relational approach.
A \textit{network} can be defined as a set of items, referred to as \textit{nodes}, 
and \textit{links} connecting them. It is a concept borrowed from the
graph theory, a subfield of combinatorics in mathematics.

The study of complex networks pervades various areas of science ranging from sociology
to statistical physics \cite{ALBERT02,DOROGOVTSEV02,NEWMAN03}. 
Many of our important technological, information, and infrastructure
systems can be considered complex networks 
\cite{BARABASI99,FALOUTSOS99,LERNER03,KORNISS03} with a large number of components. 
The links between the nodes in these networks facilitate some kind of effective
interaction/dynamics between the nodes. Examples (with the processes
inducing the interaction between the nodes) include high-performance
scalable parallel or grid-computing networks (synchronization
protocols for massive parallelization) \cite{KORNISS03}, diffusive load-balancing
schemes (relocating jobs among processors) \cite{RABANI98}, the Internet
(protocols for sending/receiving packets) \cite{BARABASI99,FALOUTSOS99,PETERSON00}, 
the World Wide Web (hyperlinks in the web pages for other web pages) \cite{ALBERT99},
the electric power grid (generating/transmitting power between generators
and buses) \cite{LERNER03}, metabolic networks (reactions between molecules) 
\cite{JEONG00} or social networks (acquaintance or social contacts) 
\cite{EUBANK04,SENTURK05}. Many of these systems are autonomous (by design or
historical evolution), i.e., they lack a central regulator.
Thus, fluctuations in the ``load'' in the respective network
(data/state savings or task allocation in parallel simulations, traffic
in the Internet, voltage/phase in the electric grid etc.) are determined
by the collective result of the individual decisions of many interacting
``agents'' (nodes). As the number of processors on parallel architectures
increases to hundreds of thousands \cite{TOP500_IBM},
grid-computing networks proliferate over the
Internet \cite{KIRKPATRICK03,GRID}, or the electric power-grid covers,  e.g., the
North-American continent \cite{LERNER03}, fundamental questions on the
corresponding dynamical processes on the respective underlying
networks must be addressed.

Regular lattices are commonly used to study physical systems
with short-range interactions. Earlier studies focused mostly on the topological
properties of the networks. Recent works, motivated by a large number of
natural and artificial systems, such as the ones listed above, have turned the focus
to processes on networks, where the interaction and dynamics between the
nodes are facilitated by a complex network. The question then
is how this possibly complex interaction topology influences 
the collective behavior of the system. 

\begin{figure}[htbp]
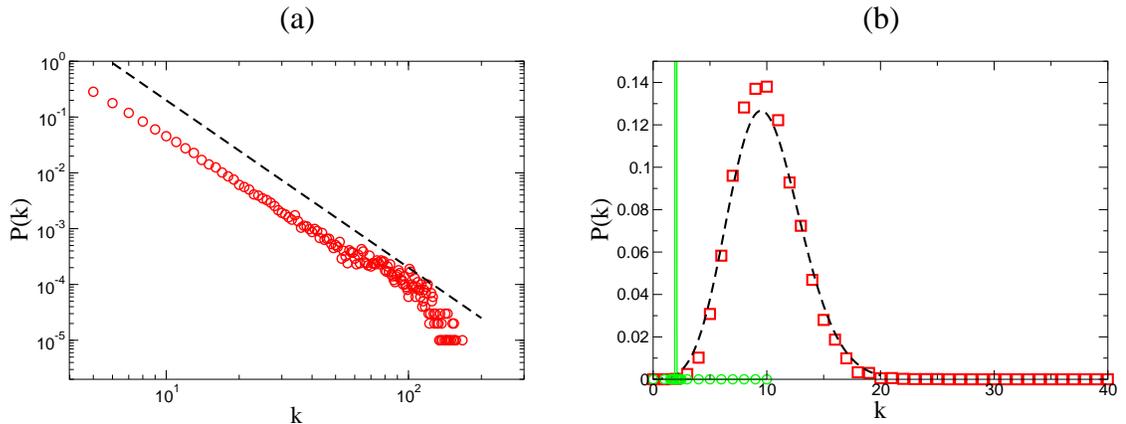

\vspace{6cm}
\includegraphics{SF_m5_pk.eps}
\includegraphics{ER_pk.eps}
\vspace{-0.5cm}
\caption[Degree distributions for Barab\'asi-Albert and Erd\H{o}s-R\'enyi networks]{(a) Degree distribution of the scale-free Barab\'asi-Albert network with $m$$=$$5$,
yielding an average degree of $\langle k \rangle$$=$$10$.
The dashed line shows the power-law behavior in log-log scale.
(b) Degree distribution of a SW network (Erd\H{o}s-R\'enyi Network on a 1D ring) with
$p$$=$$8$. This $p$ value with the additional two nearest-neighbor links yields
an average degree $\langle k \rangle$$=$$10$. The dashed curve is a Poissonian.
The delta-function with circles around 
$\langle k \rangle$$=$$2$ is
the degree distribution of the regular one-dimensional short-range network 
with only two nearest-neighbor links.}
\vspace{-0.3cm}
\label{fig_sf_sw_degree}
\end{figure}

A common property of many real-life networks is that the 
degree or connectivity (number of connections of a node)
follow a scale-free (power-law) distribution. Examples include
world wide web, router level Internet, movie actors collaboration
network, science collaboration network, cellular networks and
linguistic networks \cite{ALBERT02}. Barab\'asi and Albert \cite{BARABASI99}
introduced a growth model with preferential attachment producing scale-free networks. 
They added one node at every time step with $m$
links and connected this node to existing nodes with a probability
proportional to the degree of the existing nodes. This method leads to a
power-law degree distribution function (having a heavier tail
compared to an exponential one), $P(k)$$=$$\frac{2m^2}{k^3}$, shown in
Fig.~\ref{fig_sf_sw_degree}(a). The consequence of the power-law tail in the
degree distribution is the existence of hubs, i.e., a few nodes with a large
number of connections, often observed in real-life networks.

%%%%%%%%%%%%%%%%%%%%%%%%%%%%%%%%%%%%%%%%%%%%%%%%%%%%%%%%%%%%%%%%%%%%%%%%%%%%%%%%%%%%%%%%
%%%%%%%%%%%%%%%%%%%%%%%%%%%%%%%%%%%%%%%%%%%%%%%%%%%%%%%%%%%%%%%%%%%%%%%%%%%%%%%%%%%%%%%

Watts and Strogatz,
inspired by a sociological experiment \cite{MILGRAM67}, have
proposed a network model known as the small-world (SW)
network \cite{WATTS98}. The SW concept describes
the observation that, despite their often large size, there is a relatively
short path between any two nodes in most
networks with some degree of randomness. 
The SW model was originally constructed as a network to
interpolate between regular lattices and completely random
networks \cite{ERDOS60}. Watts and Strogatz considered a regular
short-range network with $k$ nearest links per node. Then they randomly
visited the links and \textit{rewired} them to randomly chosen nodes with
probability $p$. Thus, by varying the parameter $p$ they were able to
interpolate between a regular ($p$$=$$0$) and a completely random ($p$$=$$1$) network.

Another way of constructing the SW network, instead of rewiring,
is visiting every pair of nodes 
and adding a link between them with probability $p/N$, where $N$ is the 
number of nodes. This construction on top of the regular network,
also called random graph and first introduced by Erd\H{o}s and R\'enyi \cite{ERDOS60}, 
have been traditionally used to describe the networks of random topology.
The degree distribution of this SW graph is a Poissonian centered at the mean
degree, $\langle k \rangle$$\simeq$$p+2$, as shown
in Fig.~\ref{fig_sf_sw_degree}(b) with $p$$=$$8$. For $p$$=$$0$, we obtain the short-range 
regular network with a Kronecker-delta degree distribution, $P(k)$$=$$\delta_{kz}$ where $z$
is the coordination number (in 1D, $z$$=$$2$, see Fig.~\ref{fig_sf_sw_degree}(b)).

\begin{figure}[htbp]
\vspace{7.5cm}
\includegraphics{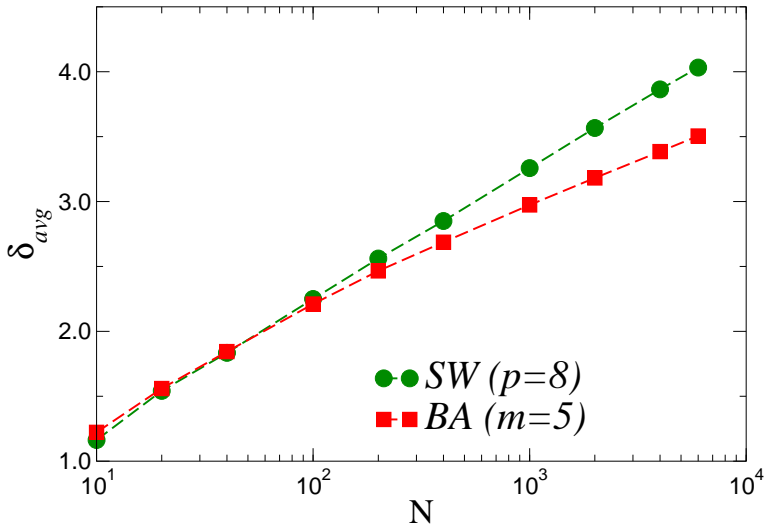}
\vspace{-0.5cm}
\caption[Average shortest path scaling for BA and ER networks]{The average shortest path length $\delta_{avg}$ as a function of system size $N$
for SW ($p$$=$$8$) and BA ($m$$=$$5$). Note normal-log scale.}
\vspace{-0.3cm}
\label{fig_sf_sw_pathN}
\end{figure}

Another important characteristic of networks is the average shortest path 
length $\delta_{avg}$. The shortest path length can be defined as the
minimum number of intermediary nodes between two nodes. All networks
with some degree of randomness has the property that $\delta_{avg}$ is
much smaller than that of regular network with the same number of
nodes and with the same average degree. This very short separation
between any pair of nodes is commonly referred to as the
``low degree of separation''. Typically the average shortest path
length increases no faster than the logarithm of the number of
nodes $N$. For illustration we show the average shortest path length
as a function of system size $N$ for SW ($p$$=$$8$) and BA ($m$$=$$5$)
network in Fig.~\ref{fig_sf_sw_pathN}. Note that on a $d$-dimensional
regular network $\delta_{avg} \sim N_{nodes}^{1/d} \sim N_{linear}$ where $N_{nodes}$
is the number of nodes, $N_{linear}$ is the linear system size 
($N_{nodes}$$=$$N_{linear}^d$) and $d$ is the dimension. 
%%%%%%%%%%%%%%%%%%%%%%%%%%%%%%%%%%%%%%%%%%%%%%%%%%%%%%%%%%%%%%%%%%%%%%%%%%%%%%%%%%%%%%%%%
%%%%%%%%%%%%%%%%%%%%%%%%%%%%%%%%%%%%%%%%%%%%%%%%%%%%%%%%%%%%%%%%%%%%%%%%%%%%%%%%%%%%%%%%

Systems and models (with well-known behaviors on regular
lattices) have been studied on SW networks, such as the Ising model
\cite{SCALETTAR91,GITTERMAN00,BARRAT00,NOVOTNY04}, the XY model \cite{KIM01}, phase
ordering \cite{HONG02_2}, the Edwards-Wilkinson model \cite{KOZMA04,KOZMA05,KOZMA05b} and diffusion
\cite{KOZMA04,KOZMA05,KOZMA05b,MONASSON99,BLUMEN_2000a,BLUMEN_2000b,ALMAAS_2002,HASTINGS04}.
Closely related to phase transitions and collective phenomena is
synchronization in coupled multi-component systems \cite{WIESENFELD96}. SW networks have been shown to
facilitate autonomous synchronization which is an important feature of these networks
from both fundamental and system-design points of view \cite{STROGATZ01,BARAHONA02,HONG02}.
In this thesis we study a synchronization problem which emerges \cite{KORNISS00} in
certain parallel/distributed algorithms referred to as  parallel
discrete-event simulation (PDES) \cite{FUJIMOTO90,NICOL94,LUBACHEVSKY00,KOLA_REV_2005}.
First, we find that constructing a SW-like synchronization network for PDES can
have a huge impact on the scalability of the algorithm \cite{KORNISS03}.
Secondly, since the particular problem is effectively ``local" relaxation in a noisy environment
in a SW network, our study also contributes to the understanding of collective phenomena on
these networks.

\section{Parallel Discrete-Event Simulation}

Simulation of large spatially extended complex systems in physics, engineering,
computer science, or military applications require vast amount of
CPU-time on serial machines using sequential algorithms. PDES
enabled researchers to implement faithful simulations on
parallel/distributed computer systems, namely, systems composed of
multiple interconnected computers
\cite{FUJIMOTO90,NICOL94,LUBACHEVSKY00,KOLA_REV_2005}. 
Developing and implementing massively parallel algorithms is among
the most challenging areas in computer/computational science 
and engineering \cite{HWANG93}. While there
are numerous technological and hardware-related points, e.g.,
concerning efficient message passing and fast communications
between computer nodes, the theoretical algorithmic challenge is
often as important.

PDES is a subclass
of parallel and distributed simulations in which changes in the
components of the system occur instantaneously from one state to
another. In physics, chemistry and biology communities these types of
simulations are most commonly referred to as dynamic or kinetic
Monte Carlo simulations \cite{BINDER98}. Examples of such simulation systems
include cellular communication networks \cite{LUBACHEVSKY00,GREENBERG94},
magnetic systems \cite{KORNISS99,KORNISS01_3}, spatial epidemic
models \cite{DEELMAN96}, thin-film growth
\cite{AMAR_PRB_2005a,AMAR_PRB_2005b}, battle-field models
\cite{NICOL87}, and internet traffic models \cite{COWIE99}. In
these simulations the discrete events are call arrivals, spin-flip
attempts, infections, monomer depositions, troop movements, and packet
transmissions/receptions respectively. In these 
simulations the algorithm must faithfully and 
reproducibly keep track of the asynchrony of the
local updates in the system's configuration. For example standard
random-sequential Monte Carlo simulations naturally produce
Poisson asynchrony. In fact, such continuous-time simulations
(e.g., single spin-flip Glauber dynamics \cite{GLAUBER63}) were long believed to be
inherently serial until Lubachevsky's illuminating work
\cite{LUBACHEVSKY87,LUBACHEVSKY88} on the  parallelization of these simulations
without altering the underlying dynamics. The essence of the
problem is to algorithmically parallelize ``physically''
non-parallel dynamics of the underlying system
while enforcing causality between events and reproducibility. 
This requires some kind of synchronization to ensure causality
between events processed by different processing elements (PEs).

The two basic
ingredients of PDES are the set of local simulated times (or
virtual times \cite{JEFFERSON85}) and a synchronization scheme \cite{FUJIMOTO90}. 
The difficulty in PDES is that the discrete
events are not synchronized by a global clock, since the dynamic
is usually asynchronous. There are two main
approaches in PDES: ({\it i}) \textit{conservative
synchronization}, which avoids the possibility of any type of
causality errors by checking if each event is safe to
process \cite{CHANDY79,CHANDY81} and ({\it ii})
\textit{optimistic synchronization}, which allows causality
errors, then initiates rollbacks to correct the erroneous
computations \cite{JEFFERSON85,DEELMAN97}. Innovative
methods have also been introduced to make optimistic
synchronization more efficient, such as reverse computation
\cite{CAROTHERS99}. Other recent improvements to exploit
parallelism in discrete event systems are the ``lookback" method
\cite{CHEN02} and the freeze-and-shift algorithm \cite{SHCHUR04}.

A PDES should have the following properties to be scalable \cite{GREENBERG94}:
First, a scalable PDES scheme must ensure that the average progress rate
of the simulation approaches a nonzero constant in the long-time
limit as the number of PEs, $N_{PE}$, goes to
infinity (computational scalability) \footnote{The current largest
supercomputer is the IBM/DOE Blue Gene/L with 32K nodes \cite{TOP500_IBM}. 
As a matter of fact the largest natural supercomputer is
the brain, which does an immense parallel computing task to sustain the individual.
In particular the human brain has $10^{11}$ PEs (neurons) each with an average of $10^4$
synaptic connections, creating a bundle on the order of $10^{15}$ ``wires'' 
jammed into a volume of approximately $1400\;\;\mbox{cm}^3$.}. 
Second, the ``width'' of the simulated time horizon (the
spread of the progress of the individual PEs) should be bounded as
$N_{PE}$ goes to infinity (measurement scalability) \cite{GREENBERG96}. The second requirement is
crucial for the measurement phase of the simulation to be
scalable to avoid long delays while waiting for ``slow"
nodes \cite{AMAR_PRB_2005a} or, alternatively, to eliminate the
need to reserve a large amount of memory for temporary data storage: 
a large width of the virtual time horizon hinders
scalable data management. Temporarily storing a large amount of
data on each PE (being accumulated for ``on-the-fly''
measurements) is limited by available memory while frequent global
synchronizations can get costly for large $N_{PE}$. Thus, one aims to
devise a scheme where the PEs make a nonzero and close-to-uniform
progress without global synchronization. In such a scheme, the PEs
autonomously learn the global state of the system (without
receiving explicit global messages) and adjust their progress rate
accordingly. In this thesis we study regular and SW network communication
topologies and show a possible way to construct \textit{fully}
scalable parallel algorithms for underlying systems with
\textit{asynchronous} dynamics and short-range interactions on
regular lattices.

Since one is interested in the dynamics of the underlying
complex system, the PDES scheme must simulate the ``physical time''
variable of the complex system. When the simulations are performed on a
single processor machine, a single (global) time stream is
sufficient to ``label'' or time-stamp the updates of the local
configurations, regardless whether the dynamics of the underlying
system is synchronous or asynchronous. When simulating
asynchronous dynamics on distributed architectures, however, each
PE generates its own physical, or virtual time, which is the
physical time variable of the particular computational domain
handled by that PE. As a result of the local stochastic time increments
and the synchronization dynamics, at a given wall-clock instant the
simulated virtual times of the PEs can differ, a phenomenon
called ``time horizon roughening". We denote the simulated, or
virtual time at PE $i$ measured at wall-clock time $t$, by
$\tau_i(t)$. The wall-clock time 
$t$ is directly proportional to
the (discrete) number of parallel steps simultaneously performed
on each PE, also called the number of Monte-Carlo steps (MCS)
in dynamic Monte Carlo simulations. Without altering the meaning,
$t$ from now on will be used to denote the number of discrete
steps performed in the parallel simulation. The set of virtual
times $\{\tau_i(t)\}_{i=1}^{N_{PE}}$ forms the virtual time
horizon (synchronization landscape) of the PDES scheme after $t$ parallel updates.

The design of efficient parallel-update schemes is a rather
challenging problem, due to the fact that the
dynamics of the simulation scheme itself is a
complex system where the specific synchronization rules
correspond to the ``microscopic dynamics'', and its properties are hard to deduce
using classical methods of algorithm analysis.
Here we present a less conventional approach to the analysis
of efficiency and scalability for the class of
massively parallel conservative PDES schemes, by
mapping the parallel computational process itself
onto a non-equilibrium surface growth model \cite{KORNISS00}.
Then, using methods from statistical mechanics to study the dynamics of
such surfaces (in a completely different context), we
solve the scalability problem of the computational
PDES scheme \cite{KORNISS03,KORNISS00}.
Similar connections between phase transitions and computational complexity have recently
been made \cite{SLOOT01,SCHONEVELD99} for rollback-based (or optimistic)
PDES algorithms \cite{JEFFERSON85} and self-organized criticality \cite{BAK87,BAK88}.
These connections have turned out to be highly fruitful to
gain more insight into traditionally hard computational problems
\cite{COMPLEXITY,MONASSON99_2}. In this thesis we consider the scalability
of conservative synchronization schemes for self-initiating
processes \cite{NICOL91,FELDERMAN91}, where update attempts on each
node are modeled as independent Poisson streams and are 
independent of the configuration of the underlying system
\cite{LUBACHEVSKY87,LUBACHEVSKY88}. We study the
morphological properties of the virtual time horizon. 
Although these properties simplify the
analysis of the corresponding PDES schemes, they can be highly
efficient \cite{KORNISS99} and are readily applicable to a large number
problems in science and engineering. Further, the performance and
the scalability of these PDES schemes become independent of the
specific underlying system i.e., we learn the generic behavior of
these complex computational schemes.
Through our study one also gains some
insight into the effects of SW-like interaction topologies on the
critical fluctuations in interacting systems.

This thesis is organized as follows. In Chapter 2 we show detailed results 
for the short-range model on one and two-dimensional regular networks with nearest-neighbor
communication, which we refer to as the basic conservative 
synchronization (BCS) scheme \cite{KORNISS00}.
In Chapter 3 we extend our study to SW networks, constructed by
adding random links to regular networks \cite{KORNISS03}.
Chapter 4 presents the results on scaling and distributions 
of the extreme fluctuations in 
regular and SW networks. In Chapter 5 we summarize our work 
and discuss future directions.

%% file: rpichap2.tex
%%%%%%%%%%%%%%%%%%%%%%%%%%%%%%%%%%%%%%%%%%%%%%%%%%%%%%%%%%%%%%%%%%% 
%                                                                 %
%                            CHAPTER TWO                          %
%                                                                 %
%%%%%%%%%%%%%%%%%%%%%%%%%%%%%%%%%%%%%%%%%%%%%%%%%%%%%%%%%%%%%%%%%%% 
 
\chapter{SYNCHRONIZATION IN REGULAR NETWORKS}
%\resetfootnote %this command starts footnote numbering with 1 again.

First, we briefly summarize the
basic observables relevant to our analysis of synchronization and the scaling
relations borrowed from non-equilibrium surface growth theory.
The set of local simulated times for the PEs, $\{\tau_i(t)\}_{i=1}^{N_{PE}}$,
constitutes the simulated time horizon. Here $N_{PE}$ is the number of PEs and $t$
is the discrete number of parallel steps, directly related to real (wall-clock) time.
On a regular $d$-dimensional hypercubic lattice $N_{PE}$=$N^d$, where $N$ is the
linear size of the lattice and $d$ is the dimension. For a one-dimensional system
$N_{PE}$=$N$. In the rest of the thesis we will use the term ``height'',
``simulated time'', or ``virtual time'' interchangeably, since we refer to the
same local observable (local field variable).

\begin{figure}[htbp]
\vspace{4.5cm}
\includegraphics{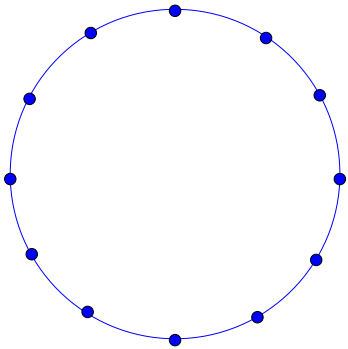}
\caption[1D regular network]{One-dimensional (1D) regular network (with periodic boundary conditions), where nodes
are connected to their nearest neighbors.}
\vspace{-0.5cm}
\label{fig_1dmodel}
\end{figure}

\begin{figure}[htbp]
\vspace{4cm}
\includegraphics{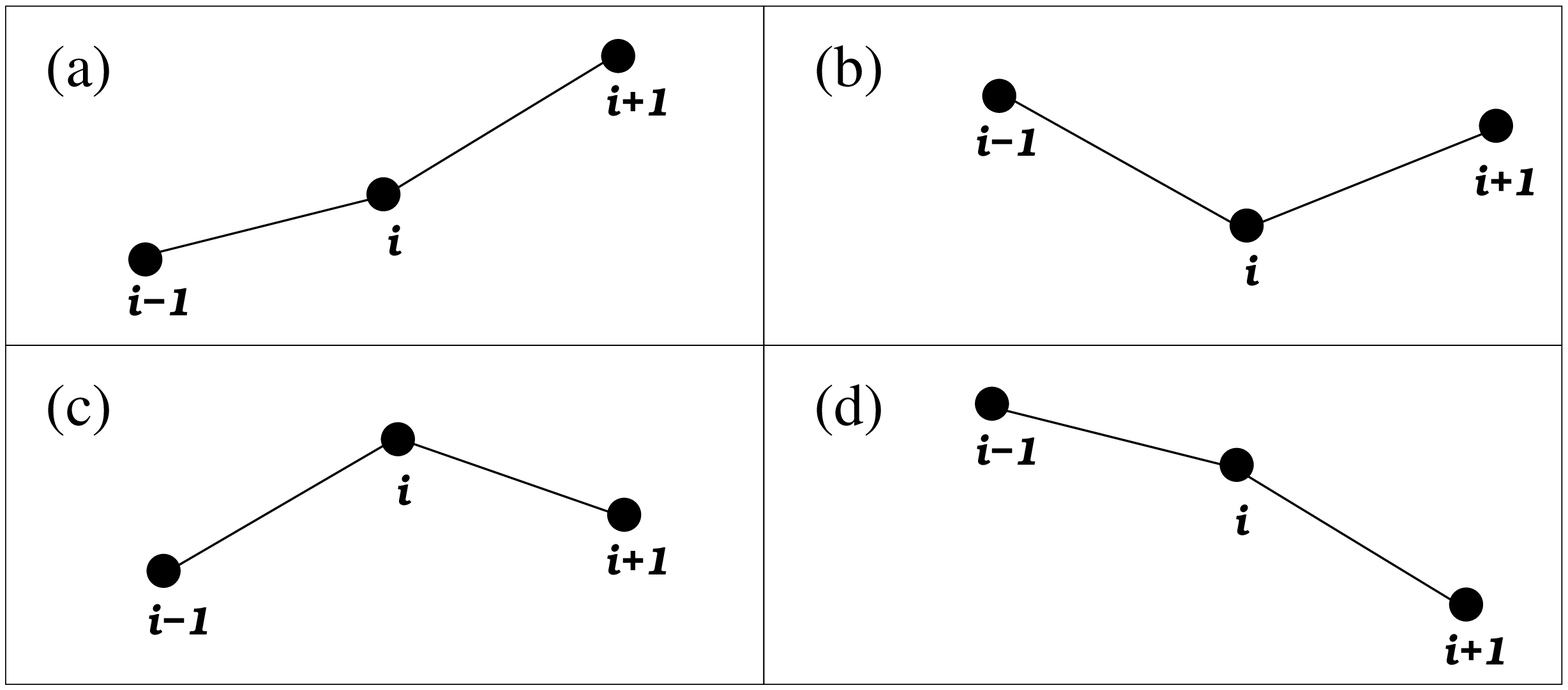}
\caption[Virtual time configurations for successive nodes for BCS in 1D]{Possible simulated-time configurations for three successive nodes
(involving two successive slopes) in the basic conservative scheme (BCS) in one dimension.
From the perspective of node $i$, only configuration (b) allows it to proceed
(node $i$ is a local minimum). In all other cases causality could be violated if an
update occurs at site $i$, because the local field variables of the neighboring nodes
are not known at the instant of update attempt.}
\vspace{-0.5cm}
\label{fig_slopes}
\end{figure}

Since the discrete events in PDES are not synchronized by a global clock, the
processing elements have to communicate with others for synchronization.
One of the first approaches to this problem for self-initiating processes is the
basic conservative synchronization (BCS) 
scheme proposed by Lubachevsky \cite{LUBACHEVSKY87,LUBACHEVSKY88}
by using only nearest neighbor interactions mimicking \cite{KORNISS00} the
interaction topology of the underlying physical system. His basic model associated
each component or site with one PE (worst-case scenario) under periodic boundary
conditions. In this BCS scheme, at each time step only those PEs
whose local simulated time is not larger than the local simulated times of their next
nearest neighbors are incremented by an exponentially distributed random amount so
that the discrete events exhibit Poisson asynchrony. Namely, a PE will only
perform its next update if it can obtain the correct information
to evolve the local configuration (local state) of the underlying
physical system it simulates, without violating causality.
Hence, the evolution equation for site $i$ simply becomes
\begin{equation}
\tau_i(t+1) = \tau_i(t) + \eta_{i}(t)\Theta(-\phi_{i}(t))\Theta(\phi_{i+1}(t))
\;,
\label{evolution}
\end{equation}
where $\eta_{i}(t)$ is an exponentially distributed random number, $\Theta(...)$
is the Heaviside step-function and $\phi_i(t) = \tau_{i}(t)-\tau_{i-1}(t)$ is the local
slope. In one-dimension with periodic boundary conditions, the network has a ring
topology as shown in Fig.~\ref{fig_1dmodel}(a), 
so each node is connected to the nearest
left and right neighbors. The nearest-neighbor interaction in the BCS
scheme implies that in order to ensure causality, PEs need to exchange information on their local
simulated (virtual) times only with neighboring PEs in the virtual network topology.
The possible configurations for the local simulated times for the successive nodes
are shown in Fig.~\ref{fig_slopes}. 
In these configurations update occurs only if
the node we are considering (node $i$) is a \textit{local minimum}. In the other three cases
the node $i$ idles. In analyzing the performance of the above scheme, it is
helpful that the progress of the simulation itself is decoupled from
the possibly complex behavior of the underlying system. This is contrary to
optimistic approaches, where the evolution of the underlying system and the progress
of the PDES simulation are strongly entangled \cite{SLOOT01}, making scalability
analysis a much more difficult task.

One of the important aspects of conservative PDES is the theoretical efficiency
or \textit{utilization} which can be defined simply as the average fraction of
non-idling PEs. It also determines the average progress rate of the simulation.
In the BCS where only nearest-neighbor interactions are present,
the utilization is equal to the density of local minima in the simulated time
horizon. On a regular one-dimensional lattice, it can be defined as
\begin{equation}
\left\langle u(N,t) \right\rangle = 
\frac{1}{N}\sum_{i=1}^{N}
\left\langle
\Theta(\tau_{i-1}-\tau_i)\Theta(\tau_{i+1}-\tau_i) \right\rangle =
\frac{1}{N}\sum_{i=1}^{N}
\left\langle \Theta(-\phi_i) \Theta(\phi_{i+1}) \right\rangle \;,
\label{utilization}
\end{equation}
where $\phi_i$$=$$\tau_{i}-\tau_{i-1}$ is the local slope, $\Theta(...)$ is the Heaviside
step function, and $\left\langle ... \right\rangle$ denotes an ensemble average.
Note that the individual terms in the sum in Eq.~(\ref{utilization}),
$\left\langle \Theta(-\phi_i) \Theta(\phi_{i+1}) \right\rangle$, become
independent of $i$ for a system of identical PEs due to
translational invariance.

Another important observable of PDES is the statistical spread or \textit{width} of
the simulated time surface. The measurement scalability of the PDES scheme, 
is characterized by the width. Instead of dealing with the actual spread 
(difference between the maximum and minimum values) we shall consider
the average ``width'', $w$. It is defined as the root-mean-square fluctuation 
of the virtual times measured from the mean, $w=\sqrt{\langle w^2 \rangle}$, where
\begin{equation}
\left\langle w^{2} \right\rangle \equiv \left\langle w^{2}(N,t) \right\rangle =
\left\langle \frac{1}{N^d}\sum_{i=1}^{N^d}
\left[ \tau_i(t)-\bar{\tau}(t) \right]^2 \right\rangle \;
\;,
\label{width_def}
\end{equation}
with $\bar{\tau}(t)$$=$$(1/N^d)\sum_{i=1}^{N^d} \tau_i(t)$ being the mean progress
(``mean height") of the time surface, and $d$ is the dimension.

As we mentioned in Chapter 1, for the
PDES scheme to be fully scalable, the following two criteria
must be met: (i) the virtual time horizon must progress on average
at a nonzero rate, and (ii) the typical spread of the time horizon
should be finite, as the number of PEs $N$ goes to infinity. When
the first criterion is ensured for large enough times $t$, the
simulation is said to be computationally scalable, 
meaning that when increasing the size of the network to
infinity, while keeping the average computational domain/load on a
single  PE the same, the simulation will progress at a nonzero
rate. However, as we will show below, while increasing the system size,
the spread in the time horizon can diverge, severely
hindering frequent data collection about the state of the
simulated system. Specifically, when one requires to take a
measurement of some physical property of the simulated system at
virtual time $\tau$, PEs have to wait
(in wall-clock time) until all the virtual simulated times
at \textit{all} the PEs pass through the value of $\tau$. Thus, in order
to collect system-wide measurements from the simulation, we incur
a waiting time proportional to the spread, or width of the
fluctuating time horizon. For PDES schemes for which the spread
diverges with system size, the waiting time for the measurements
will also diverge, and the scheme is not measurement
scalable. When condition (ii) is fulfilled for large enough times
$t$, we say that the PDES scheme is measurement scalable.

\section{Scaling in non-equilibrium surfaces}

Since we use the formalism and terminology of non-equilibrium surface growth phenomena,
we briefly review scaling concepts for self-affine or rough surfaces. The scaling behavior
of the width, $\langle w^{2}(N,t) \rangle$ where $N$ is the linear system size and
$t$ is the time, alone typically captures and
identifies the universality class of the non-equilibrium growth process
\cite{BARABASI95,HEALY95,KRUG97}. In a finite system, the width initially grows as
$\left\langle w^2(N,t)\right\rangle\sim t^{2\beta}$. After a system-size
dependent cross-over time $t_{\times}\sim{N}^{z}$, it reaches a steady-state
$\left\langle w^2(N,t)\right\rangle\sim{N}^{2\alpha}$ for $t \gg t_x$. 
In expressions above
$\alpha$, $\beta$ and $z$=$\alpha/\beta$ are called the roughness, 
the growth, and the dynamic exponents, respectively. The above behavior can be summarized
as follows
\begin{equation}
\left\langle w^2(N,t)\right\rangle \sim
\left\{
\begin{array}{ll}
t^{2\beta}    & \mbox{for $t$$\ll$$t_x$} \\
N^{2\alpha} & \mbox{for $t$$\gg$$t_x$}
\end{array}
\right. \;,
\label{width_scaling}
\end{equation}
where $t_x$$\sim$$N^z$ is the cross-over time. From this scaling, one can
also extract a length-scale, known as lateral correlation length, $\xi \sim t^{1/z}$
for times less than $t_x$, reaching the system size at the cross-over time. 
The temporal and system-size
scaling of the width exhibited by Eq.~(\ref{width_scaling}) can be captured by the Family-Vicsek \cite{FAMILY85} relation,
\begin{equation}
\left\langle w^2(N,t)\right\rangle = N^{2\alpha}f(t/N^{z})
\;.
\label{vicsek}
\end{equation}
Note that the scaling function $f(x)$ depends on $t$ and the linear system-size $N$
only through the specific combination, $t/N^z$, reflecting the importance of the
crossover time $t_\times$. For small values of its argument $f(x)$ behaves as a power
law, while for large arguments it approaches a constant
\begin{equation}
f(x) \sim
\left\{
\begin{array}{ll}
x^{2\beta}    & \mbox{if $x$$\ll$$1$} \\
\mbox{const.} & \mbox{if $x$$\gg$$1$}
\end{array}
\right. \;,
\label{f_scaling}
\end{equation}
yielding the correct scaling behavior of the width for early times and for the
steady-state, respectively.

A somewhat less frequently studied quantity is the growth rate of a
growing surface. This quantity is typically non-universal 
\cite{KORNISS00,TOROCZKAI00,KRUG90,KORNISS02,KOLAKOWSKA03,KOLAKOWSKA03_2,KOLAKOWSKA04}, 
but as was shown by Krug and Meakin \cite{KRUG90}, on
$d$-dimensional regular lattices, the finite-size corrections to it
are. In the context of the basic PDES scheme, the growth rate of the
simulated time surface corresponds to the progress rate (or
utilization) of the simulation, hence our special interest in this observable. 
For the finite-size behavior of the steady-state growth rate, one has \cite{KRUG90}
\begin{equation}
\langle u(N)\rangle \simeq \langle u(\infty)\rangle + \frac{const.}{N^{2(1-\alpha)}} \;,
\label{utild}
\end{equation}
where $\langle u(\infty) \rangle$ is the value of the growth rate
in the asymptotic infinite system-size limit and
$\alpha$ is the dimension-dependent roughness exponent of the growth process.

\section{One-Dimensional Basic Conservative Synchronization Network}

\begin{figure}[htbp]
\vspace{9cm}
\includegraphics{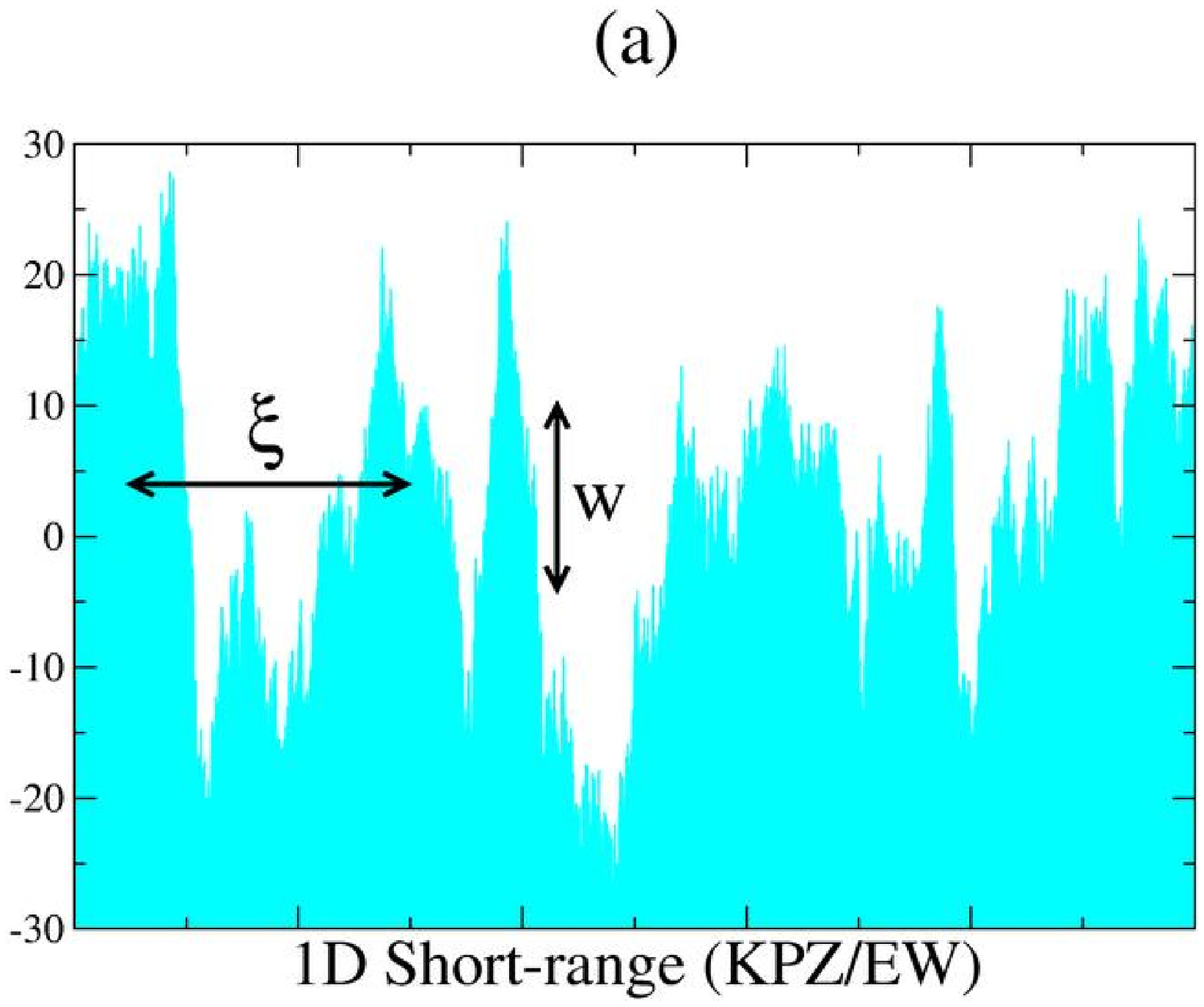}
\includegraphics{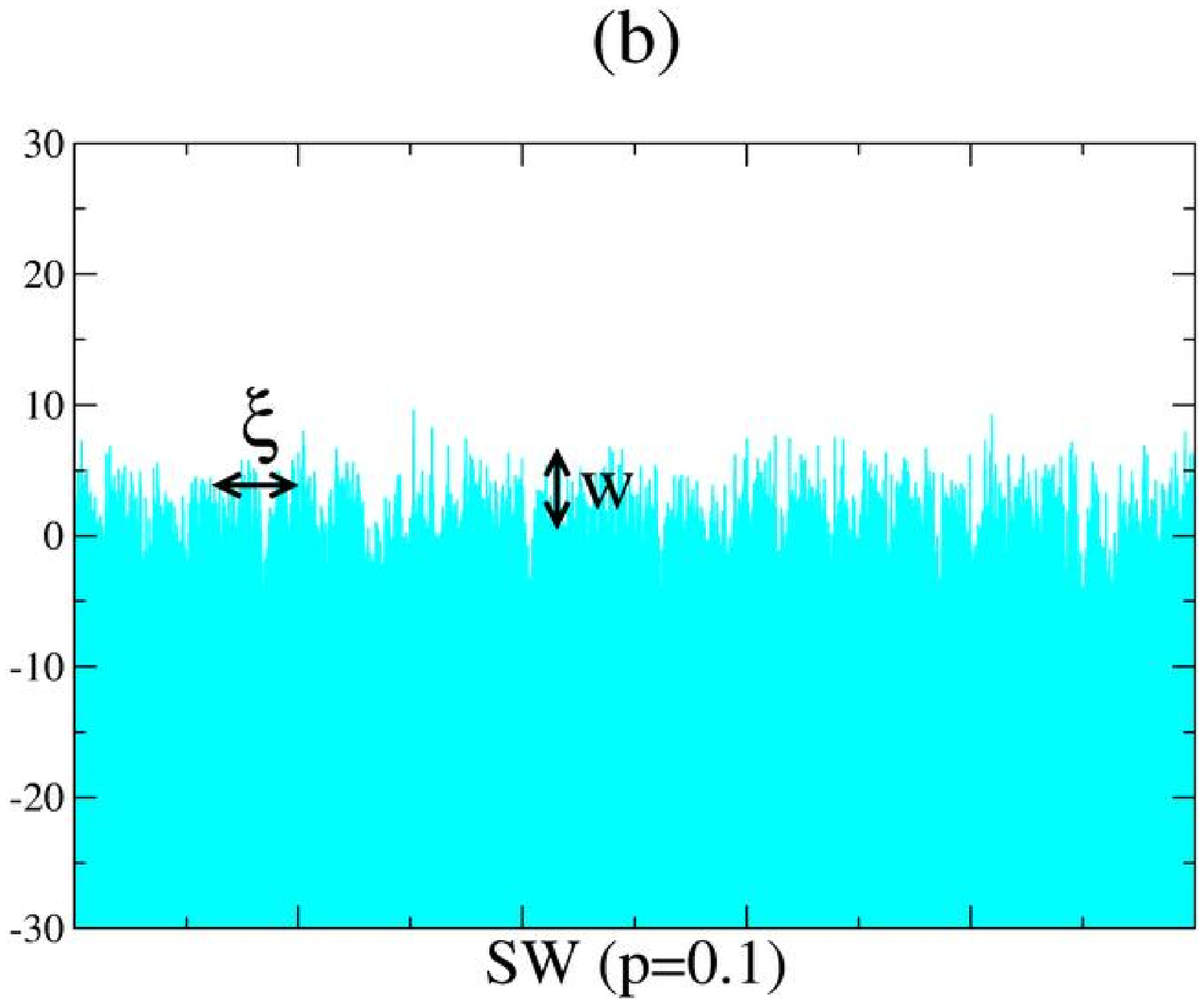}
\vspace{-3cm}
\caption[Virtual time horizon snapshots]{Virtual time horizon snapshots for 10,000 sites in 1D.
(a) For the regular network with nearest-neighbor connections ($p$=$0$).
The lateral correlation length $\xi$ and width $w\equiv \sqrt{w^2}$ are
shown for illustration in the graph. The rough steady-state surface belongs to the
KPZ/EW universality class.
(b) For the SW synchronization network. The heights are effectively decorrelated and both
the correlation length and the width are reduced, and approach system-size independent values
for sufficiently large systems. The resulting surface is macroscopically smooth.
Note that the heights are relative to the average height.}
\vspace{-0.5cm}
\label{figsn}
\end{figure}

Based on a mapping between virtual times and surface site heights \cite{KORNISS00} and 
on the analogy with the single-step surface growth model \cite{PLISCHKE87},
in the coarse-grained description \cite{TOROCZKAI00}, the virtual time horizon
of the BCS is proposed to be governed by
the Kardar-Parisi-Zhang (KPZ) equation \cite{KARDAR86}, well-known in surface growth
phenomena
\begin{equation}
\partial_t \hat{\tau}_i = \nabla^2\hat{\tau}_i - \lambda (\nabla \hat{\tau}_i)^2
+ ... + \eta_i(t)
\;,
\label{KPZ}
\end{equation}
where $\nabla^2\hat{\tau}_i$ is the discretized Laplacian,
$\nabla^2\hat{\tau}_i = \hat{\tau}_{i+1} + \hat{\tau}_{i-1} - 2\hat{\tau}_i$,
$\nabla \hat{\tau}_i$ is the discretized gradient,
$\nabla \hat{\tau}_i = \hat{\tau}_{i+1}-\hat{\tau}_i$,
$\hat{\tau}_i(t)=\tau_i-\bar{\tau}$ is the surface height fluctuation
(or virtual time) measured
from the mean, $\eta_i(t)$ is Gaussian noise delta-correlated in space and time,
$\langle\eta_i(t)\eta_j(t')\rangle = 2D\delta_{ij}\delta(t-t')$,
$\lambda$ is a positive constant, and $...$ stands for higher order irrelevant terms.
Equation~(\ref{KPZ}) can also give an account of a number
of other nonlinear phenomena such as Burgers turbulence \cite{BURGERS74} and directed polymers in
random media \cite{BARABASI95}. When the simple update rule of the basic synchronization scheme is implemented
on a one-dimensional network, one can observe a simulated time surface governed by the KPZ equation,
and in the steady-state, by an Edwards-Wilkinson Hamiltonian \cite{EDWARDS82} [Fig.~\ref{figsn}(a)].

%%%%%%%%%%%%%%%%%%%%%%%%%%%%%%%%%%%%%%%%%%%%%%%%%%%%%%%%%%%%%%%%%%%%%%%%%%%%%%%%%%%%%%%%%
%%%%%%%%%%%%%%%%%%%%%%%%%%%%%%%%%%%%%%%%%%%%%%%%%%%%%%%%%%%%%%%%%%%%%%%%%%%%%%%%%%%%%%%%

When analyzing the statistical and morphological properties of the
stochastic landscape of the simulated times, it is convenient to
study the height-height correlation or its Fourier transform, the
height-height structure factor. The equal-time height-height
structure factor $S(k,t)$ in one-dimension is defined through
\begin{equation}
N \delta_{k,-k^\prime} S(k,t) = 
\langle \tilde{\tau}_k(t)\tilde{\tau}_{k^\prime}(t) \rangle
\;,
\label{str_factor}
\end{equation}
where $\tilde{\tau}_k = \sum_{j=1}^{N}e^{-ikj}\hat\tau_j$ is the
Fourier transform of the virtual times with the wave number
$k$=$2\pi n/N$, $n$=$1,2,...,N-1$ and $\delta_{k,-k^\prime}$ is
the Kronecker delta. The structure factor essentially contains all the
``physics'' needed to describe the scaling behavior of the time
surface. Here we focus on the steady-state properties
($t$$\to$$\infty$) of the time horizon where the structure factor
becomes independent of time, $\lim_{t\to\infty}S(k,t)$$=$$S(k)$. In
the long-time limit, in one dimension, for a KPZ surface described
by Eq.~(\ref{KPZ}) one has (see Appendix A) \cite{TOROCZKAI00}
\begin{equation}
S(k) = \frac{D}{2[1-\cos(k)]} \sim \frac{1}{k^2}\;,
\label{sf_toro}
\end{equation}
where $D$ is a constant and the latter approximation holds for small values of $k$. By
performing the inverse Fourier transformation of
Eq.~(\ref{sf_toro}), we can also obtain the spatial two-point
correlation function, 
\begin{equation}
G(l) = \frac{1}{N}\sum_{i=1}^N G_{i,i+l} = \frac{1}{N}\sum_{k\neq 0}e^{ikl}S(k)
\;, 
\label{Gl_sf}
\end{equation}
where $G_{i,i+l}$$=$$\langle(\tau_i-\bar\tau)(\tau_{i+l}-\bar\tau)\rangle$ is the
site-dependent two point function,
yielding
\cite{TOROCZKAI00,TOROCZKAI03}
\begin{equation}
G(l)\simeq\frac{D}{2}\left(\frac{N}{6} - l \right)
\label{corr_toro}
\end{equation}
for $1\ll l \ll L$. In particular, for the steady-state width one
finds
\begin{equation}
\langle w^2 \rangle = \frac{1}{N}\sum_{k \not= 0} S(k) =G(0)
\simeq \frac{D}{12}N \sim N
\label{sf_w}
\end{equation}
in one dimension \cite{TOROCZKAI00}. This divergent width is caused by
a divergent length scale, $\xi$, the ``lateral" correlation length
in the KPZ-like synchronization landscape.

\begin{figure}[htbp]
\vspace{6cm}
\includegraphics{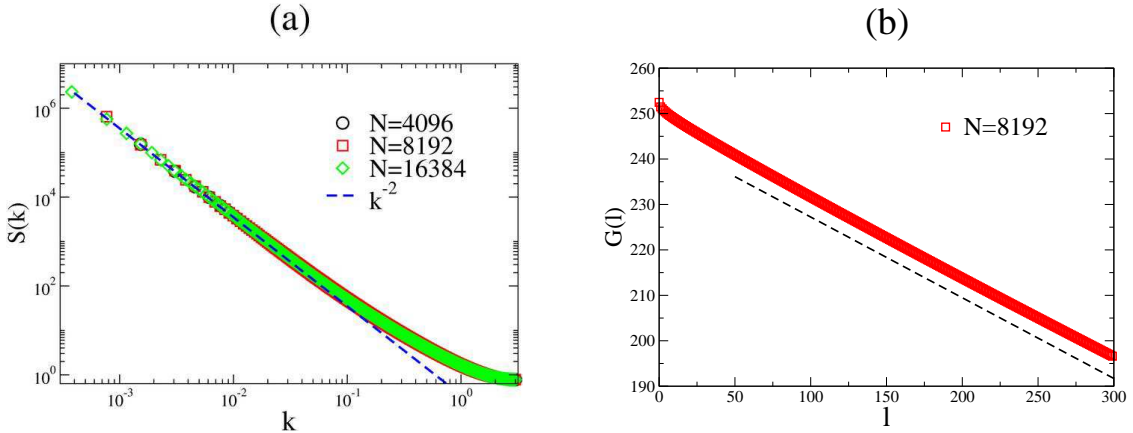}
\includegraphics{1D-KPZ-corr-L8192.eps}
\vspace{-0.3cm}
\caption[Structure factor and two-point correlation function for BCS in 1D]{(a) The steady-state structure factor as a function of the wave number for the BCS
scheme in 1D. The small-$k$ course-grained prediction (consistent with the steady-state EW/KPZ
universality class in 1D) is indicated by a dashed line [Eq.~(\ref{sf_toro})].
Note the log-log scales.
(b) Steady-state spatial two-point correlation function. The straight line again indicates
the asymptotic EW/KPZ behavior in one dimension [Eq.~(\ref{corr_toro})].}
\vspace{-0.3cm}
\label{fig_kpz1dsf}
\end{figure}

\begin{figure}[htbp]
\vspace{6cm}
\includegraphics{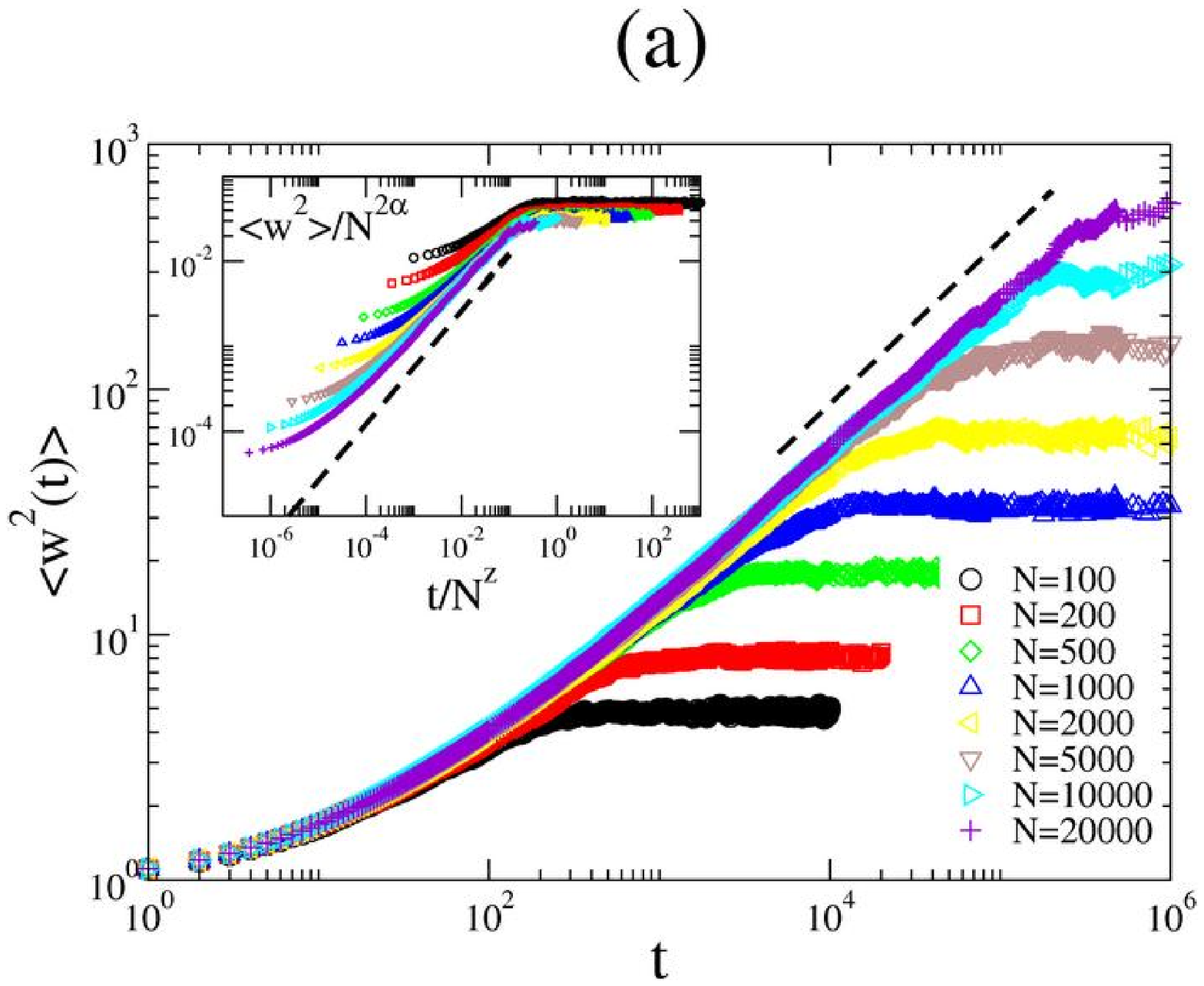}
\includegraphics{1D-KPZ-w2L.eps}
\vspace{-0.3cm}
\caption[Time evolution and scaling of the width for BCS in 1D]{(a) Time evolution of the width for different system-sizes
in the BCS scheme in 1D. The dynamics of the evolution of the virtual
times is governed by the KPZ equation. The inset shows the same data on scaled axes,
$\langle w^2\rangle/N^{2\alpha}$ versus $t/N^z$. Each curve has been obtained by averaging 
over at least fifty realizations.
(b) The steady-state width
of the time horizon for the one-dimensional BCS as a function
of system-size. The dashed straight line represents the asymptotic one-dimensional
KPZ/EW behavior, $\left\langle w^2(N)\right\rangle\sim N^{2\alpha}$ with $\alpha$=$1/2$.}
\vspace{-0.5cm}
\label{fig_kpz1dw2Lt}
\end{figure}

The measured steady-state structure factor
[Fig.~\ref{fig_kpz1dsf}(a)], obtained by simulating the BCS based
on the exact rules for the evolution of the synchronization
landscape confirms the coarse-grained prediction for small $k$
values, $S(k) \sim 1/k^2$. Figure~\ref{fig_kpz1dsf}(b) shows the
corresponding spatial two-point correlation function, $G(l)$.
%%%%%%%%%%%%%%%%%%%%%%%%%%%%%%%%%%%%%%%%%%%%%%%%%%%%%%%%%%%%%%%%%%%%%%%%%%%%%%%%%%%%%
%%%%%%%%%%%%%%%%%%%%%%%%%%%%%%%%%%%%%%%%%%%%%%%%%%%%%%%%%%%%%%%%%%%%%%%%%%%%%%%%%%%%%
Simulation of the BCS scheme in one dimension yields scaling
exponents that agree within error of the predictions of the KPZ
equation \cite{BARABASI95,HEALY95,KARDAR86}. The time evolution of the width
[Fig.~\ref{fig_kpz1dw2Lt}(a)] shows that the growth exponent
$\beta\simeq 1/3$.  
Looking at the system-size dependence of the
steady-state width [Figure~\ref{fig_kpz1dw2Lt}(b)], we find the
roughness exponent $\alpha\simeq 1/2$, consistent with the
one-dimensional KPZ value, $\langle w^2 \rangle \sim N^{2\alpha}
\sim  N$. The dynamic exponent values found from the width as a
function of the cross-over time and $z$=$\alpha/\beta$ are the same,
about $3/2$. The inset in Fig.~\ref{fig_kpz1dw2Lt}(a) shows
that the scaled version of the width evolution by using the
scaling exponents is consistent with the Family-Vicsek relation
[Eq.~(\ref{vicsek})], although with relatively large corrections
to scaling.

The steady-state width distributions, $P(w^2)$, have been introduced to provide
a more detailed characterization of surface growth processes
\cite{FOLTIN94,PLISCHKE94,RACZ94,ANTAL96} and have been used
to identify universality classes \cite{KORNISS00}. Note that the width
\begin{equation}
w^2 = \frac{1}{N^d}\sum_{i=1}^{N^d}[\tau_i(t)-\bar\tau(t)]^2
\;,
\label{width_def2}
\end{equation}
itself is a fluctuating quantity. The width
distribution for the EW (or a steady-state one-dimensional KPZ)
class is characterized by a universal scaling function, $\Phi(x)$,
such that $P(w^2)=\langle w^2\rangle^{-1}\Phi(w^2/\langle
w^2\rangle)$, where $\Phi(x)$ can be calculated analytically for a
number of models, including the EW class \cite{FOLTIN94}. The width
distribution for the basic synchronization scheme is shown in
\begin{figure}[htbp]
\vspace{7.5cm}
\includegraphics{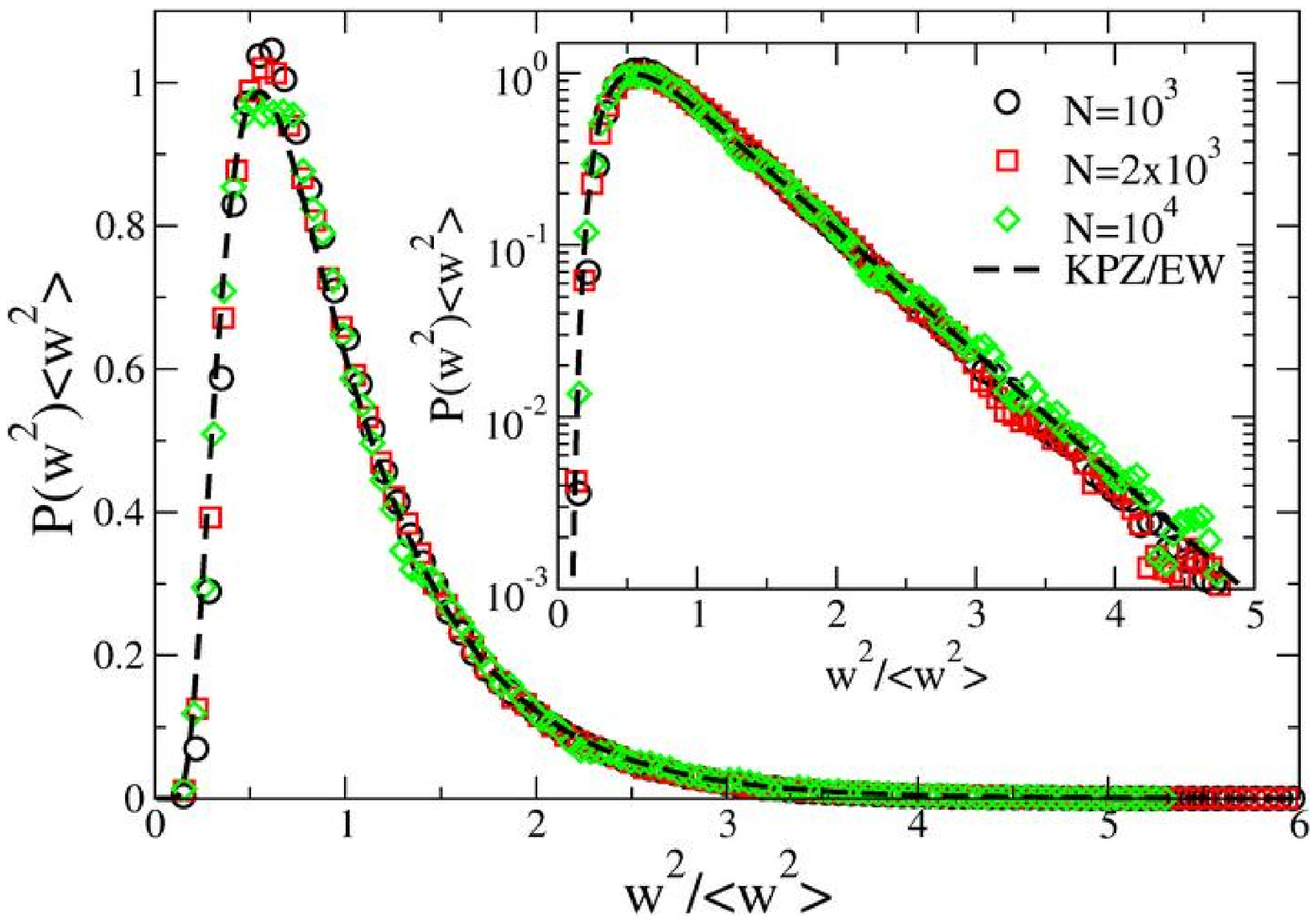}
\vspace{-0.3cm}
\caption[Scaled width distribution for BCS in 1D]]{Scaled width distributions for the BCS scheme in 1D.
The exact asymptotic EW/KPZ width distribution \cite{FOLTIN94} is shown with a dashed line.
The inset shows the same distributions on log-normal scales.}
\vspace{-0.5cm}
\label{fig_kpz1dw2dist}
\end{figure}
Fig.~\ref{fig_kpz1dw2dist}. Systems with $N\geq 10^3$ show
convincing data collapse onto this exact scaling function. The
inset in Fig.~\ref{fig_kpz1dw2dist} shows the same graph in
log-normal scale to show the collapse at the tail of the
distribution. The convergence to the limit distribution is very
slow when compared to other microscopic models (such as the
single-step model \cite{BARABASI95,ANTAL96}) belonging to the same
KPZ universality class.
%%%%%%%%%%%%%%%%%%%%%%%%%%%%%%%%%%%%%%%%%%%%%%%%%%%%%%%%%%%%%%%%%%%%%%%%%%%%%%%%%%%
%\begin{equation}
%\Phi(x) = \frac{\pi^2}{3}\sum_{n=1}^{\infty}(-1)^{n-1}n^2e^{-(\pi^2/6)n^2x}
%\;.
%\label{pw2_phi}
%\end{equation}
%%%%%%%%%%%%%%%%%%%%%%%%%%%%%%%%%%%%%%%%%%%%%%%%%%%%%%%%%%%%%%%%%%%%%%%%%%%%%%%%%%%

\begin{figure}[htbp]
\vspace{6cm}
\includegraphics{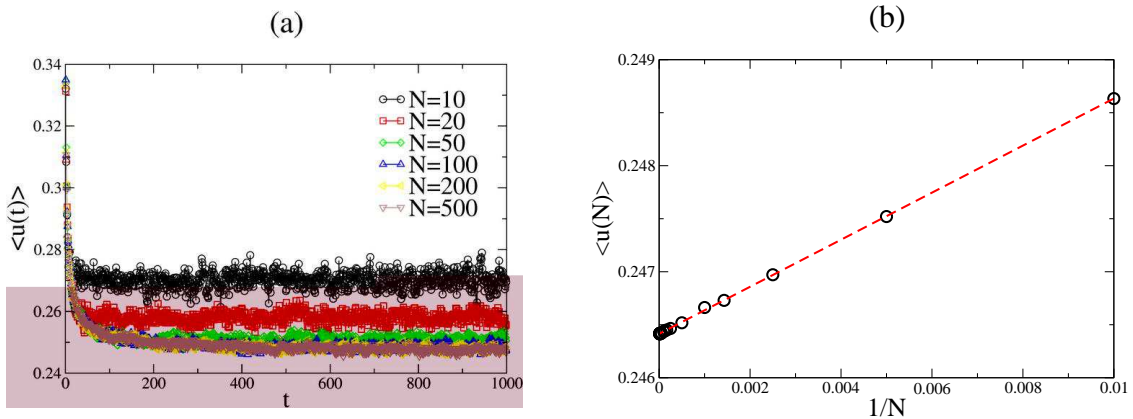}
\includegraphics{1D-KPZ-utilN.eps}
\vspace{-0.3cm}
\caption[The evolution and scaling of steady-state utilization for BCS in 1D]{(a) Utilization in the 1D BCS scheme as a function of time for various
system sizes. (b) Steady-state utilization as a function of 
the $1/N^{2(1-\alpha)}$ as suggested by Eq.~(\ref{utild}) 
with the 1D KPZ roughness exponent $\alpha$$=$$1/2$. 
The dashed line is a linear fit, 
$\langle u(N)\rangle \approx 0.2464 + 0.2219/N$}
\vspace{-0.5cm}
\label{fig_kpz1dutil}
\end{figure}

\begin{figure}[htbp]
\vspace{8.8cm}
\includegraphics{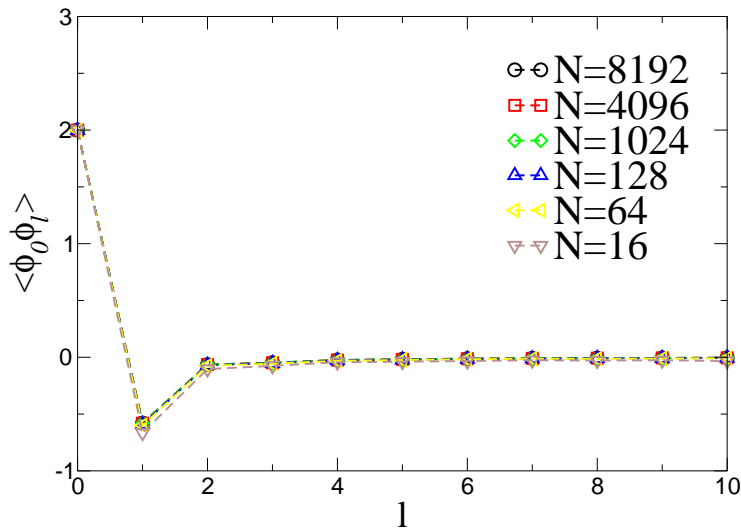}
\vspace{-0.5cm}
\caption[Slope-slope correlation function for BCS in 1D]{Slope-slope correlation function for the BCS scheme on
a 1D short-range network for various system sizes.}
\vspace{-0.5cm}
\label{fig_kpz1dscf}
\end{figure}

Now we discuss our findings for the steady-state utilization of the BCS
scheme. As stated above, the synchronization landscape of the virtual
times belongs to the EW universality class in one dimension. This
implies that the {\em local slopes} in the steady-state landscape
are short-range correlated \cite{TOROCZKAI00}. Hence the density of
local minima in the synchronization landscape, and in turn the
utilization, remains {\em nonzero} in the infinite system-size
limit \cite{KORNISS00,TOROCZKAI00}. 
For a fixed $N$, the utilization drops from relatively higher
initial value at early times to its steady-state value in a 
very short time [Fig.~\ref{fig_kpz1dutil}(a)].
Further, the steady-state utilizations for various systems converge
to the asymptotic system-size independent value.
In 1D, since $\alpha$$=$$0.5$ the utilization,
by using Eq.~(\ref{utild}) as a function of system size, becomes 
\begin{equation}
\langle u(N)\rangle \simeq \langle u(\infty)\rangle + \frac{const.}{N}
\label{util_1d_kpz}
\end{equation}
as shown in Fig.~\ref{fig_kpz1dutil}(b).  For the KPZ
model [Eq.~(\ref{KPZ})] $\langle u(\infty)\rangle = 1/4$, since
in the steady state the slopes are delta-correlated, resulting in
a probability $1/4$ for the configuration in
Fig.~\ref{fig_slopes}(b), corresponding to a local minimum. For
the actual BCS synchronization profile $\langle u(\infty)\rangle
\simeq 0.2464$ \cite{KORNISS00,TOROCZKAI00}, as a result of
non-universal short-range correlations present for the slopes in
the specific microscopic model \cite{TOROCZKAI03} as can be seen
in Fig.~\ref{fig_kpz1dscf}. 

In summary, we have shown that the 1D BCS time horizon belongs to the
KPZ universality class as $N$ goes to infinity, then the measurement part of
the 1D BCS scheme is not scalable.

\begin{section}{Two-Dimensional Basic Conservative Synchronization Network}

A natural generalization to pursue is the synchronization dynamics
and the associated landscapes on the networks in higher
dimensions. One might ask whether PDES of two-dimensional
phenomena exhibit kinetic roughening of the virtual time horizon.
Preliminary results indicated that this is the case
\cite{KIRKPATRICK03,KORNISS01}. In this section we give detailed
results when the BCS scheme is extended into a two-dimensional
lattice in which each node has four nearest neighbors. We consider
a system with periodic boundary conditions in both axes as can be
seen in Fig.~\ref{fig_2d-model}(a). 
%%%%%%%%%%%%%%%%%%%%%%%%%%%%%%%%%%%%%%%%%%%%%%%%%%%%%%%%%%%%%%%%%%%%%%%%%%%%%%%%%%%
\begin{figure}[htbp]
\vspace{5.5cm} 
\includegraphics{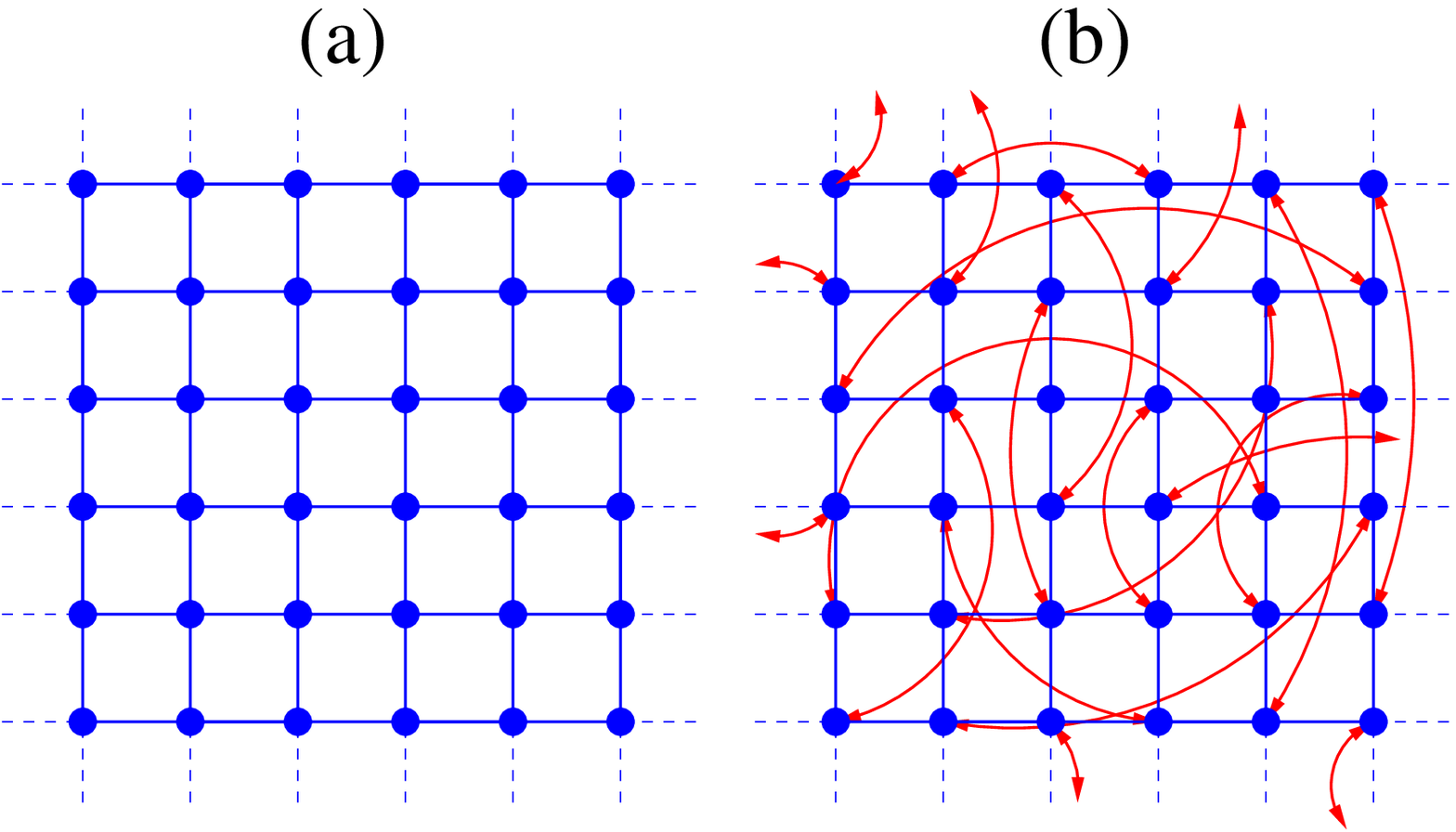} 
\vspace{-0.2cm}
\caption[Communication topologies in 2D]{Communication
topologies in 2D. (a) 2D regular network, where each node is connected to its four nearest neighbors.
(b) SW synchronization network. One random link per node is added on top
of the 2D regular network. Arrowed lines show the bidirectional random links between the nodes.}
\label{fig_2d-model}
\end{figure}
%%%%%%%%%%%%%%%%%%%%%%%%%%%%%%%%%%%%%%%%%%%%%%%%%%%%%%%%%%%%%%%%%%%%%%%%%%%%%%%%%%%%%%%%

The same microscopic rules, i.e., 
each node increments its local simulated time by an
exponentially distributed random amount when it is a local minima
among its nearest neighbors, are applied to this lattice. As in
the one-dimensional case, during the evolution of the local
simulated times correlations between the nodes develop in the system. 
One observes a rough time surface in the steady-state of the 2D 
BCS network. Figure~\ref{fig_2d-surface}(a) shows the
\begin{figure}[htbp]
\vspace{7cm}
\includegraphics{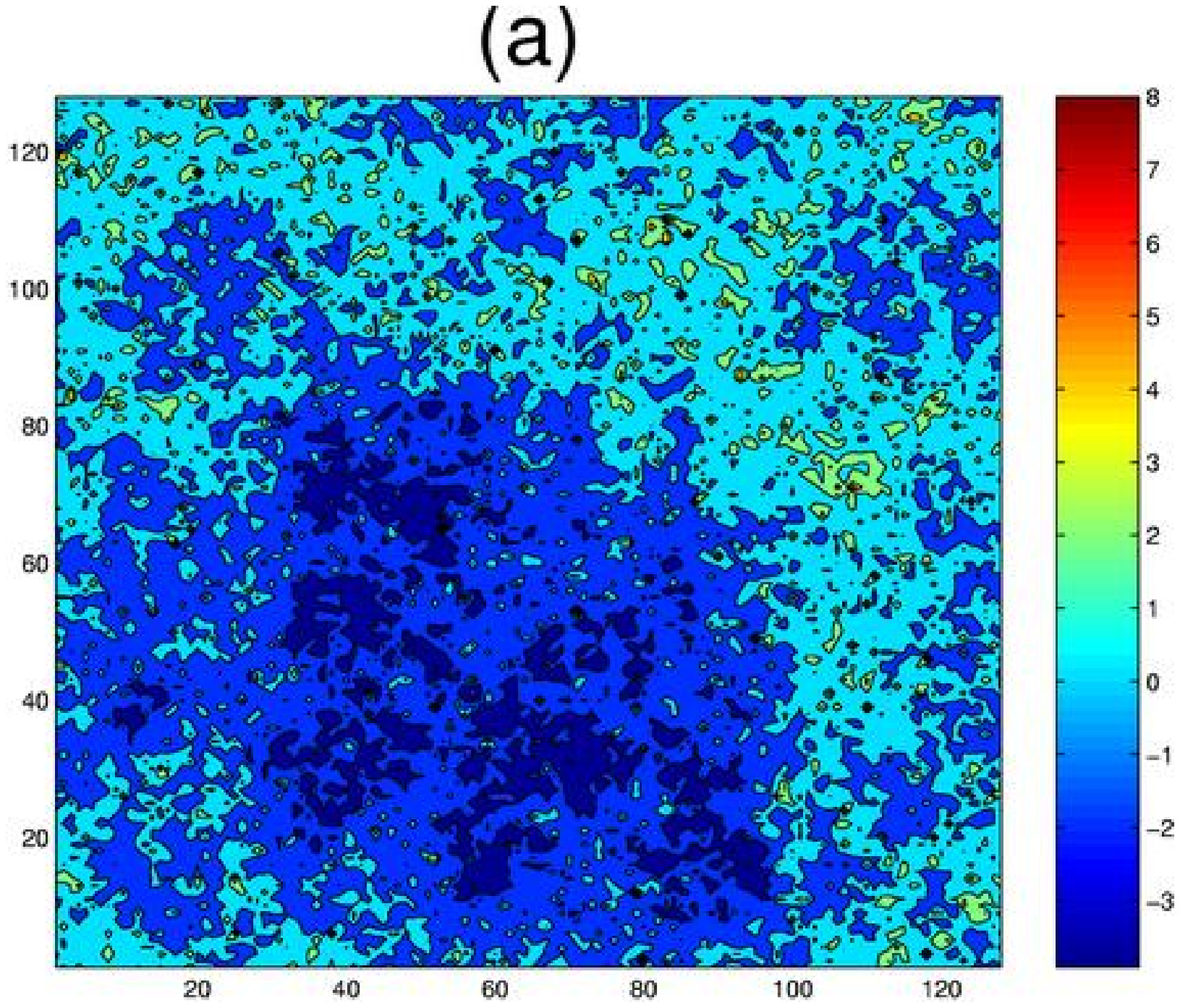}
\includegraphics{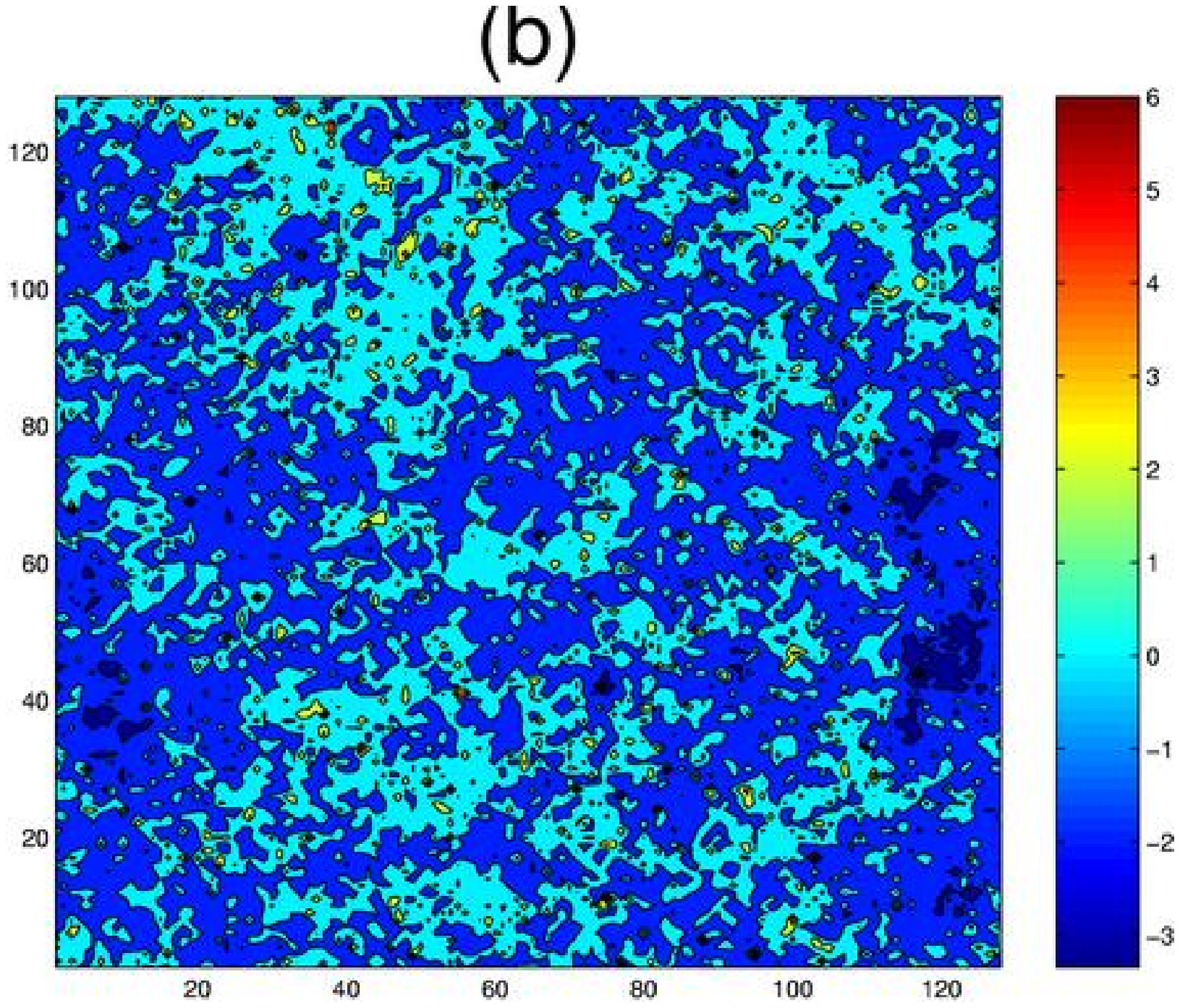}
\includegraphics{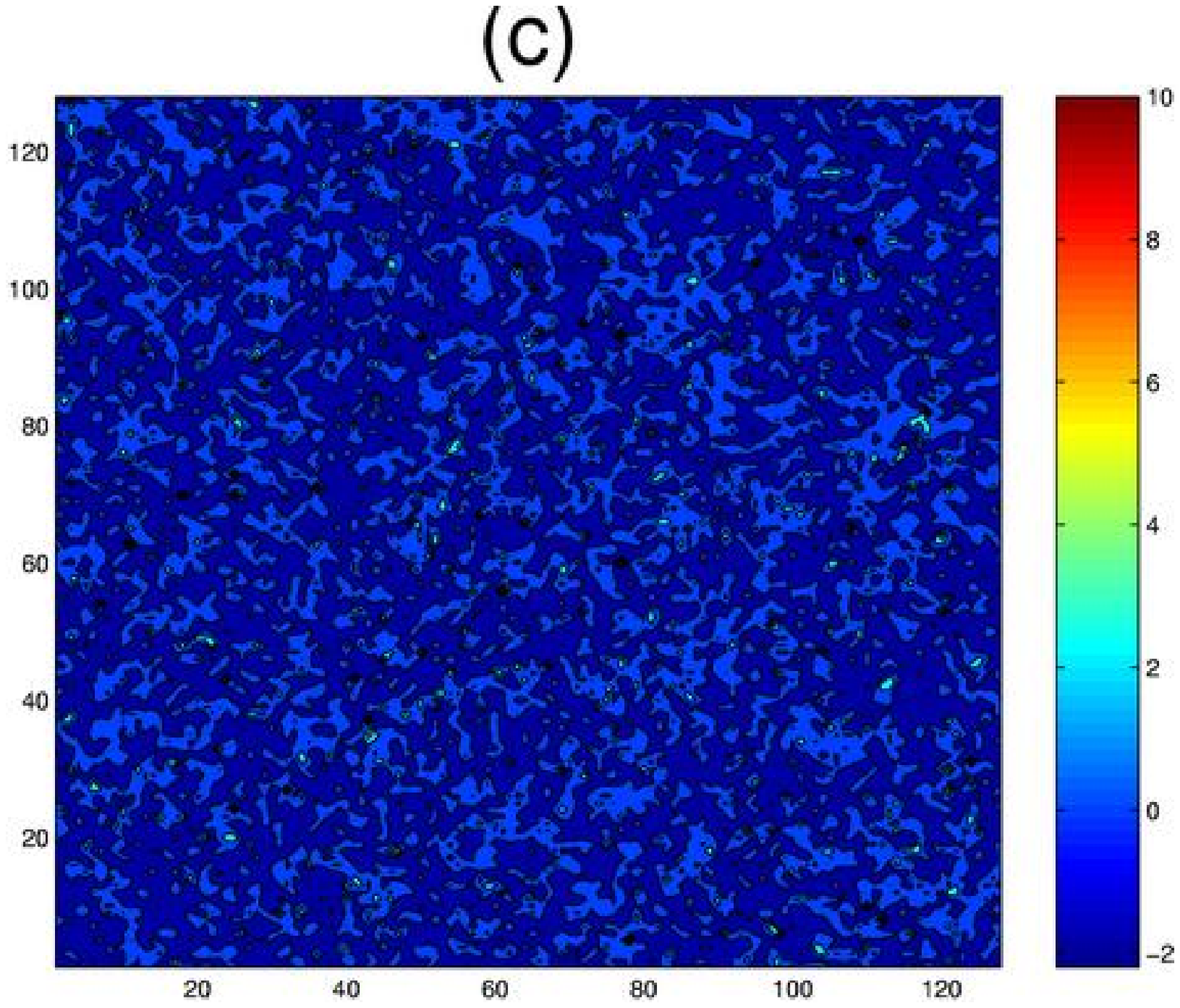}
\vspace{-2cm}
\caption[Synchronization surfaces for BCS in 2D]{Synchronization landscapes as contour plots for the 2D BCS on regular and SW
networks of 128x128 nodes. (a) for the BCS scheme on a regular lattice with only
nearest-neighbor connections (equivalent to $p$=$0$); (b) for $p$=$0.1$; (c) for $p$=$1.0$.}
\vspace{-0.5cm}
\label{fig_2d-surface}
\end{figure}
%%%%%%%%%%%%%%%%%%%%%%%%%%%%%%%%%%%%%%%%%%%%%%%%%%%%%%%%%%%%%%%%%%%%%%%%%%%%%%%%%%
contour plot of the simulated time surface for BCS scheme in 2D.
In 2D as well, we observe kinetic roughening of the BCS scheme.
The simulated time surface for a finite system roughens with time
in a power-law fashion. It then saturates after some system-size
dependent crossover time to its system-size dependent steady-state
value, as shown in [Fig.~\ref{fig_2d-w2}(a)]. Our estimate for the
growth exponent in the early-time regime is $\beta$$=$$0.125$,
significantly smaller than that of one dimension.

%%%%%%%%%%%%%%%%%%%%%%%%%%%%%%%%%%%%%%%%%%%%%%%%%%%%%%%%%%%%%%%%%%%%%%%%%%%%%%%%%%%%%%
\begin{figure}[htbp]
\vspace{6.5cm}
\includegraphics{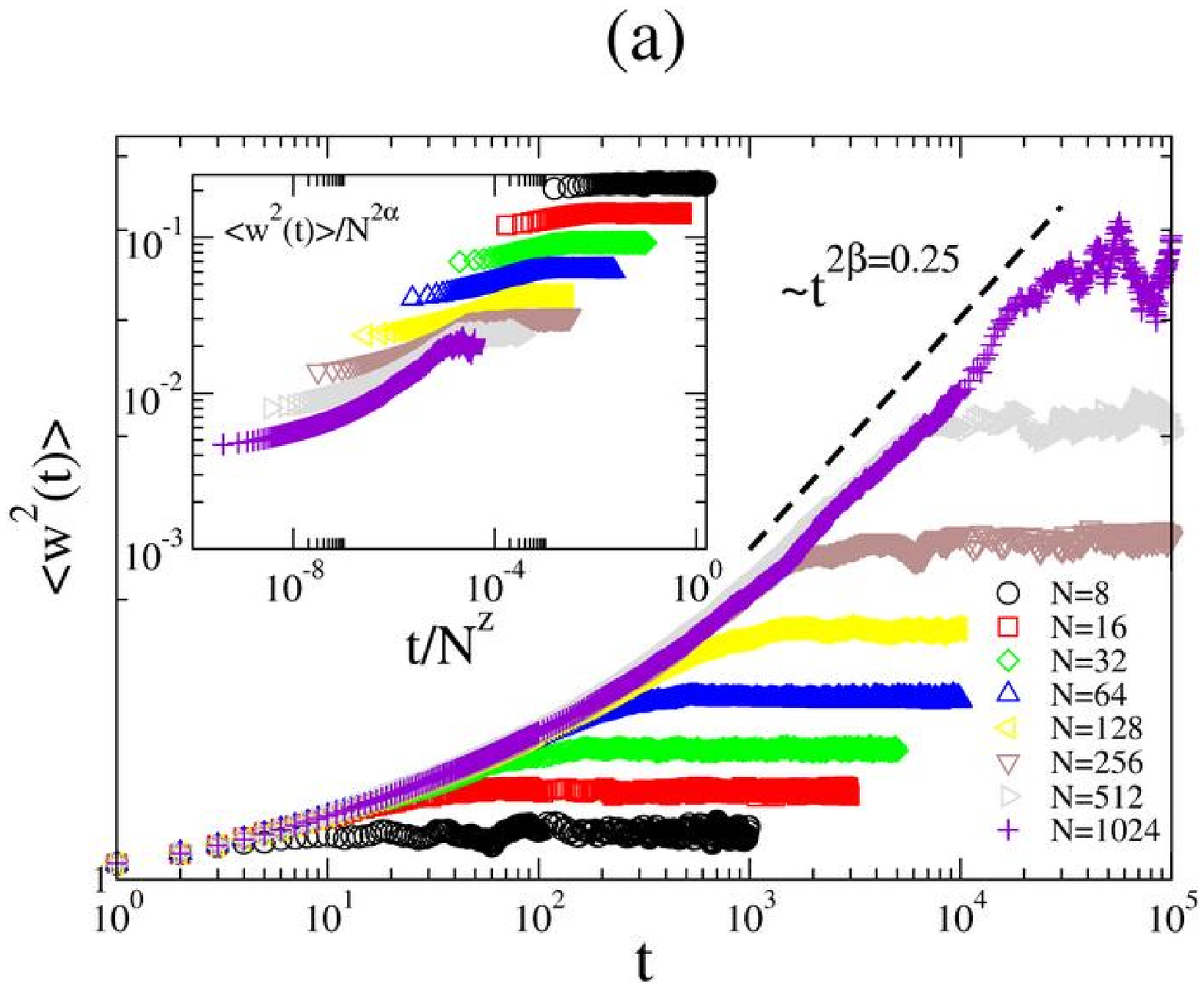}
\includegraphics{2D-KPZ-w2L.eps}
\vspace{-0.3cm}
\caption[Time evolution and scaling of the width for BCS in 2D]{(a) Time evolution of the width in 2D BCS scheme. The
dashed line indicates the power-law behavior of the width before saturation with a
growth exponent $\beta$$\approx$$0.125$. The inset shows the scaled plot
$\left\langle w^2\right\rangle/N^{2\alpha}$ vs. $t/N^z$.
(b) Steady-state width of the 2D BCS scheme as a function of the linear system-size.
The dashed line corresponds to the asymptotic 2D KPZ scaling with roughness exponent
$2\alpha$$=$$0.78$ as obtained by high-precision simulations of the RSOS model
\cite{MARINARI00}. Note the log-log scales.}
\vspace{-0.5cm}
\label{fig_2d-w2}
\end{figure}

The roughness exponent $\alpha$ for KPZ-like systems have been
measured and estimated in a number of experiments and simulations
\cite{BARABASI95}. Since exact exponents for the higher-dimensional
KPZ universality class are not available, for reference, we
compare our results to a recent high-precision simulation study by
Marinari et al. \cite{MARINARI00} on the restricted solid-on-solid
(RSOS) model \cite{KIM89}, a model believed to belong to the KPZ
class. They found in \cite{MARINARI00} that $\alpha$$\simeq$$0.39$ for the 2D
RSOS roughness exponent. While our simulations of the virtual time
horizon show kinetic roughening in Fig.~\ref{fig_2d-w2}(a), the
scaled plot, suggested by Eq.~(\ref{vicsek}), indicates very
strong corrections to scaling  for the BCS in 2D (inset). Figure
~\ref{fig_2d-w2}(b) and Fig.\ref{fig_2d-w2dist} also indicates that the (KPZ) scaling
regime is approached very slowly in the steady-state, which is not completely
unexpected: for the 1D BCS scheme as well, convergence to the
steady-state roughness exponent [Fig.~\ref{fig_kpz1dw2Lt}(a)] and to
the KPZ width distribution [Fig.~\ref{fig_kpz1dw2Lt}(b)] only
appears for linear system sizes $N>{\cal O}(10^3)$. Here, for the
2D case, the asymptotic roughness scaling [Fig.~\ref{fig_2d-w2}(b)]
and width distribution [Fig.~\ref{fig_2d-w2dist}] has not been
%%%%%%%%%%%%%%%%%%%%%%%%%%%%%%%%%%%%%%%%%%%%%%%%%%%%%%%%%%%%%%%%%%%%%%%%%%%%%%%%%%%%%%
\begin{figure}[htbp]
\vspace{8.3cm}
\includegraphics{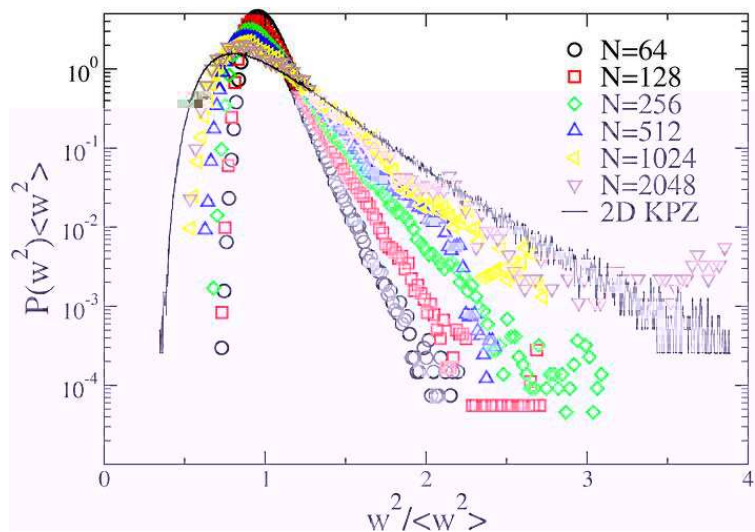}
\vspace{-0.5cm}
\caption[Scaled width distributions for BCS in 2D]{The scaled width distributions for the 2D BCS scheme.
The solid curve is the asymptotic 2D KPZ scaled width distribution, again from
high-precision RSOS simulations \cite{MARINARI02}. Note the log-normal scales.}
\vspace{-0.5cm}
\label{fig_2d-w2dist}
\end{figure}
reached for the system sizes we could simulate (up to linear
system size $N$$=$$4096$). Nevertheless the trend in the finite-size
behavior, and the identical microscopic rules (simply extended to
2D) suggest that 2D BCS landscape belongs to the 2D KPZ universality
class.

For further evidence, we also constructed the structure factor for the 2D BCS
steady-state landscape. As shown in Fig.~\ref{fig_2d-kpz-sf}(a),
$S(k_x,k_y)$ exhibits a strong singularity about ${\bf k}$$=$${\bf 0}$. 
%%%%%%%%%%%%%%%%%%%%%%%%%%%%%%%%%%%%%%%%%%%%%%%%%%%%%%%%%%%%%%%%%%%%%%%%%%%%%%%%%%%%%%%%%%%%
\begin{figure}[htbp]
\vspace{5.5cm} 
\includegraphics{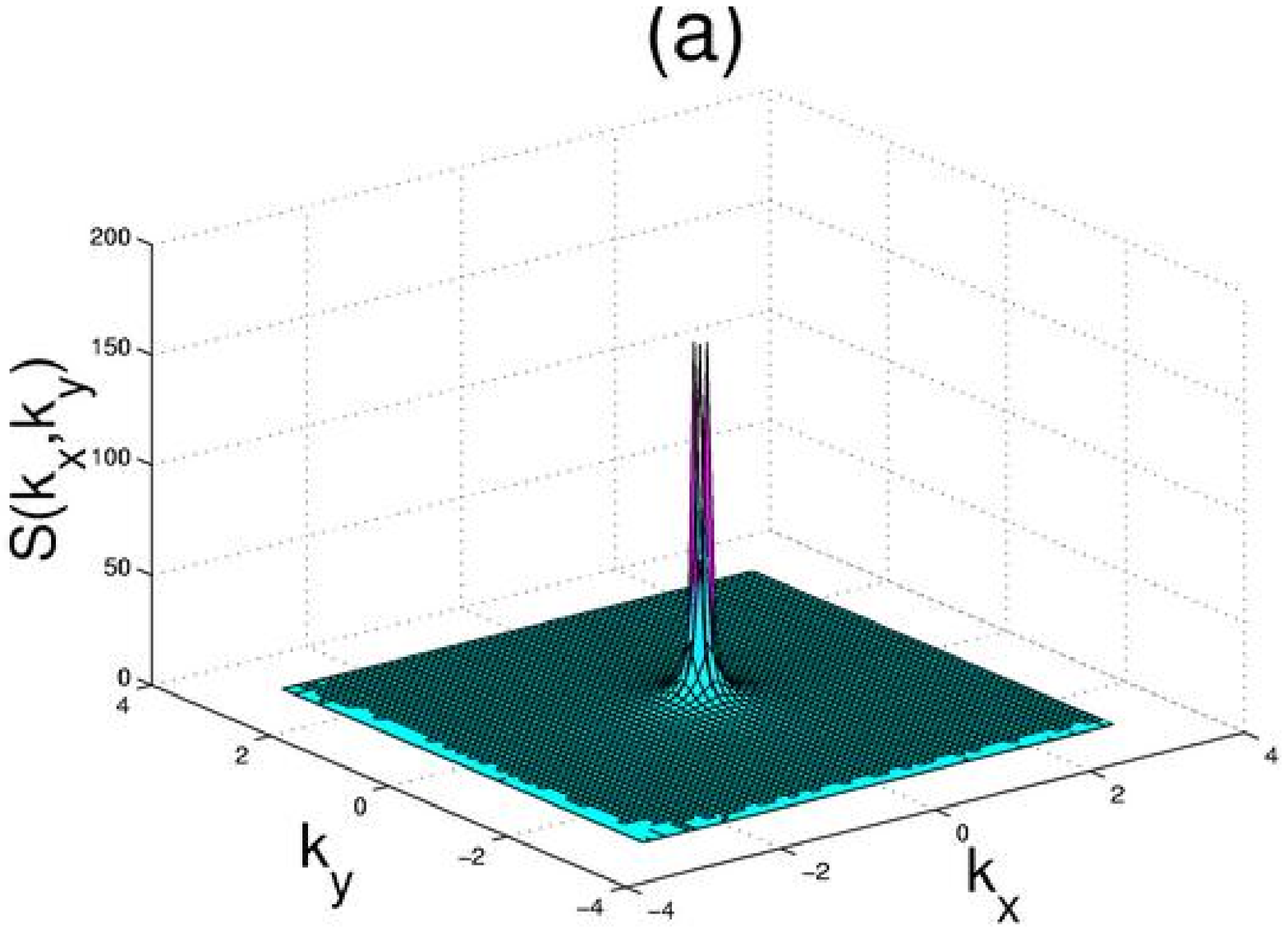}
\includegraphics{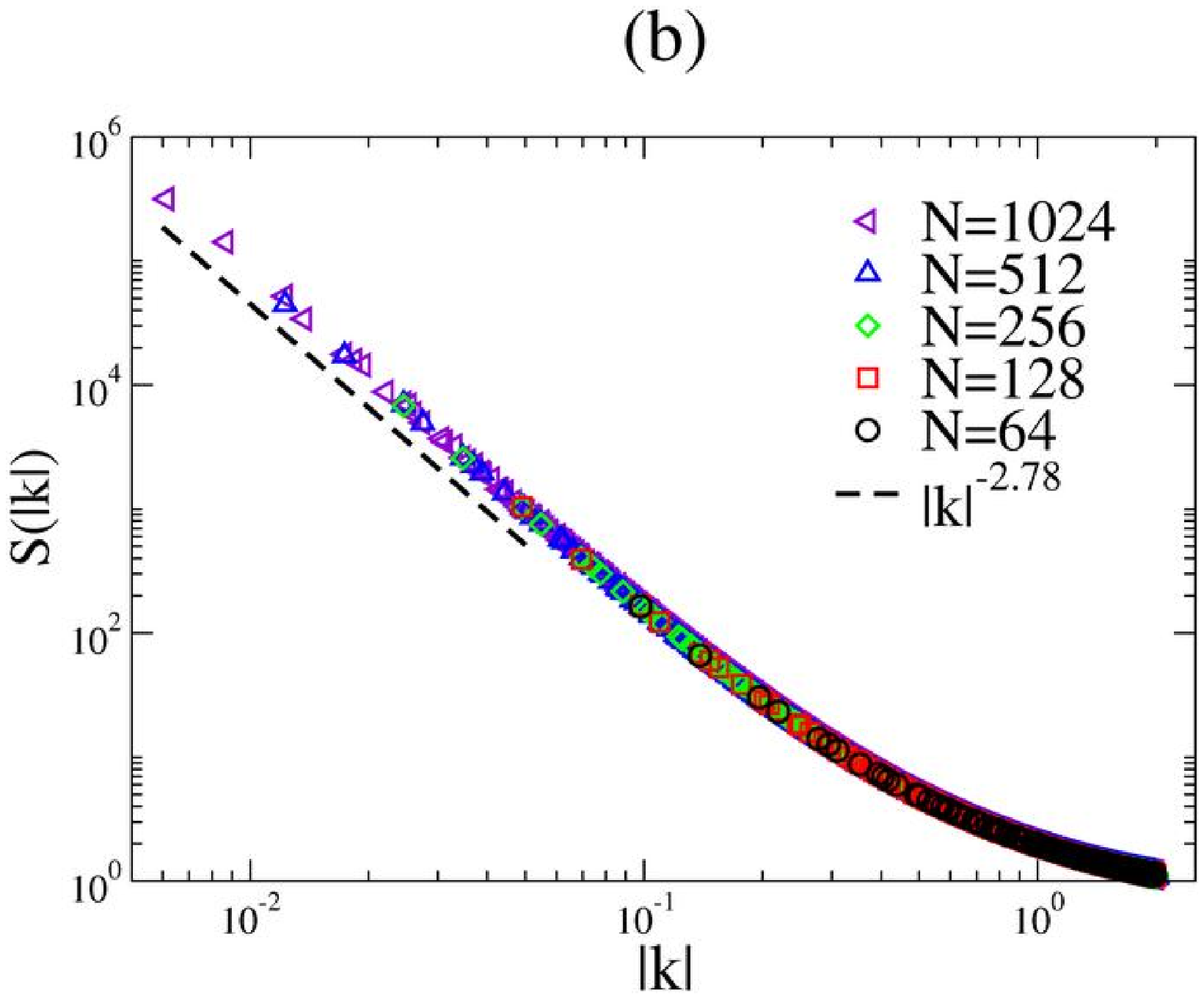}
\vspace{-0.4cm}
\caption[Structure factor for BCS in 2D]{Structure factor for the BCS scheme in 2D.
(a) as a function of the wave number components $k_x$ and $k_y$.
(b) as a function of the magnitude of the wave number for
different system-sizes. The dashed line shows the asymptotic 2D KPZ behavior
for small values of $|\textbf{k}|$ [Eq.~(\ref{S_k_2D_KPZ})] with $\alpha$$=$$0.39$ \cite{MARINARI00}.
Note the log-log scales.}
\vspace{-0.5cm}
\label{fig_2d-kpz-sf}
\end{figure}
%%%%%%%%%%%%%%%%%%%%%%%%%%%%%%%%%%%%%%%%%%%%%%%%%%%%%%%%%%%%%%%%%%%%%%%%%%%%%%%%%%%%%%%%%%%%
For further analysis, we exploited the symmetry of
$S(k_x,k_y)$ that it can only depend on $|{\bf k}|=\sqrt{k_{x}^2 +
k_{y}^2}$. Hence, we averaged over all directions having the same
wave number $|{\bf k}|$ to obtain $S(|{\bf k}|)$.  For small wave
numbers we found that it diverges as
%%%%%%%%%%%%%%%%%%%%%%%%%%%%%%%%%%%%%%%%%%%%%%%%%%%%%%%%%%%%%%%%%%%%%%%%%%%%%%%%%%%%%%%%%%%%%
\begin{equation}
S(|{\bf k}|) \sim \frac{1}{|{\bf k}|^{2+2\alpha}} \;,
\label{S_k_2D_KPZ}
\end{equation}
%%%%%%%%%%%%%%%%%%%%%%%%%%%%%%%%%%%%%%%%%%%%%%%%%%%%%%%%%%%%%%%%%%%%%%%%%%%%%%%%%%%%%%%%%%%%%
with $\alpha=0.39$, as shown in Fig.~\ref{fig_2d-kpz-sf}(b). This
is consistent with the small-$k$ behavior of the structure factor
of the 2D KPZ universality class with roughness exponent
$\alpha$$\simeq$$0.39$ \cite{MARINARI00}. As noted above, the scaling of the width
and its distribution exhibited very slow convergence to those of our reference-KPZ system, the
RSOS model \cite{MARINARI00,MARINARI02}. This is likely the effect of the non-universal and
surprisingly large contributions coming from the large-$k$ modes, leading to very strong
corrections to scaling for the system sizes we were able to study in 2D.
Looking directly at the small-$|{\bf k}|$ behavior of $S(|{\bf k}|)$ is
undisturbed by the larger-$|{\bf k}|$ modes,
hence the agreement with the 2D KPZ scaling is relatively good.

%%%%%%%%%%%%%%%%%%%%%%%%%%%%%%%%%%%%%%%%%%%%%%%%%%%%%%%%%%%%%%%%%%%%%%%%%%%%%%%%%%%%%%%%%%%%
\begin{figure}[htbp]
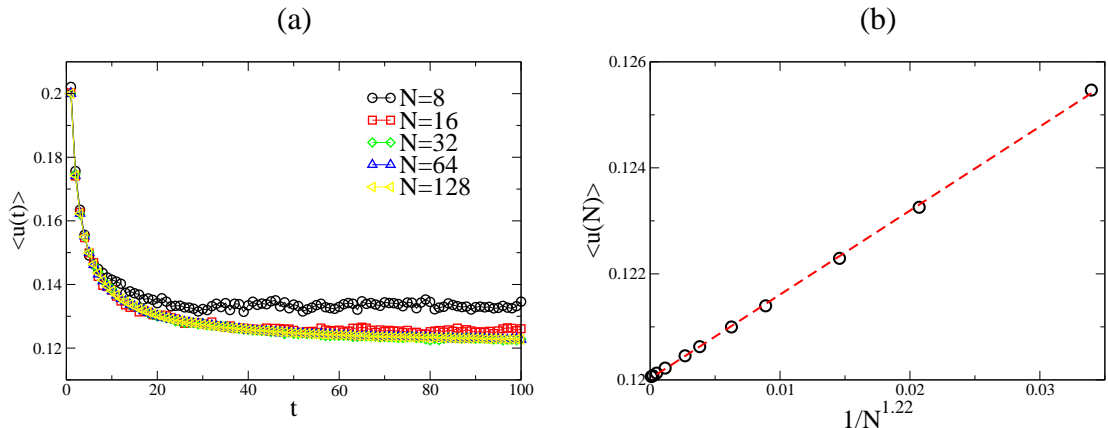

\vspace{5.5cm} 
\includegraphics{2D-KPZ-utilt.eps}
\includegraphics{2D-KPZ-utilN.eps}
\vspace{-0.5cm}
\caption[Time evolution and scaling of the utilization for BCS in 2D]{(a) The time evolution of the steady-state utilization in 2D BCS scheme
for various system sizes.
(b) The steady-state utilization in the 2D BCS scheme as a function of 
$1/N^{2(1-\alpha)}$ as suggested by Eq.~(\ref{utild}) with the 2D KPZ
roughness exponent $\alpha$$=$$0.39$. The dashed line is a linear fit,
$\langle u(N)\rangle \approx 0.1201 + 0.1585/N^{1.22}$.}
\vspace{-0.5cm}
\label{fig_2d-kpzutil}
\end{figure}
%%%%%%%%%%%%%%%%%%%%%%%%%%%%%%%%%%%%%%%%%%%%%%%%%%%%%%%%%%%%%%%%%%%%%%%%%%%%%%%%%%%%%%%%%%%%

The steady-state utilization (density of local minima) in the 2D
BCS synchronization landscape approaches a nonzero value in the
limit of infinite number of nodes, $\langle u(\infty) \rangle
\simeq 0.1201$ as can be seen in Fig.~\ref{fig_2d-kpzutil}(a). 
This is consistent with the general {\em approximate}
behavior $\langle u(\infty) \rangle\simeq const./d$ on hypercubic
lattices in $d$ dimension \cite{GREENBERG96,KORNISS01}, i.e.,
$\langle u(\infty) \rangle$ is approximately inversely proportional to the
coordination number. The system-size dependence of the steady state 
utilization in the 2D BCS also follows Eq.~(\ref{utild}). As shown 
in Fig.~\ref{fig_2d-kpzutil}(b), for a two-dimensional BCS scheme, the utilization
becomes 
\begin{equation}
\langle u(N)\rangle \simeq \langle u(\infty)\rangle + \frac{const.}{N^{1.22}}
\label{2dkpzutil}
\end{equation}
where we have used the 2D KPZ roughness exponent $\alpha$$=$$0.39$ \cite{MARINARI00}.

We have seen that similar to the 1D case, the 2D BCS scheme also exhibits a finite
progress rate but the width diverges as the system size goes to infinity, hindering
measurement scalability.

\end{section}

\section{The K-random Synchronization Network}

\begin{figure}[htb]
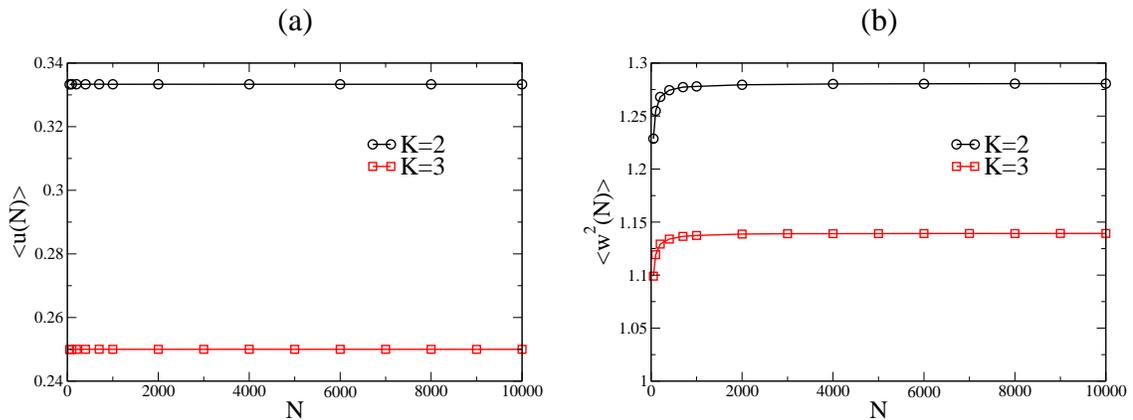

\vspace{6cm} 
\includegraphics{1D-Krandom-utilN.eps}
\includegraphics{1D-Krandom-w2N.eps}
\vspace{-0.5cm}
\caption[Utilization and width for K-random network]{The steady-state observables in $K$-random network.
(a) utilization as a function of system size. (b) width as a function of system size.}
\vspace{-0.5cm}
\label{fig_1d-krandom}
\end{figure}

In order to obtain an analytically tractable scalability model for
the BCS, Greenberg et al introduced the $K$-random interaction
network model \cite{GREENBERG96}. In this model at each update attempt
PEs compare their local simulated times to those of a set of $K$
\textit{randomly} chosen PEs. This set is rechosen for each update
attempt (i.e., the network is ``annealed"), even if a previous
update attempt has failed. It was shown that in the limit of
$t$$\to$$\infty$ and $N$$\to$$\infty$, the utilization (or the
average rate of progress) converges to a \textit{non-zero} constant,
$1$$/$$(K$$+$$1)$ [Fig.~\ref{fig_1d-krandom}(a)]. They also suggested that the scaling properties of
$K$-random model as $t$$\to$$\infty$ and $N$$\to$$\infty$ are
universal and hold for regular lattices as well. But changing the
interaction topology from the nearest neighbor PEs on a regular lattice to
randomly chosen PEs changes the universality class of the time
horizon. Simply, the underlying topology has a crucial
effect on the universal behavior of the time horizon. The random
(annealed) interaction topology of the $K$-random model results in
a mean-field-like behavior, where the simulated time surface is
uncorrelated and has a finite width in the limit of an infinite
number of PEs [Fig.~\ref{fig_1d-krandom}(b)]. Their conjecture for the width does not hold, thus,
the BCS scheme for \textit{regular lattices} cannot be
equivalently described by the $K$-random model (at least not below the
upper critical dimension of the KPZ universality class
\cite{MARINARI02}). 
%%%%%%%%%%%%%%%%%%%%%%%%%%%%%%%%%%%%%%%%%%%%%%%%%%%%%%%%%%%%%%%%%%%%%%%%%%%%%%%%%%%%%%%%%%%%
%%%%%%%%%%%%%%%%%%%%%%%%%%%%%%%%%%%%%%%%%%%%%%%%%%%%%%%%%%%%%%%%%%%%%%%%%%%%%%%%%%%%%%%%%%%%

However, we were inspired by \cite{GREENBERG96} to
change the communication topology of the PEs by introducing random
links \textit{in addition} to the necessary short-range
connections. In the next chapter we present our modification to
the original conservative scheme on regular lattices to achieve a
fully scalable algorithm where both scalability conditions are
satisfied.

%% file: rpichap3.tex
%%%%%%%%%%%%%%%%%%%%%%%%%%%%%%%%%%%%%%%%%%%%%%%%%%%%%%%%%%%%%%%%%%% 
%                                                                 %
%                            CHAPTER THREE                        %
%                                                                 %
%%%%%%%%%%%%%%%%%%%%%%%%%%%%%%%%%%%%%%%%%%%%%%%%%%%%%%%%%%%%%%%%%%% 
 
\chapter{SYNCHRONIZATION IN SMALL-WORLD NETWORKS}

The divergent width of the synchronization landscapes in regular networks 
for very large systems, as discussed in the previous chapter,
is the result of the divergent lateral correlation length $\xi$ of the virtual
time surface reaching the system size $N$ in the steady-state
\cite{BARABASI95,KORNISS02,KOLAKOWSKA03,KOLAKOWSKA03_2,KOLAKOWSKA04}. To de-correlate the
simulated time horizon, first, we modify the virtual communication topology of the
PEs. The resulting communication network must include the original short-range
(nearest-neighbor) connections to faithfully simulate the dynamics of the
underlying system. In the modified network, the connectivity of the nodes
(the number of links of a node) should remain non-extensive (i.e., only a finite number
of virtual neighbors per node is allowed). This is in accordance with our desire
to design a PDES scheme where no global intervention or synchronization is employed
(PEs can only have $O(1)$ communication exchanges per step). It is clear that the
added synchronization links (or at least some of them) have to be long range.
Short range links alone would not change the universality class and the scaling
properties of the width of the time horizon. One can satisfy this condition
by selecting the additional links (called small-world links) randomly among all
the nodes in the network. Also, fluctuations in the individual
connectivity should be avoided for load balancing purposes, i.e., requiring
the same number of added links (e.g., one) for each node is a reasonable constraint.

\begin{figure}[htbp]
\vspace{4.5cm}
\includegraphics{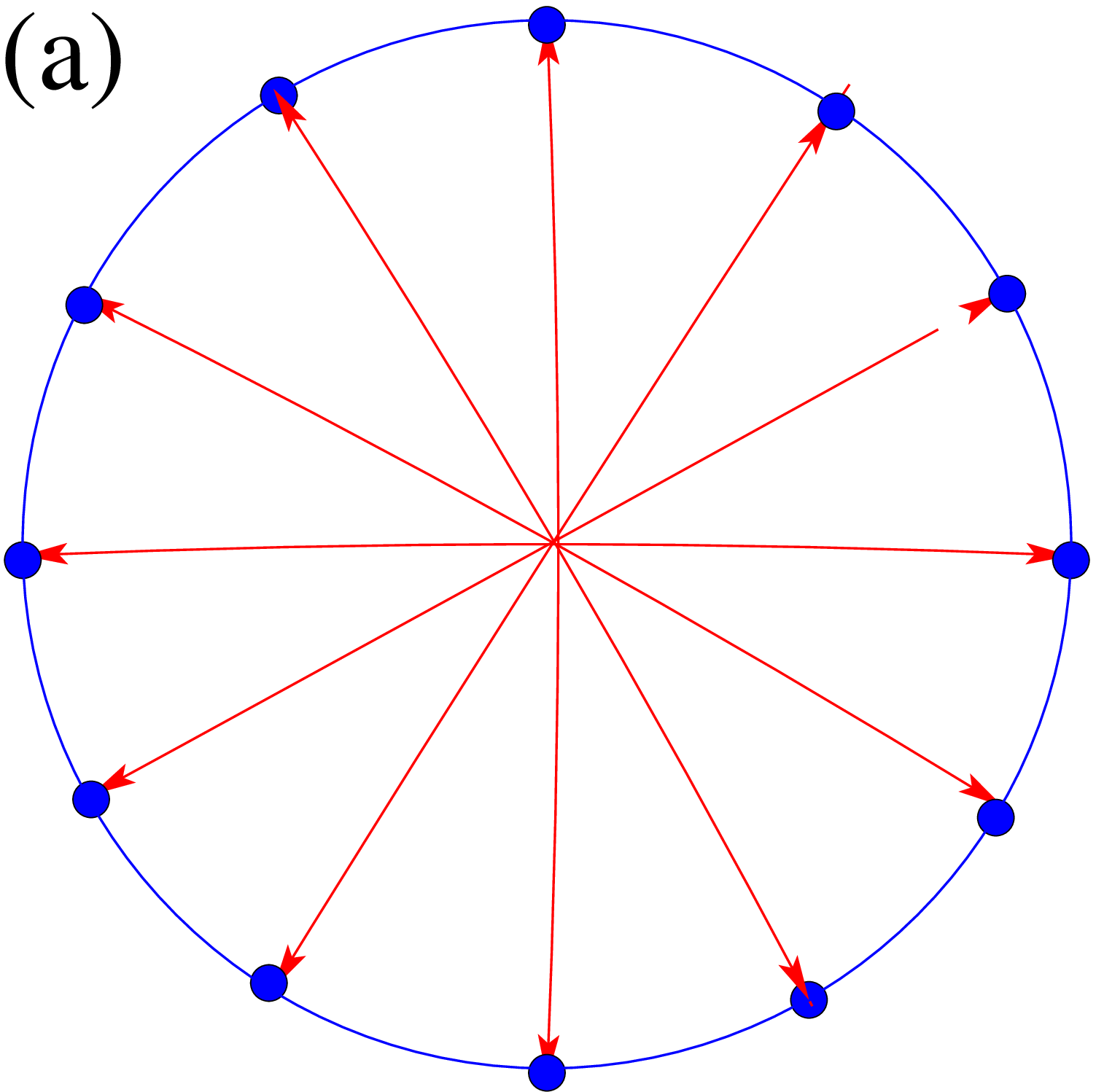}
\includegraphics{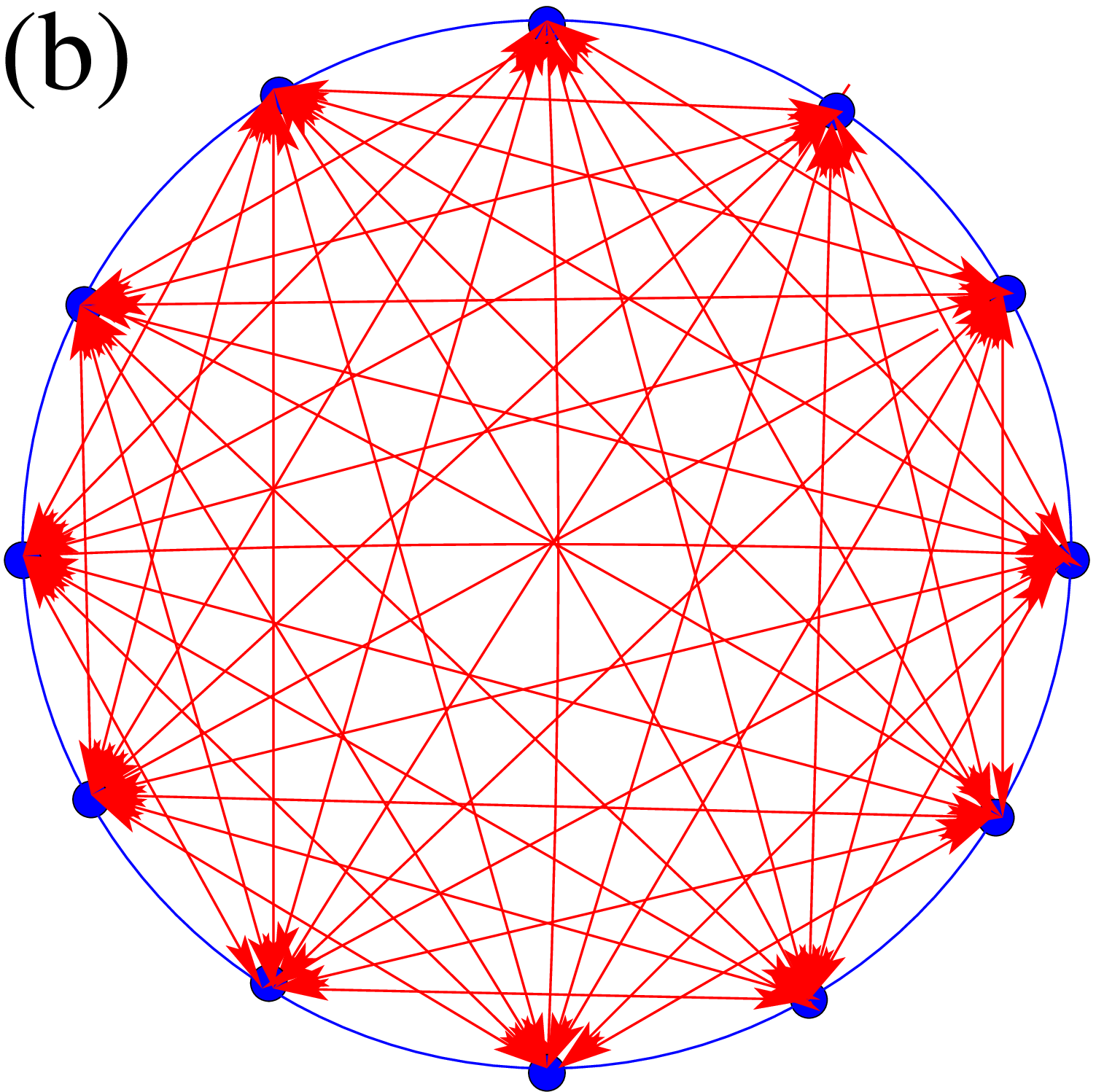}
\includegraphics{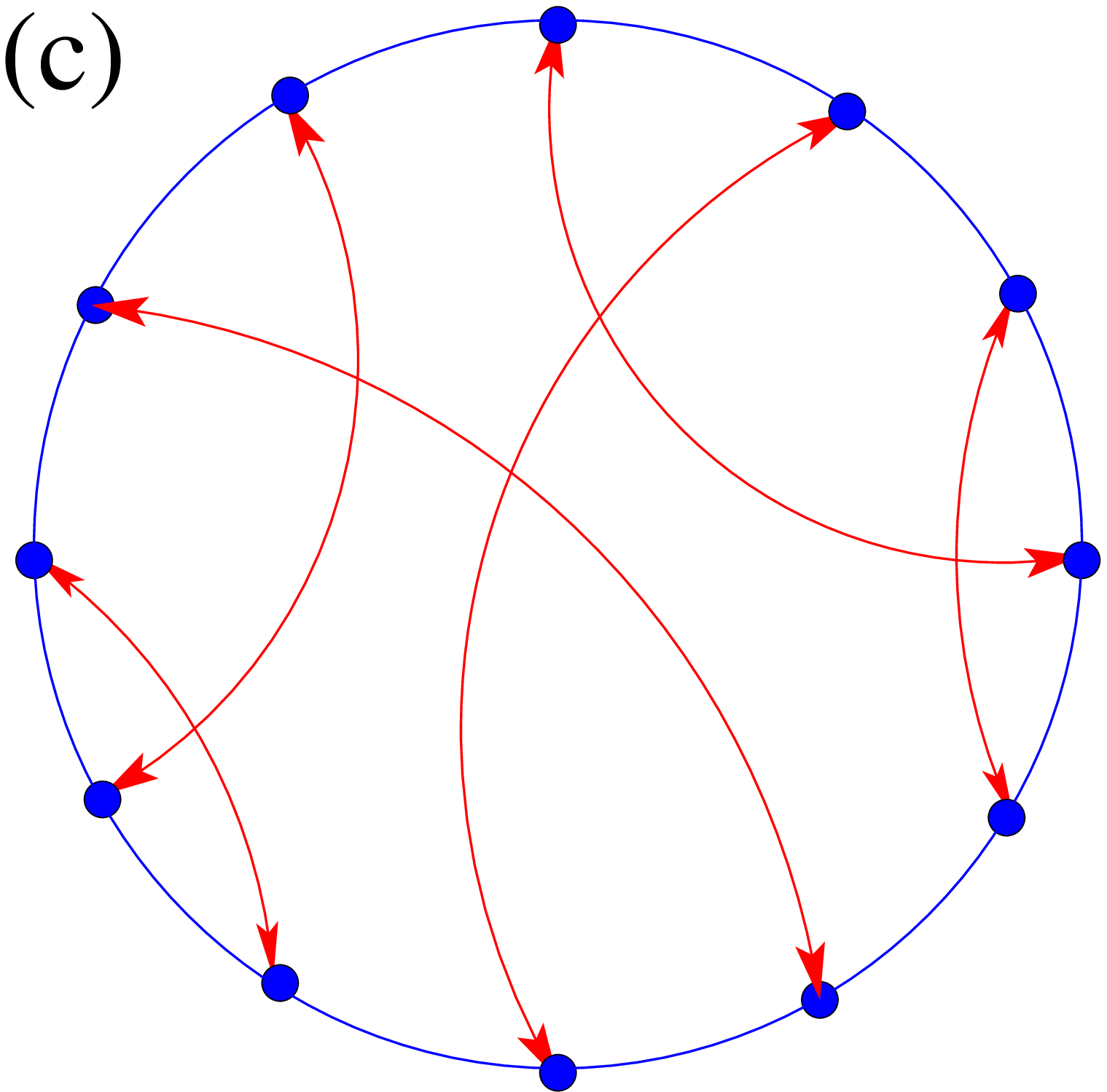}
\caption[Different communication topologies in 1D]{(a) Maximal-distance network in 1D where each node is connected to 
nearest neighboring nodes and to the node which is at the maximum distance.
(b) Fully connected network where each node is connected to every other node in the network.
(c) Small-world (SW) synchronization network in 1D, where each node is connected to a randomly chosen
one in addition to the nearest neighbors.}
\vspace{-0.5cm}
\label{fig_qrm_1dmodel}
\end{figure}

One may wonder how the collective behavior of the PDES scheme would change if
each node was connected to the one located at the ``maximum''
possible distance away from it ($N/2$ on a ring)
[Fig.~\ref{fig_qrm_1dmodel}(a)]. 
Consider a linear
coarse-grained Langevin equation with Gaussian noise where the
effective strength of the added long-range links is $\gamma$,
%%%%%%%%%%%%%%%%%%%%%%%%%%%%%%%%%%%%%%%%%%%%%%%%%%%%%%%%%%%%%%%%%%%%%%
\begin{equation}
\partial_{t} \tau_{i}(t) =
(\tau_{i+1}+\tau_{i-1}-2\tau_{i}) - \gamma(\tau_{i}-\tau_{i+N/2}) + \eta_{i}(t)
\;,
\label{maxLangeq}
\end{equation}
%%%%%%%%%%%%%%%%%%%%%%%%%%%%%%%%%%%%%%%%%%%%%%%%%%%%%%%%%%%%%%%%%%%%%%
with periodic boundary conditions. Since Eq.~(\ref{maxLangeq}) is translationally
invariant, Fourier transformation decouples the equations for different wave numbers $k$
and one obtains for the steady-state structure factor (see Appendix A)
\begin{equation}
S(k) = \frac{D}{2[1-\cos(k)]+\gamma[1-\cos(kN/2)]} 
\;,
\label{sf_maxdist}
\end{equation}
where $k=(2\pi n)/N$, $n=0,1,2,\ldots,N\!\!-\!\!1$ as before (and N is even for
simplicity). Then for the average width we find
%%%%%%%%%%%%%%%%%%%%%%%%%%%%%%%%%%%%%%%%%%%%%%%%%%%%%%%%%%%%%%%%%%%%%%
\begin{equation}
\langle w^2 \rangle =
\frac{1}{N}\sum_{k\neq 0} S(k) =
\frac{1}{N}\sum_{k\neq 0}\frac{D}{2[1-\cos(k)]+\gamma[1-\cos(kN/2)]} \;.
\label{sf_width}
\end{equation}
%%%%%%%%%%%%%%%%%%%%%%%%%%%%%%%%%%%%%%%%%%%%%%%%%%%%%%%%%%%%%%%%%%%%%%
 Separating the terms with even and odd
$n$ values above, we find
%%%%%%%%%%%%%%%%%%%%%%%%%%%%%%%%%%%%%%%%%%%%%%%%%%%%%%%%%%%%%%%%%%%%%%
\begin{eqnarray}
\langle w^2 \rangle & = &
\frac{1}{N}\sum_{n = {\rm odd}} \frac{D}{2[1-\cos(2\pi n/N)]+2\gamma}
\nonumber \\
& + & \frac{1}{N}\sum_{n = {\rm even}}\frac{D}{2[1-\cos(2\pi n/N)]} \;.
\end{eqnarray}
%%%%%%%%%%%%%%%%%%%%%%%%%%%%%%%%%%%%%%%%%%%%%%%%%%%%%%%%%%%%%%%%%%%%%%
The first sum yields a finite $N$-independent value in the
$N$$\to$$\infty$ limit. The second sum, on the other hand, is identical to
the width of the EW model on a regular network of size $N/2$. Thus, in
the large $N$ limit the width for the ``maximal-distance'' connected
network [Fig.~\ref{fig_qrm_1dmodel}(a)] diverges as $\langle w^2(N)
\rangle$$\simeq$$DN/24$. Indeed, one can realize, that such
regularly patterned long-range links make the network equivalent to a
$2$$\times$$(N/2)$ quasi one-dimensional system with only
nearest-neighbor interactions and helical boundary conditions.
The above extreme case suggests, that the maximal-distance
synchronization network cannot work either.
%%%%%%%%%%%%%%%%%%%%%%%%%%%%%%%%%%%%%%%%%%%%%%%%%%%%%%%%%%%%%%%%%%%%%

Now instead, consider the scenario where each node is connected to every other node in the network 
by a ``weak'' link, i.e.,
constructing a ``fully'' connected network as shown in Fig.~\ref{fig_qrm_1dmodel}(b). In this
case we can rewrite the Langevin equation in Eq.~(\ref{maxLangeq}) by using an effective
strength of links, $\gamma/N$,
\begin{equation}
\partial_{t} \tau_{i}(t) =
(\tau_{i+1}+\tau_{i-1}-2\tau_{i}) - \frac{\gamma}{N} \sum_{j=1}^{N}(\tau_{i}-\tau_{j}) + \eta_{i}(t)
\;.
\label{fcnLangeq}
\end{equation}
Performing the summation above yields an exact mean-field-like coupling, where
each node is coupled to the average height:
\begin{equation}
\partial_{t} \tau_{i}(t) =
(\tau_{i+1}+\tau_{i-1}-2\tau_{i}) - \gamma(\tau_{i}-\bar{\tau}) + \eta_{i}(t)
\;,
\label{fcnLangeq2}
\end{equation}
where $\bar{\tau}=\sum_{j=1}^{N}\tau_{j}$ is the average height. For the steady-state 
structure factor one finds (see Appendix A)
\begin{equation}
S(k) = \frac{D}{2[1-\cos(k)]+\gamma} \;.
\end{equation}
Then by using the relation between the structure factor and the width [Eq.~\ref{sf_width}]
one obtains 
\begin{equation}
\langle w^2 \rangle =
\frac{1}{N}\sum_{k\neq 0} S(k) =
\frac{1}{N}\sum_{k\neq 0}\frac{D}{2[1-\cos(k)]+\gamma}
\simeq \int_{-\infty }^{\infty}\frac{dk}{2\pi}\frac{D}{k^2+\gamma}=\frac{D}{2\sqrt{\gamma}} \;.
\label{sf_width2}
\end{equation}
The above relation between the mean-field coupling constant $\gamma$ and the width shows that for
a non-zero $\gamma$ the width is finite in the thermodynamic limit.
But connecting each node in PDES to every other node
would be cost-inefficient and cumbersome in terms of communication times. 
As we discuss in the next section, one can construct
an ``effectively'' fully connected and yet cost-efficient network which has a finite width and 
relatively high progress rate by only employing a few random
links.

\section{One-Dimensional Small-World-Connected Synchronization Network}

As we have seen in Chapter 2 our attempts
to make the PDES fully scalable have failed because the PDES on short-range
network is not measurement scalable (width is infinite for an infinite system).
One of the proposed networks discussed in the previous section, the maximal-distance network,
fails as a candidate for a fully scalable synchronization scheme because it is effectively
equivalent to a short-range network. On the other hand, the fully connected network, is very
inefficient in performance although it is measurement scalable. Motivated by
the social networks we propose
a network topology in which each node is connected to exactly one another 
randomly chosen node in addition to the nearest neighbors, resulting in a SW-like synchronization network. 
As we shall see, adding one random link
to every node is cost-efficient and makes the network an ``effectively''
fully-connected one. 

\begin{figure}[htbp]
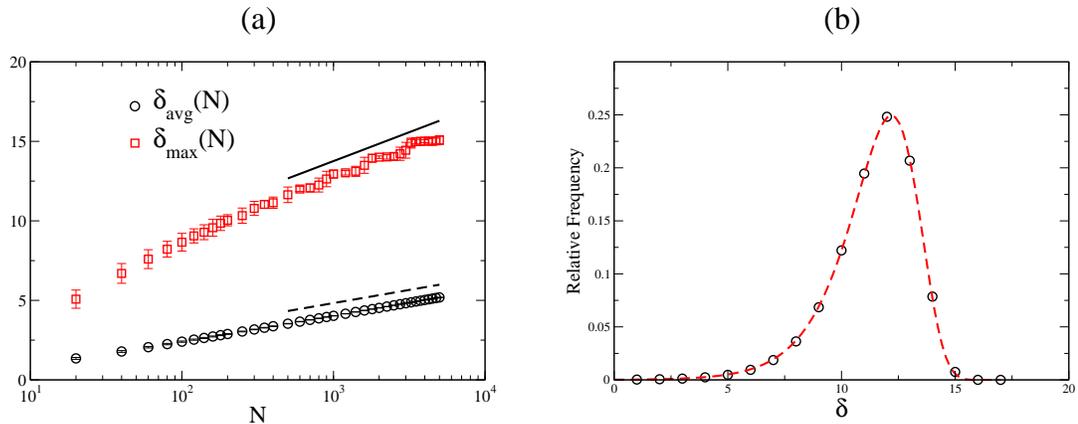

\vspace{5.5cm}
\includegraphics{1D-QRM-sp.eps}
\includegraphics{1D-QRM-sp-hist.eps}
\vspace{-0.2cm}
\caption[Shortest path scaling and distribution for SW network in 1D]{(a) Average and maximum shortest path lengths as a function of
the number of nodes for the SW synchronization network in 1D, as described in the text.
The latter is also referred to as the diameter of the network. The solid and dashed lines both
indicate the logarithmic dependence. Note the normal-log axes.
(b) Histogram for the length of the shortest paths in one realization of the network
with system size $N$$=$$10^4$. The dashed curve is appropriately fitted Gumbel distribution
for minima in the form $f(\delta)=\frac{1}{b}e^{\frac{\delta-a}{b}-e^{\frac{\delta-a}{b}}}$ where 
$a$$\protect\cong $$12.2$ and $b$$\protect\cong $$1.48$.}
\vspace{-0.4cm}
\label{fig_1d-qrm-sp}
\end{figure}

One of the basic structural characteristics of SW-like networks is the ``low degree
of separation" between the nodes. The most commonly used observables to analyze this
property are the average shortest path length, $\delta_{avg}(N)$, and the maximum
shortest path length, $\delta_{max}(N)$. The shortest path length between two nodes is
defined as the minimum number of nodes one has to visit in order to go from one of
the nodes to the other.
The average shortest path length is the average of all these possible shortest 
paths between the nodes in the network.
The maximum shortest path length, also known as \textit{diameter} of the network, is the length of
the longest among the shortest paths in the network. Both of these observables scale logarithmically
with the system-size $N$ in SW-like networks \cite{BOLLOBAS01}. The system-size
dependence of these path lengths for our one-dimensional SW network,
in which we have both nearest neighbors and random SW links, is
logarithmic as expected, see Fig.~\ref{fig_1d-qrm-sp}(a)-(b).

%%%%%%%%%%%%%%%%%%%%%%%%%%%%%%%%%%%%%%%%%%%%%%%%%%%%%%%%%%%%%%%%%%%%%%%%%%%%%%%%%%%%%%%%
%%%%%%%%%%%%%%%%%%%%%%%%%%%%%%%%%%%%%%%%%%%%%%%%%%%%%%%%%%%%%%%%%%%%%%%%%%%%%%%%%%%%%%%%

We now describe the modified algorithmic steps for the SW-connected 
PEs \cite{KORNISS03}. In the PDES on SW synchronization network, 
in every parallel time step each PE with probability $p$
compares its local simulated time with its \textit{full} virtual
neighborhood, and can only advance if it is the minimum in this
neighborhood, i.e., if $\tau_i(t) \leq
\min\{\tau_{i-1}(t),\tau_{i+1}(t),\tau_{r(i)}(t)\}$, where $r(i)$
is the random connection of PE $i$. With probability $(1-p)$ each
PE follows the original scheme, i.e., the PE then can advance if
$\tau_i(t) \leq \min\{\tau_{i-1}(t),\tau_{i+1}(t)\}$. Our network
model including the nearest neighbors and random SW links can be
seen in Fig.~\ref{fig_qrm_1dmodel}(c). Note that the occasional extra
checking of the simulated time of the random neighbor is
\textit{not} needed for the faithfulness of the simulation. It is
merely introduced {\em to control} the width of the time horizon. The
occasional checking of the virtual time of the random neighbor
(with rate $p$) introduces an effective strength $J=J(p)$ for
these links. Note that this is a dynamic ``averaging" process
controlled by the parameter $p$ and can possibly be affected by
nonlinearities in the dynamics through renormalization effects.
The exact form of $J(p)$ is not known. The only plausible
properties we assume for $J$ is that it is a monotonically
increasing function of $p$ and is only zero when $p$=$0$.

In what follows, we focus on the characteristics of the dynamics
on the network. As we have seen for the one-dimensional ring, the
communication protocol between the nodes (up to linear terms)
leads to simple relaxation, governed by the Laplacian on the
regular grid. Random communication links give rise to analogous
effective couplings between the nodes, corresponding to the
Laplacian on the random part of the network. Thus, the large-scale
properties of the virtual time horizon of our SW scheme are
governed by the effective Langevin equation
%%%%%%%%%%%%%%%%%%%%%%%%%%%%%%%%%%%%%%%%%%%%%%%%%%%%%%%%%%%%%%%%%%%%%%%%%%%%%
\begin{equation}
\partial_t \hat{\tau}_i = \nabla^2\hat{\tau}_i
- \sum_j J_{ij}(\hat{\tau}_i-\hat{\tau}_j) + ... + \eta_i(t) \;,
\label{meaf_KPZ}
\end{equation}
%%%%%%%%%%%%%%%%%%%%%%%%%%%%%%%%%%%%%%%%%%%%%%%%%%%%%%%%%%%%%%%%%%%%%%%%%%%%%
where the ... stands for infinitely many non-linear terms
(involving non-linear interactions through the random links as
well), and $J_{ij}$ is proportional to the symmetric adjacency matrix
of the random part of the network: $J_{ij}$$=$$J(p)$ if sites $i$ and
$j$ are connected by a random link and $J_{ij}$$=$$0$ otherwise.
For our specific SW construction each node has exactly one random
neighbor, i.e., there are no fluctuations in the individual
connectivity (degree) of the nodes. Our simulations (to be
discussed below) indicate that when considering the large-scale
properties of the systems, the Laplacian on the random part of the
network generates an effective coupling $\gamma$ to the mean
\cite{KORNISS03}. At the level of the structure factor, it
corresponds to an effective mass $\gamma$ (in a field-theory
sense)
%%%%%%%%%%%%%%%%%%%%%%%%%%%%%%%%%%%%%%%%%%%%%%%%%%%%%%%%%%%%%%%%%%%%%%%%%%%%%
%%%%%%%%%%%%%%%%%%%%%%%%%%%%%%%%%%%%%%%%%%%%%%%%%%%%%%%%%%%%%%%%%%%%%%%%%%%
\begin{equation}
S(k)\propto\frac{1}{\gamma + k^2}\;,
\label{S_k}
\end{equation}
%%%%%%%%%%%%%%%%%%%%%%%%%%%%%%%%%%%%%%%%%%%%%%%%%%%%%%%%%%%%%%%%%%%%%%%%%%%%
where $\gamma$$=$$\gamma(p)$ is a monotonically increasing
function of $p$ with $\gamma(0)$=$0$.

We emphasize that the above is not a derivation of
Eq.~(\ref{S_k}), but rather a ``phenomenological" description of
our findings. It is also strongly supported by exact
asymptotic results for the (linear) EW model on SW networks, where
the effect of the Laplacian on the random part of the network is
to generate a mass \cite{KOZMA04,KOZMA05b}. The averaging over the
quenched network ensemble, however, can introduce nontrivial
scaling and corrections in the effective coupling
\cite{KOZMA04,KOZMA05,KOZMA05b}. In our case, this is further
complicated by the nonlinear nature of the interaction. The
results of ``simulating the simulation'', however, suggest that
the dynamic control of the link strength and nonlinearities only
give rise to a renormalized coupling and a corresponding
renormalized mass. Thus, the dynamics of the BCS scheme with
random couplings is effectively governed by the EW relaxation in a
small-world \cite{KOZMA04,KOZMA05b,KOZMA05}.
From Eq.~(\ref{S_k}) it directly follows that the lateral
correlation length in the infinite system-size limit
%%%%%%%%%%%%%%%%%%%%%%%%%%%%%%%%%%%%%%%%%%%%%%%%%%%%%%%%%%%%%%%%%%%%%%%%%%%%%%
\begin{equation}
\xi\sim \gamma^{-1/2} \;,
\label{xi}
\end{equation}
%%%%%%%%%%%%%%%%%%%%%%%%%%%%%%%%%%%%%%%%%%%%%%%%%%%%%%%%%%%%%%%%%%%%%%%%%%%%%%
i.e., becomes finite for all $p\not$=$0$ [Fig.~\ref{figsn}(b)].
The presence of the effective mass term in the structure factor
Eq.~(\ref{S_k}) implies that $\lim_{k\rightarrow 0}S(k)<\infty$,
that is, there are no large amplitude long-wavelength modes in the
surface. Consequently, the width $\langle w^2\rangle =
(1/N)\sum_{k \not= 0}S(k)$ is also finite. Our simulated time
landscapes indeed show that they become macroscopically smooth
when SW links are employed [Fig.~\ref{figsn}(b)], compared to the
the same dynamics with only short-range links
[Fig.~\ref{figsn}(a)].

%%%%%%%%%%%%%%%%%%%%%%%%%%%%%%%%%%%%%%%%%%%%%%%%%%%%%%%%%%%%%%%%%%%%%%%%%%%%%
\begin{figure}[htbp]
\vspace{5.8cm}
\includegraphics{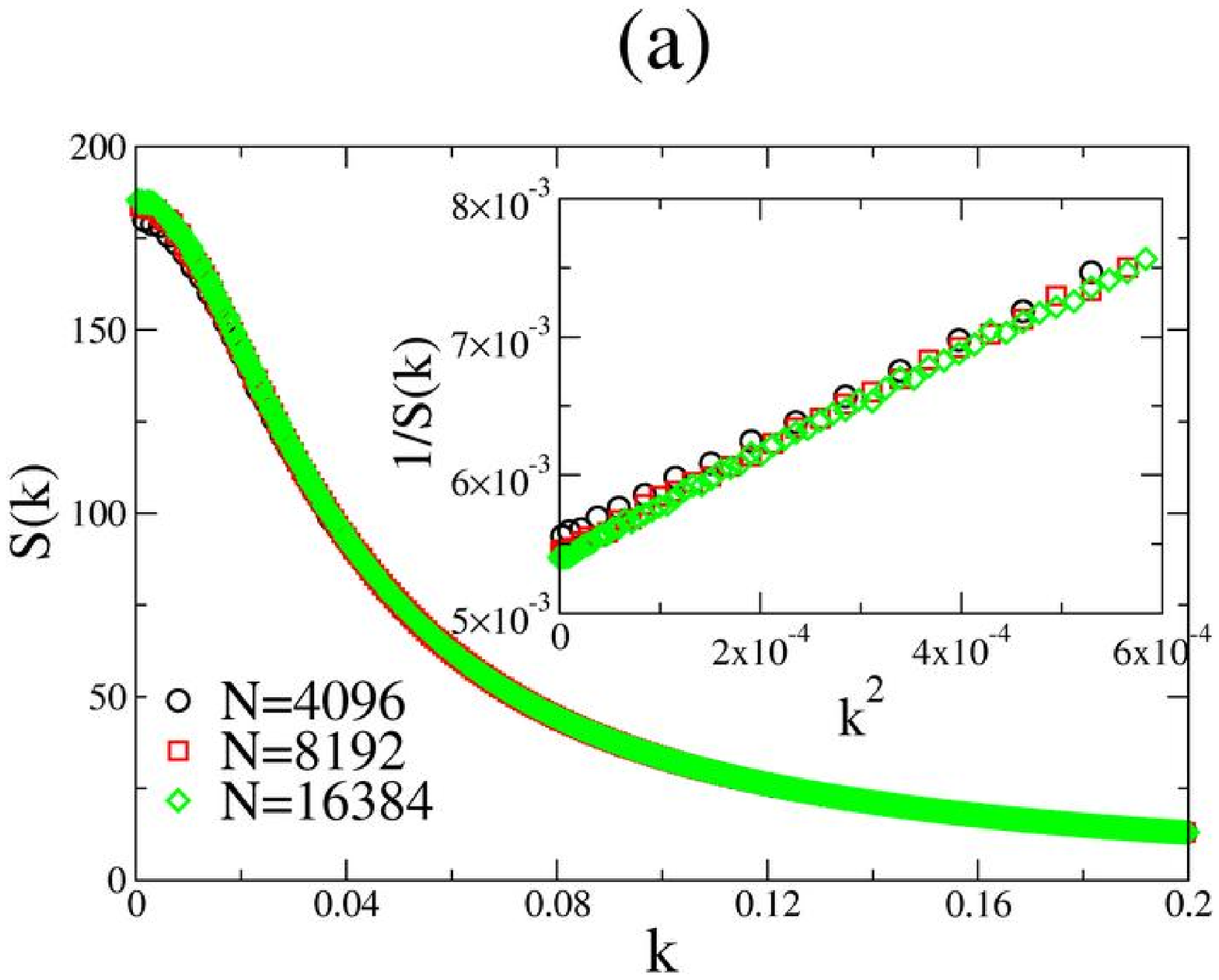}
\includegraphics{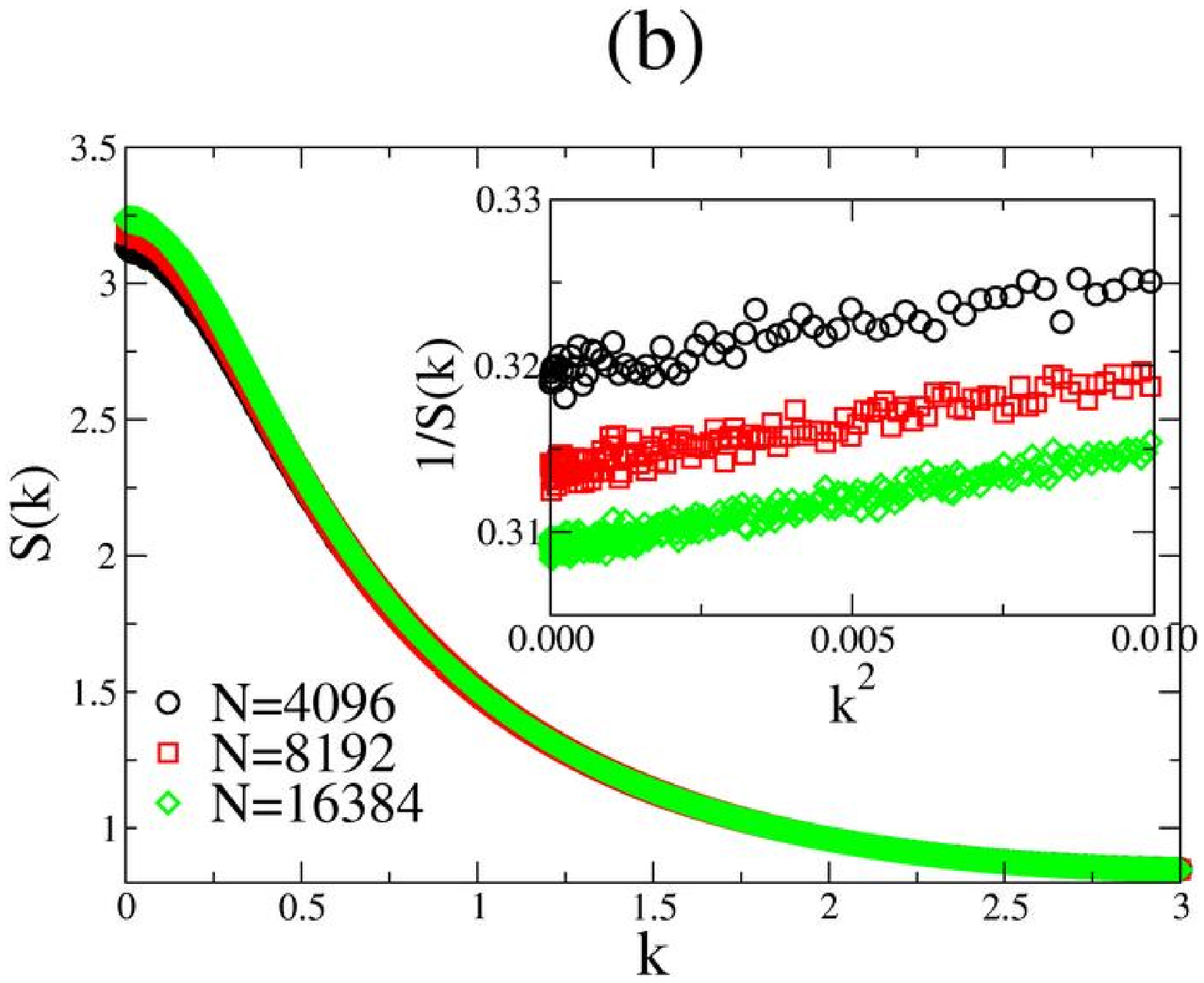}
\vspace{-0.5cm}
\caption[Structure factor for SW network 1D]{Structure factor for the 1D SW synchronization scheme with
(a) $p$=$0.1$ and (b) $p$=$1$. The insets show $1/S(k)$ vs. $k^2$ for small values of $k$, confirming the
coarse-grained prediction Eq.~(\ref{S_k}).}
\vspace{-0.5cm}
\label{fig_sw1dsf}
\end{figure}
%%%%%%%%%%%%%%%%%%%%%%%%%%%%%%%%%%%%%%%%%%%%%%%%%%%%%%%%%%%%%%%%%%%%%%%%%%%%%%%%%%%%%%

In the simulations, we typically performed averages over 10-100 network realizations,
and compared the results to those of individual ones. Our results indicate
that the observables we studied (the width and its distribution,
the structure factor, and the utilization) display strong {\em self-averaging} properties, i.e.,
for large enough systems, they become independent of the particular realization of the
\begin{figure}[htbp]
\vspace{7cm}
\includegraphics{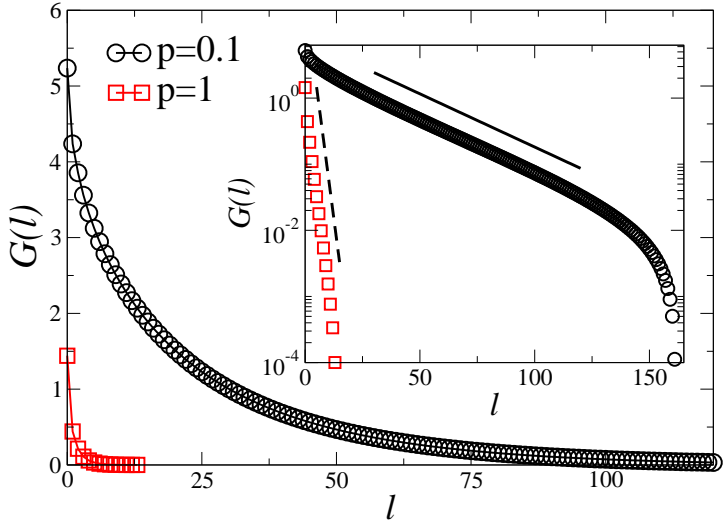}
\vspace{-0.7cm}
\caption[Two-point correlation function for SW network in 1D]{The spatial two-point correlation function as a function of the Euclidean distance $l$
between the nodes for two different values of $p$, indicating an exponential decay with
an average correlation length $\xi$$\approx$$27$ and $\xi$$\approx$$1.6$ for $p$=$0.1$ and $p$=$1.0$,
respectively. The number of nodes is $N$$=$$16$$,$$384$. The inset shows the same data in log-normal scale.}
\vspace{-0.7cm}
\label{fig_sw1dcorr}
\end{figure}
underlying SW network. Simulation results for the structure factor, $S(k)$, for the SW
synchronization scheme are shown in Fig.~\ref{fig_sw1dsf}(a) and (b). If an
infinitesimally small $p$ is chosen, $S(k)$ approaches a finite
constant in the limit of $k$$\rightarrow$$0$, and in turn, the
virtual time horizon becomes macroscopically smooth with a
\textit{finite} width.

%%%%%%%%%%%%%%%%%%%%%%%%%%%%%%%%%%%%%%%%%%%%%%%%%%%%%%%%%%%%%%%%%%%%%%%%%%%%%
\begin{figure}[htbp]
\vspace{7cm}
\includegraphics{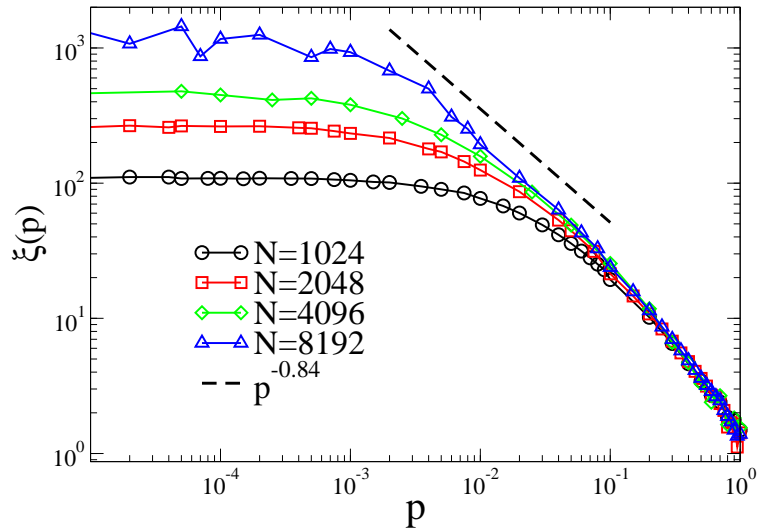}
\vspace{-0.5cm}
\caption[Correlation length for SW network in 1D]{Correlation length ($\xi$) vs. $p$ for different system-sizes. The dashed line
corresponds to the estimate of the exponent $s$, $\xi(p)\sim p^{-s}$ with $s$$\approx$$0.84$,
in the small-$p$ regime for an asymptotically infinite system.}
\vspace{-0.5cm}
\label{fig_swxi}
\end{figure}
%%%%%%%%%%%%%%%%%%%%%%%%%%%%%%%%%%%%%%%%%%%%%%%%%%%%%%%%%%%%%%%%%%%%%%%%%%%%%%%%%%%%%%

%%%%%%%%%%%%%%%%%%%%%%%%%%%%%%%%%%%%%%%%%%%%%%%%%%%%%%%%%%%%%%%%%%%%%%%%%%%%%%%%%%
\begin{figure}[htb]
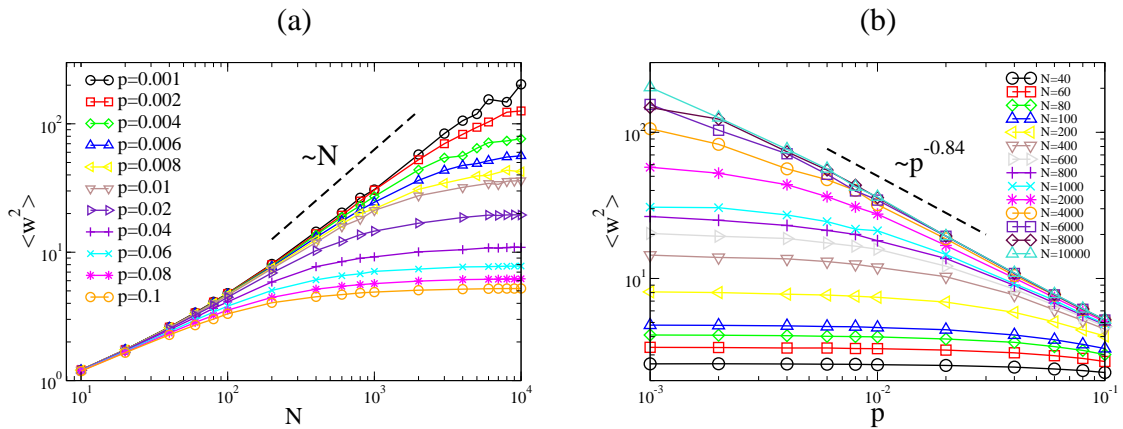

\vspace{5.2cm}
\includegraphics{1D-QRM-w2L-all.eps}
\includegraphics{1D-QRM-w2p-all.eps}
\vspace{-0.4cm}
\caption[Width dependence on $p$ and $N$ for SW network in 1D]{(a) The average steady-state width in the 1D SW synchronization landscape
as a function of the system size for different values of $p$ in the range of $[10^{-3},10^{-1}]$.
The dashed line indicates the EW/KPZ scaling, corresponding to the small system-size behavior.
(b) The average steady-state width in the 1D SW synchronization landscape
as a function of $p$ for different values of $N$ in the range of $[40,10^{4}]$.
The dashed line indicates the best-fit power law in the asymptotic large-$N$ small-$p$ regime
to extract the correlation length exponent $s$, according to
Eqs.~(\ref{new_scaling})-(\ref{xi_scale}).}
\vspace{-0.5cm}
\label{fig_1d-w2-scaling}
\end{figure}

A possible (phenomenological) way to obtain the correlation
length is to fit our structure factor data to Eq.~(\ref{S_k}),
more specifically, by plotting $1/S(k)$ versus $k^2$, which exhibits a
linear relationship. By a linear fit, $\gamma$ is then the ratio
of the intercept and the slope (insets in
Fig.~\ref{fig_sw1dsf}). Alternatively, one can confirm that the
\begin{figure}[ht]
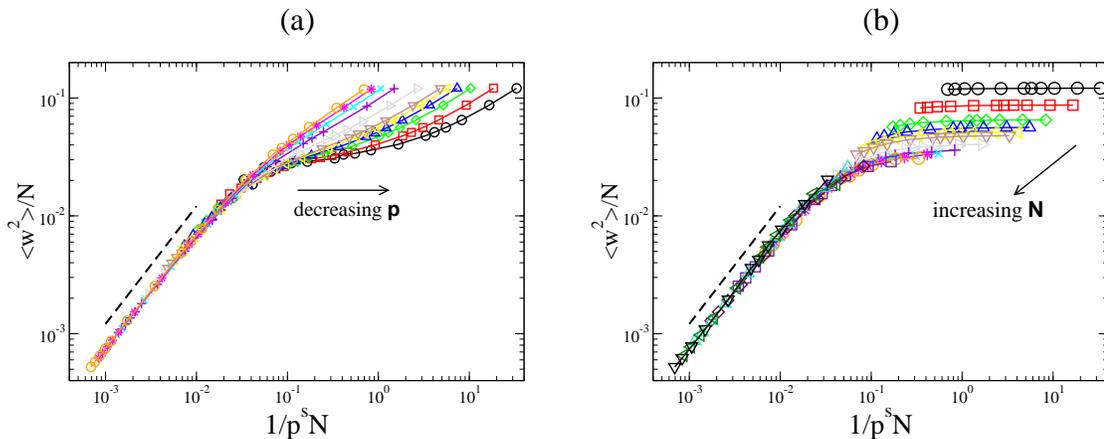

\vspace{5.2cm}
\includegraphics{1D-QRM-w2L-all-scaled.eps}
\includegraphics{1D-QRM-w2p-all-scaled.eps}
\vspace{-0.4cm}
\caption[Scaled width dependence on $p$ and $N$ for SW network in 1D]{The scaled versions of Fig.~\ref{fig_1d-w2-scaling}(a) and (b), as proposed by
Eq.~(\ref{new_scaling}) and Eq.~(\ref{xi_scale}) by plotting
$\left\langle w^2\right\rangle/N$ vs. $1/p^sN$ with $s$=$0.84$.
The data points are identical in (a) and (b),
but data points connected by a line
have the same value of $p$ in (a) and
have the same value of $N$ in (b), as obtained by rescaling
Fig.~\ref{fig_1d-w2-scaling}(b) and (b), respectively.
The dashed line corresponds to the asymptotic small-$x$ behavior of the scaling function $g(x)$
[Eq.~(\ref{f_newscaling})].}
\vspace{-0.5cm}
\label{fig_1d-w2-scaling2}
\end{figure}
massive propagator Eq.~(\ref{S_k}) indeed leads to an exponential
decay in the two-point correlation function from which the
correlation length can also be extracted
[Fig.~\ref{fig_sw1dcorr}]. In our case with a system-size
$N$=$16$$,$$384$, $\xi$$\approx$$27$ for $p$=$0.1$ and
$\xi$$\approx$$1.6$ for $p$=$1$. Figure~\ref{fig_swxi}(c) shows
the correlation length extracted from the structure factor $S(k)$
as a function of $p$ for different system-sizes.

\begin{figure}[ht]
\vspace{5.5cm}
\includegraphics{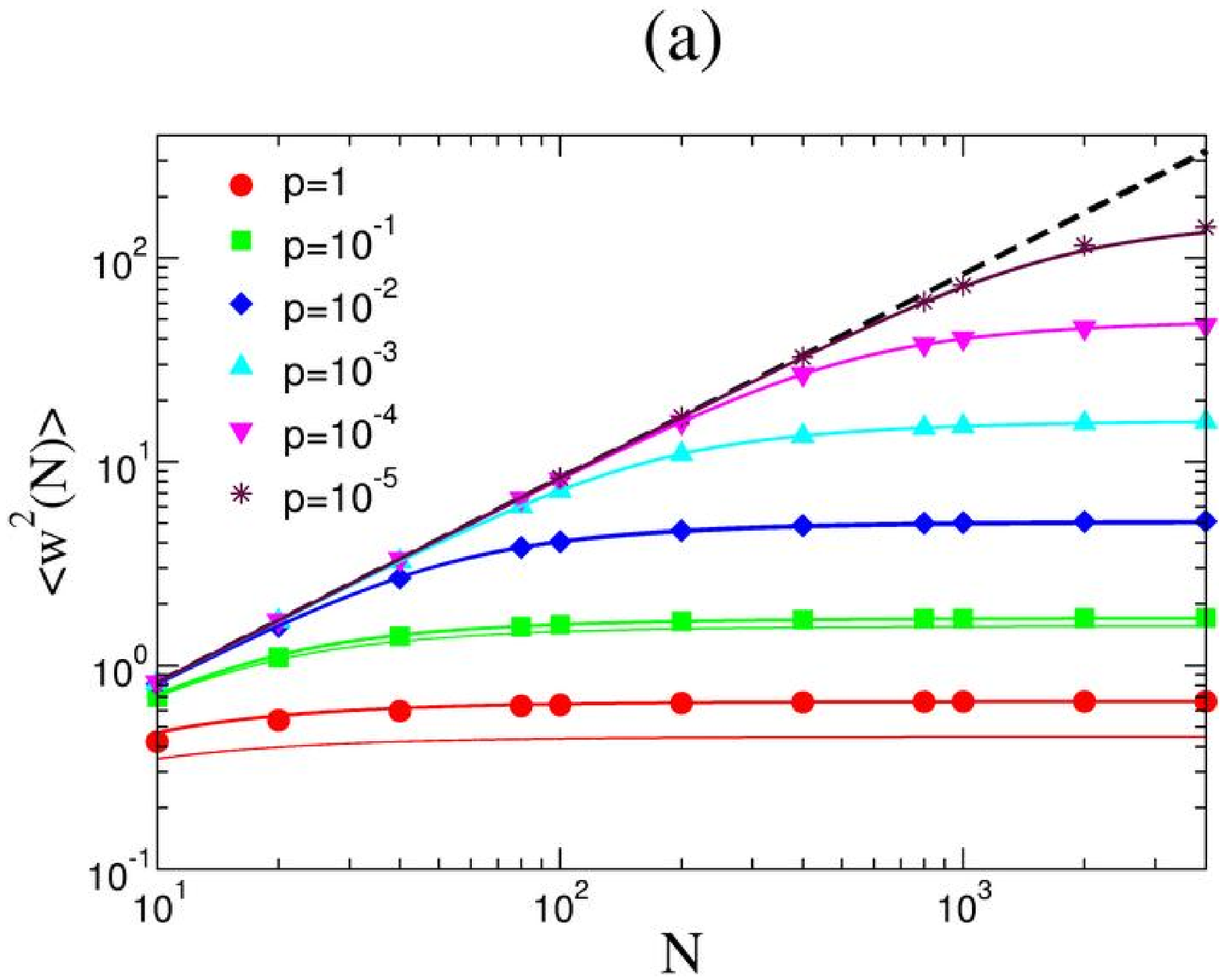}
\includegraphics{1D-QRM-EW-w2N-scaled.eps}
\vspace{-0.3cm}
\caption[Width scaling for ER network]{(a) Steady-state width for linear EW model as a function of the system-size from
\cite{KOZMA05} for comparison. 
The solid lines from are from theoretical calculations and the symbols from simulations. 
The dashed line shows the behavior
of the width when there is no SW links, namely the BCS scheme (short-range network)
(b) The scaled version of (a) by using Eq.~(\ref{new_scaling}) and Eq.~(\ref{f_newscaling}). For the
linear EW model the scaling does not have any noticeable corrections.}
\vspace{-0.5cm}
\label{fig_1d-qrm-ew-w2N}
\end{figure}
%%%%%%%%%%%%%%%%%%%%%%%%%%%%%%%%%%%%%%%%%%%%%%%%%%%%%%%%%%%%%%%%%%%%%%%%%%%%%%%%%%%%%%
%\begin{figure}[ht]
%\vspace{7cm}
%\special{psfile=1D-QRM-Pall-w2L.eps hscale=40 vscale=40 hoffset=0 voffset=0 angle=0}
%\vspace{-0.5cm}
%\caption[Width scaling for SW network in 1D]{The average steady-state width as a function of system-size for different values of $p$
%in the 1D SW synchronization scheme. The $p$$=$$0$ case corresponds to the purely 1D BCS scheme
%(exhibiting EW/KPZ scaling, indicated by a dashed line) and is also shown for
%comparison.}
%\vspace{-0.5cm}
%\label{fig_1d-kpz-qrm-w2}
%\end{figure}
%%%%%%%%%%%%%%%%%%%%%%%%%%%%%%%%%%%%%%%%%%%%%%%%%%%%%%%%%%%%%%%%%%%%%%%%%%%%%%%

An alternative way to determine the correlation length is using the
finite-size scaling of the width $\langle w^2\rangle$. From
dimensional analysis it follows that $\langle w^2\rangle$ has length
dimension in 1D. There are two length
scales in the system: the linear system size $N$ and the
correlation length $\xi$ of an infinite system. For $p=0$, $\langle
w^2\rangle$$\sim$$N$, while for $p$$>$$0$ and $N$$\to$$\infty$, $\langle
w^2\rangle$$\sim$$\xi$. For non-zero $p$ and finite $N$ the scaling
of the steady-state width can be expected \cite{KOZMA05} to follow
%%%%%%%%%%%%%%%%%%%%%%%%%%%%%%%%%%%%%%%%%%%%%%%%%%%%%%%%%%%%%%%%%%%%%%%%%%%%%%%%
\begin{equation}
\left\langle w^2\right\rangle = N g(\xi/N)\;,
\label{new_scaling}
\end{equation}
%%%%%%%%%%%%%%%%%%%%%%%%%%%%%%%%%%%%%%%%%%%%%%%%%%%%%%%%%%%%%%%%%%%%%%%%%%%%%%%%
where $g(x)$ is a scaling function such that
%%%%%%%%%%%%%%%%%%%%%%%%%%%%%%%%%%%%%%%%%%%%%%%%%%%%%%%%%%%%%%%%%%%%%%%%%%%%%%%%
\begin{equation}
g(x) \sim \left\{
\begin{array}{ll}
x    & \mbox{if $x$$\ll$$1$} \\
\mbox{const.} & \mbox{if $x$$\gg$$1$}
\end{array}
\right. \;.
\label{f_newscaling}
\end{equation}
%%%%%%%%%%%%%%%%%%%%%%%%%%%%%%%%%%%%%%%%%%%%%%%%%%%%%%%%%%%%%%%%%%%%%%%%%%%%%%%%%%
For non-zero $p$ and for sufficiently small systems ($N$$\ll$$\xi(p)$)
one can confirm that the behavior of the width follows that of the
system without random links $\langle w^2\rangle$$\sim$$N$
[Fig.~\ref{fig_1d-w2-scaling}(a)]. For large-enough systems, on
the other hand, we can extract the $p$-dependence of the
infinite-system correlation length as $\langle w^2\rangle$$\sim$ $\xi(p)$ 
[Fig.~\ref{fig_1d-w2-scaling}(b)], yielding
%%%%%%%%%%%%%%%%%%%%%%%%%%%%%%%%%%%%%%%%%%%%%%%%%%%%%%%%%%%%%%%%%%%%%%%%%%%%%%%%%%
\begin{equation}
\xi(p) \sim p^{-s} \;,
\label{xi_scale}
\end{equation}
%%%%%%%%%%%%%%%%%%%%%%%%%%%%%%%%%%%%%%%%%%%%%%%%%%%%%%%%%%%%%%%%%%%%%%%%%%%%%%%%%%
where $s$$\approx$$0.84$.

%%%%%%%%%%%%%%%%%%%%%%%%%%%%%%%%%%%%%%%%%%%%%%%%%%%%%%%%%%%%%%%%%%%%%%%%%%%%%%%%%%%%%%%%%%%%%%%%%
We then studied the data collapse as proposed by
Eq.~(\ref{new_scaling}) by plotting $\langle w^2\rangle/N$ vs.
$1/p^{s}N$. In fact, we performed this rescaling
originating from both raw data sets
Fig.~\ref{fig_1d-w2-scaling}(a) and (b). The resulting scaled
data points in Fig.~\ref{fig_1d-w2-scaling2}(a) and (b), of course, are
identical in the two figures, but the lines connect data points
with the same value of $p$ in Fig.~\ref{fig_1d-w2-scaling2}(a) and
with the same value of $N$ in Fig.~\ref{fig_1d-w2-scaling2}(b).
These scaled plots in Fig.~\ref{fig_1d-w2-scaling2} indicate
that there are very strong corrections to scaling: data for larger
$p$ or smaller $N$ values ``peel off'' from the proposed scaling form
in Eq.~(\ref{f_newscaling}) relatively quickly. These strong corrections are
possibly the result of the nonlinear nature of the interaction
between the nodes on the quenched network. We note that the
linear EW model on identical networks exhibits the scaling
proposed in Eq.~(\ref{new_scaling}) and Eq.~(\ref{f_newscaling})
{\em without} noticeable corrections \cite{KOZMA05} 
[Fig.~\ref{fig_1d-qrm-ew-w2N}(a) and (b)].

\begin{figure}[htb]
\vspace{7cm}
\includegraphics{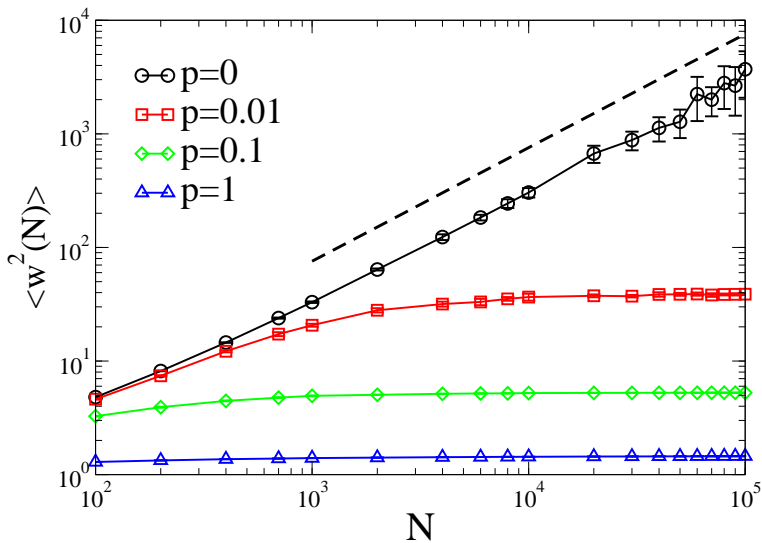}
\vspace{-0.5cm}
\caption[Width scaling for SW network in 1D]{The steady-state width as a function of system size for 1D SW synchronization
network as a function of system size. The width saturates to finite values
as long as $p\not=0$. The dashed line shows the power-law divergence
for BCS in 1D. Note the log-log scales.}
\vspace{-0.5cm}
\label{fig_1d_QRM_w2N}
\end{figure}

\begin{figure}[ht]
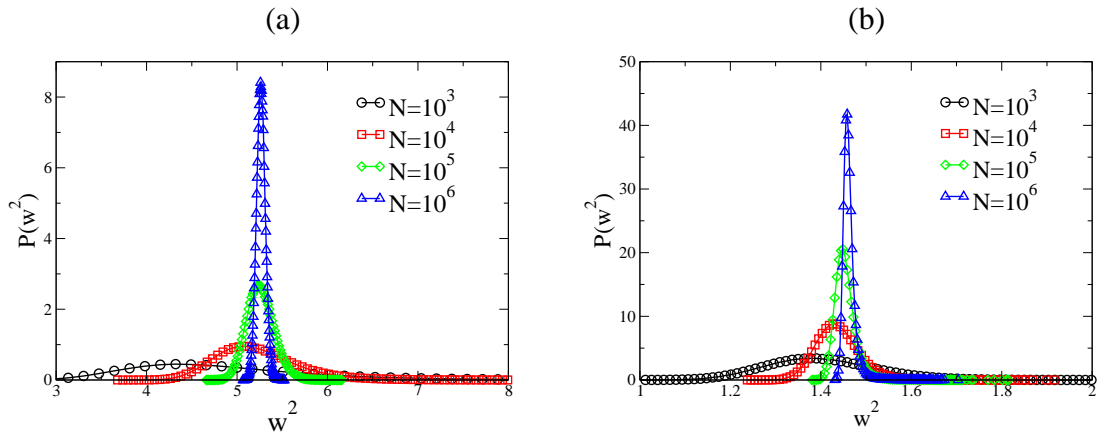

\vspace{6cm}
\includegraphics{1D-QRM-P0.10-w2-dist.eps}
\includegraphics{1D-QRM-P1.00-w2-dist.eps}
\vspace{-0.3cm}
\caption[Width distributions for SW network in 1D]{Steady-state width distributions in the 1D SW synchronization
scheme for (b) $p$=$0.1$ and for (c) $p$=$1.0$.
The distributions were constructed using ten different network realizations,
except for $N$=$10^6$, where only one realization was obtained due to
computational limitations. All width distributions, however,
indicated self-averaging.}
\vspace{-0.5cm}
\label{fig_1d-kpz-qrm-w2dist}
\end{figure}

\begin{figure}[ht]
\vspace{6.3cm}
\includegraphics{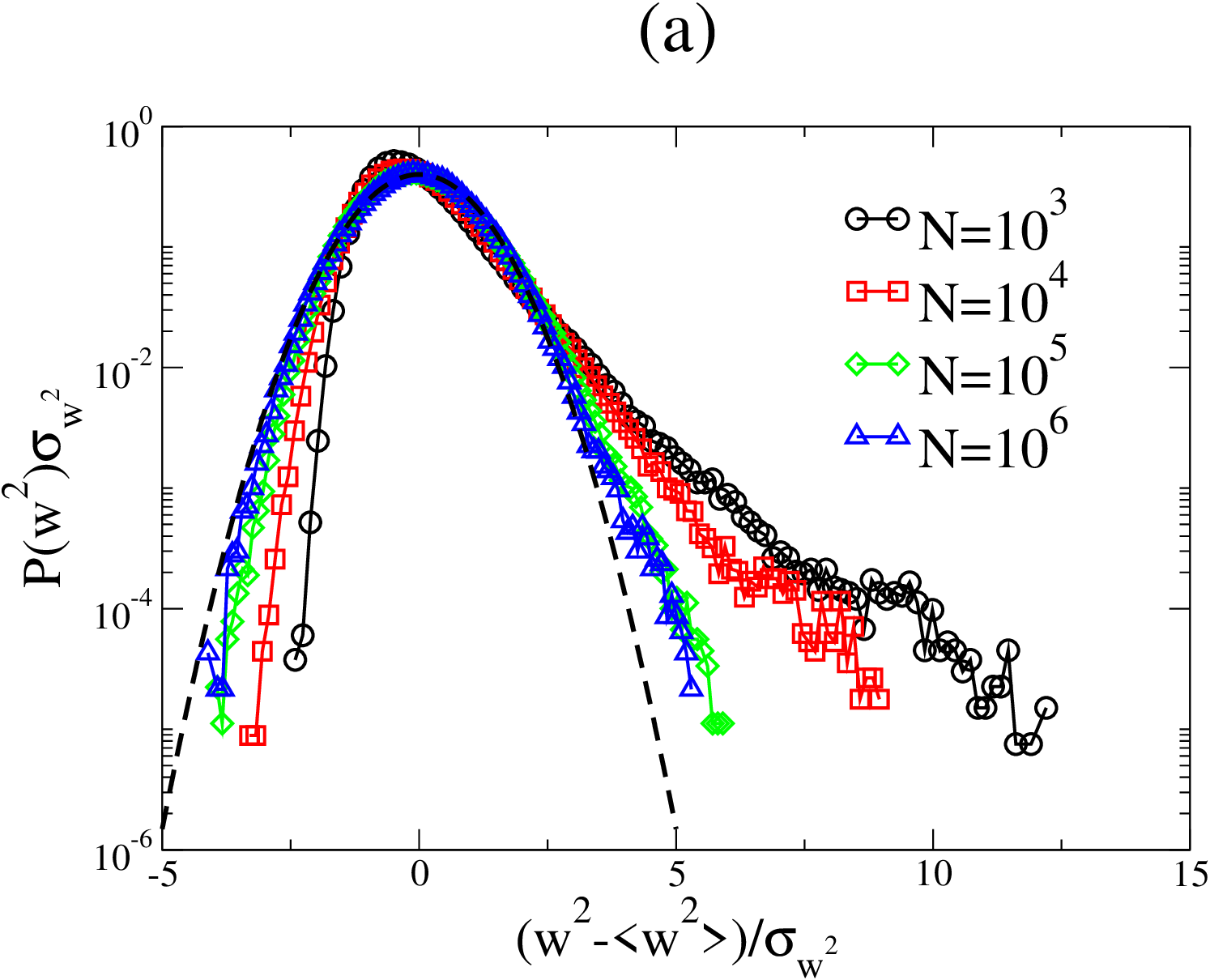}
\includegraphics{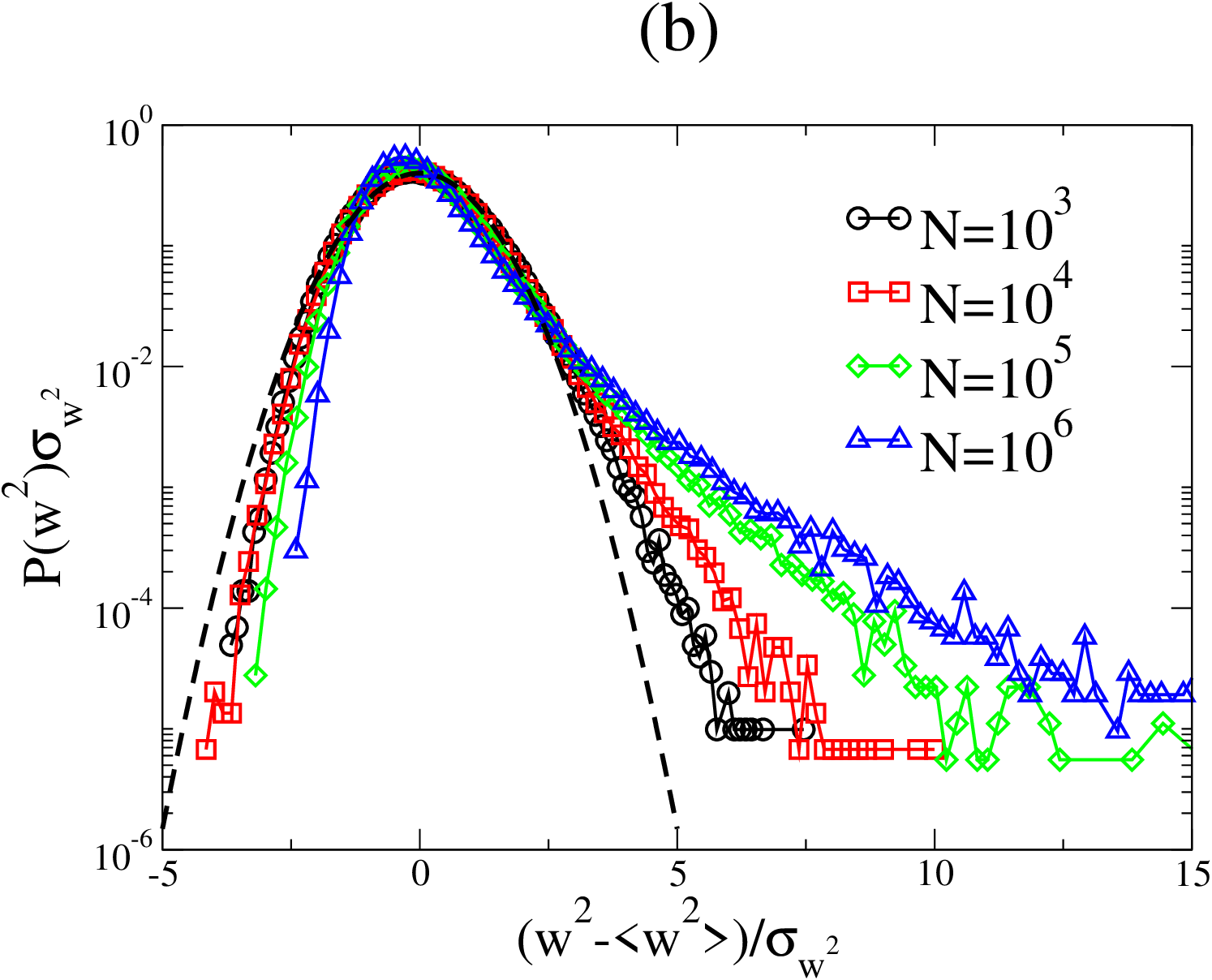}
\vspace{-0.3cm}
\caption[Scaled width distributions for SW network in 1D]{Steady-state width distributions for the 1D SW synchronization scheme scaled to zero mean
and unit variance for (a) $p$=$0.1$ and (b) $p$=$1.0$.  The dashed curves are similarly scaled
Gaussians for comparison.}
\vspace{-0.5cm}
\label{fig_1d-qrm-w2dist-scaled}
\end{figure}

The non-zero $\gamma$, leading to a finite correlation length,
$\xi$, ensures a finite width in the infinite system-size limit. Our
simulations show that the width saturates to a finite value for
$p$$>$$0$ [Fig.~\ref{fig_1d_QRM_w2N}(a)]. 
The distribution of
the width $P(w^2)$ changes from the EW/KPZ distribution to a delta
function for non-zero values of $p$ as the system size goes to
infinity. Figure~\ref{fig_1d-kpz-qrm-w2dist}(a) and (b) shows the
width distributions for $p$=$0.1$ and $p$=$1$, respectively. 
The scaled width distributions (to zero mean and unit variance),
however, exhibit the convergence to a delta function through
nontrivial shapes for different values of $p$. For $p$=$0.1$
%%%%%%%%%%%%%%%%%%%%%%%%%%%%%%%%%%%%%%%%%%%%%%%%%%%%%%%%%%%%%%%%%%%%%%%%%%%%%%%%%%%%%%
[Fig.~\ref{fig_1d-qrm-w2dist-scaled}(a)] the distributions appear
to slowly converge to a Gaussian as the system-size increases. 
For $p$=$1$  [Fig.~\ref{fig_1d-qrm-w2dist-scaled}(b)], the trend is
opposite up to the system sizes we could simulate; as the
system-size increases, the distributions exhibit progressively
non-Gaussian features (closer to an exponential) around the center
up to $N$=$10^6$. Note that not only the average width $\langle
w^2\rangle$, but also the full distribution $P(w^2)$ is
self-averaging, i.e., is independent of the particular realization
of the underlying SW network.

%%%%%%%%%%%%%%%%%%%%%%%%%%%%%%%%%%%%%%%%%%%%%%%%%%%%%%%%%%%%%%%%%%%%%%%%%%%%%%%
%%%%%%%%%%%%%%%%%%%%%%%%%%%%%%%%%%%%%%%%%%%%%%%%%%%%%%%%%%%%%%%%%%%%%%%%%%%%%%%

To get some insight into the possible role of the disorder in approaching the
limit distribution of the width, we studied the two-point function for
{\em individual} pair of nodes. Note, that by construction, the observable previously
considered, $G(l)$, is the site (or spatially) averaged two-point function over all
nodes with Euclidean distance $l$, $G(l)$$=$$\frac{1}{N}\sum_{i=1}^{N}
\langle (\tau_{i}-\bar\tau) (\tau_{i+l}-\bar\tau) \rangle$. If the height values on the
nodes for a fixed network realization are sufficiently weakly
correlated, the width distribution should converge 
to a Gaussian, governed by the central limit theorem \cite{FELLER68,FELLER71}. 
As we saw above, for larger values of $p$, this may not be
the case, at least for finite systems.

\begin{figure}[htb]
\vspace{6cm}
\includegraphics{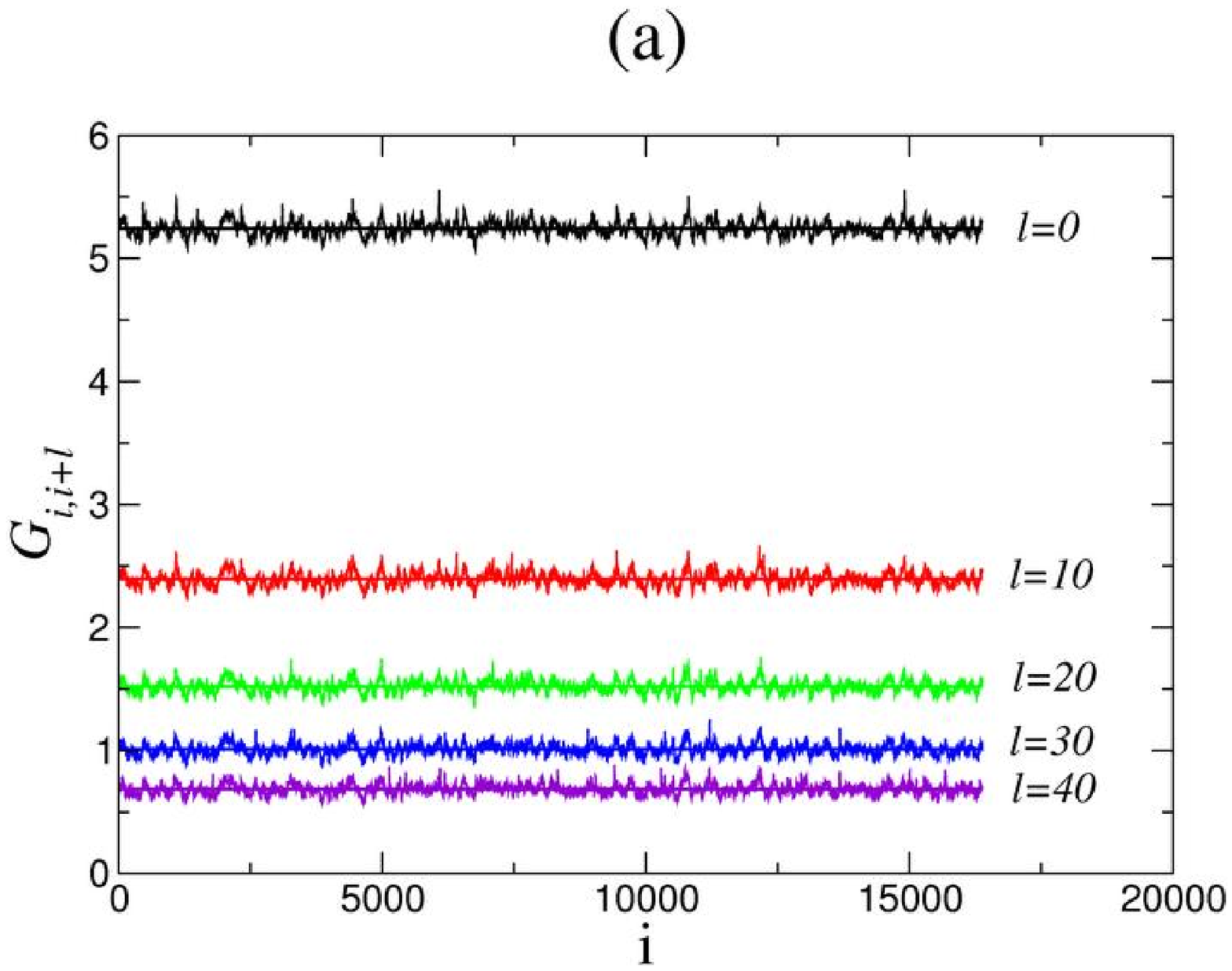}
\includegraphics{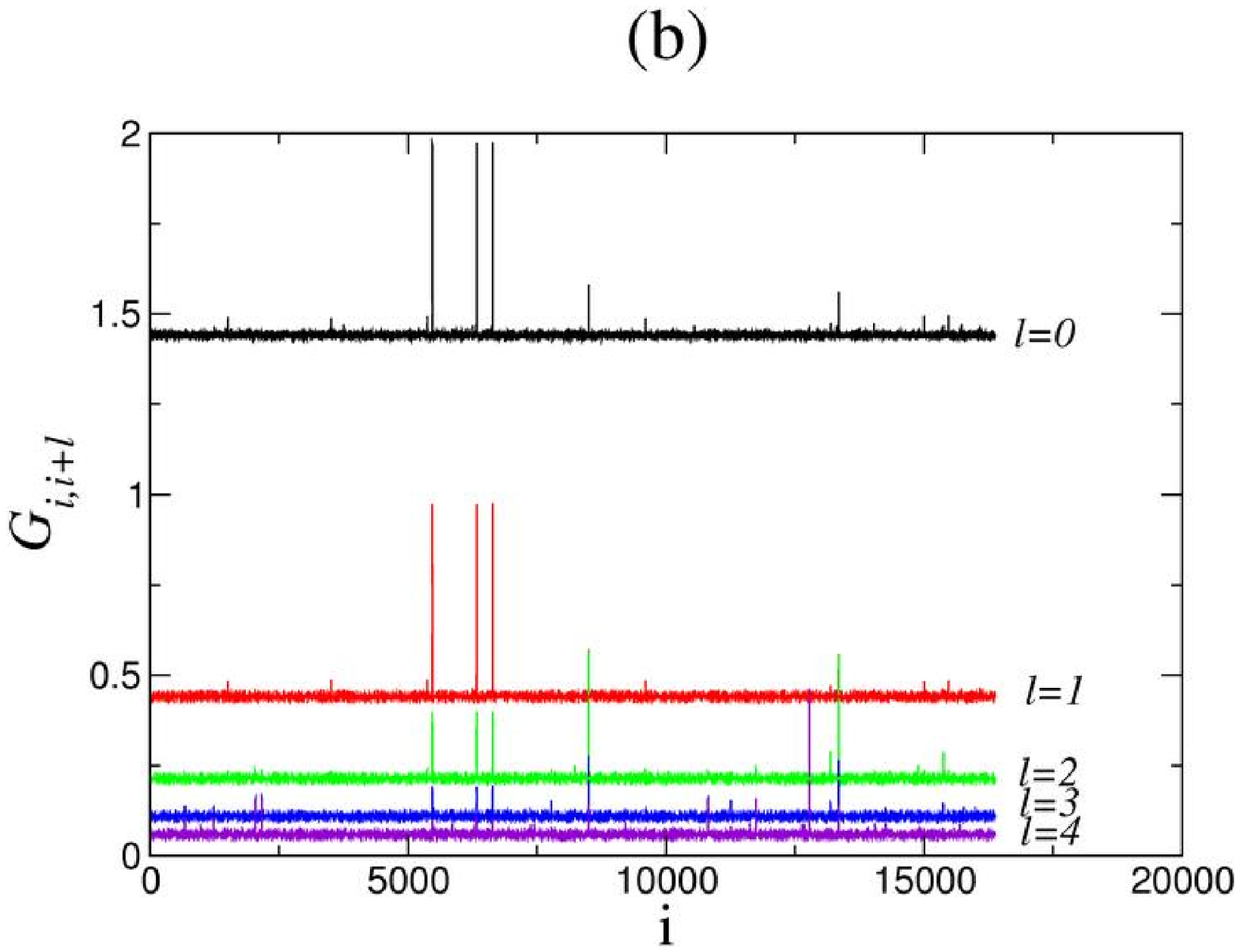}
\vspace{-0.5cm}
\caption[Two-point correlation function for SW network in 1D]{Two-point correlation function for each node $i$ for various separation 
$l$ for a fixed 1D SW synchronization network with size
$N$$=$$16,384$ (a) $p$$=$$0.1$ and (b) $p$$=$$1$. The horizontal lines correspond to
spatial average $G(l)$ for each $l$.}
\vspace{-0.5cm}
\label{fig_1d-qrm-twopoint}
\end{figure}

In order to have some measure how the individual terms in the width are correlated, 
$w^2$$=$$(1/N)\sum_{i=1}^{N}(\tau_{i}-\bar\tau)^2$,
we constructed the two-point function for
all sites $i$ for a few chosen separation $l$ [Fig.~\ref{fig_1d-qrm-twopoint}] for a {\em fixed}
network realization. Of course, as already
discussed, averaging over all $i$, will yield an exponential decay as a
function of $l$. Now, instead, we focus on the full two-point correlation
``profile'' for a given separation $l$.
As can be seen in Fig.~\ref{fig_1d-qrm-twopoint}(a), for $p$$=$$0.1$, the node-to-node fluctuations in
the two-point correlation profile, compared to their spatial average
$G(l)$, are small. With the increasing strength of the disorder ($p$$=$$1$), however,
certain sites develop abnormally large, frozen correlations as shown
in Fig.~\ref{fig_1d-qrm-twopoint}(b). The deviation of the two-point correlations for these few
nodes, from the mean $G(l)$, is much larger than those for the other
nodes, and is comparable to the mean itself. This property can work
``against'' the necessary conditions of the central limit theorem and,
thus, can have a strong effect on the convergence (or the apparent
lack of it) of the width distribution to a Gaussian.

\begin{figure}[htb]
\vspace{9cm}
\includegraphics{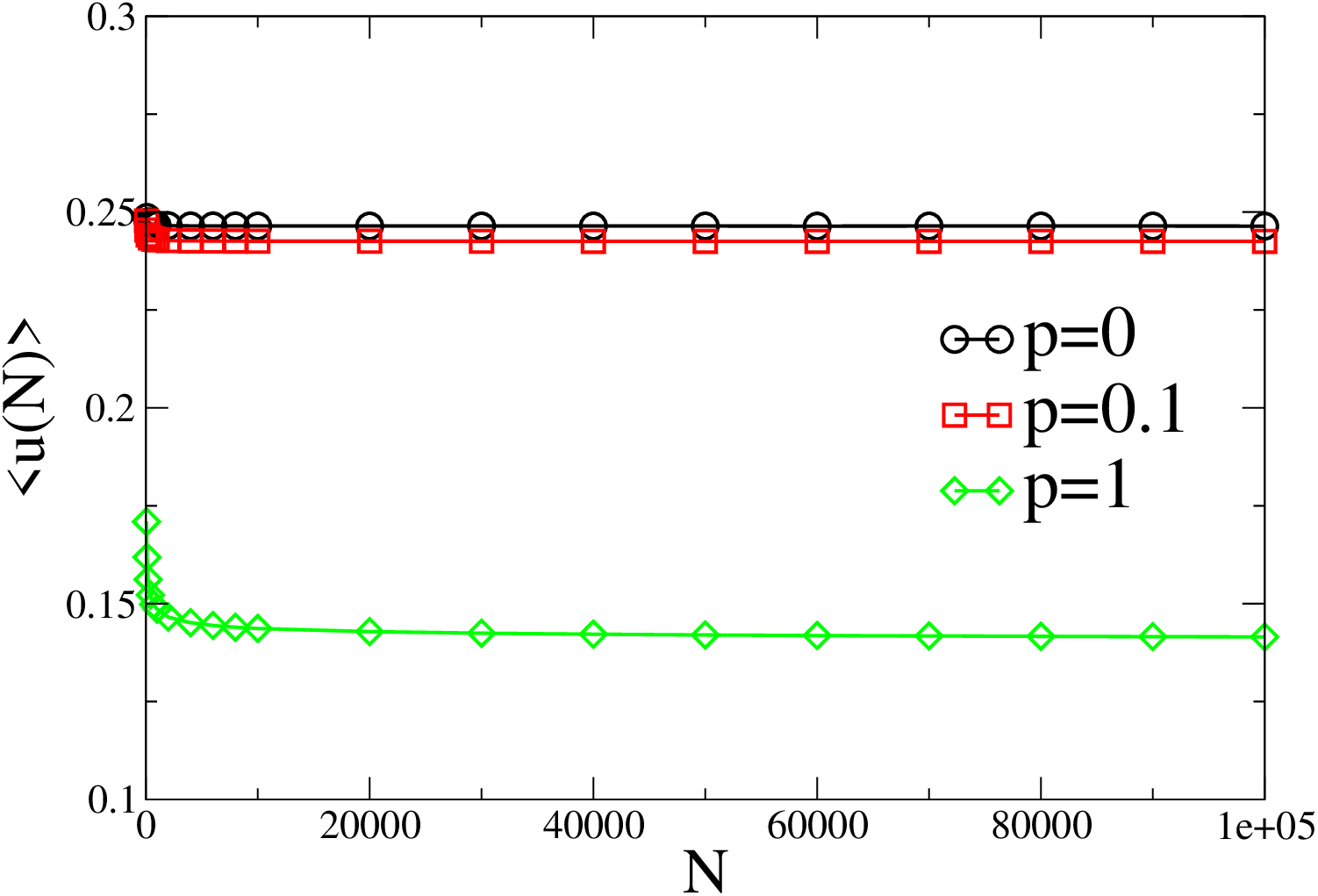}
\vspace{-0.5cm}
\caption[Utilization scaling for SW network in 1D]{Steady-state utilization of SW synchronization network in 1D 
as a function of system-size for three different
values of $p$$=$$0$ (BCS), $p$$=$$0.1$ and $p$$=$$1$.}
\vspace{-0.5cm}
\label{fig_1d-qrm-util}
\end{figure}

The effect of the random communication links on the utilization
can be understood as follows. According to the algorithmic rules,
the virtual times of the full network neighborhood (including the
random neighbor) are checked with probability $p$, while with
probability $(1$$-$$p)$ only short-ranged synchronization is employed.
Thus, the average progress rate of the simulated times becomes
%%%%%%%%%%%%%%%%%%%%%%%%%%%%%%%%%%%%%%%%%%%%%%%%%%%%%%%%%%%%%%%%%%%%%%
\begin{equation}
\langle u \rangle = (1-p) \langle\Theta(-\phi_{i-1})
\Theta(\phi_{i}) \rangle + p \langle\Theta(-\phi_{i-1})
\Theta(\phi_{i}) \Theta(\tau_{r(i)}-\tau_{i}) \rangle \;.
\label{u_SW}
\end{equation}
%%%%%%%%%%%%%%%%%%%%%%%%%%%%%%%%%%%%%%%%%%%%%%%%%%%%%%%%%%%%%%%%%%%%%%%
Note that the disorder (network) averaging makes the right hand
side independent of $i$. In the presence of the SW links the
regular density of local minima $\langle\Theta(-\phi_{i-1})
\Theta(\phi_{i}) \rangle$ {\em remains nonzero} (in fact,
increases compared to the short-range synchronized BCS scheme)
\cite{KORNISS03,TOROCZKAI03,GUCLU04_2}. Thus, for an infinitesimally
small $p$, the utilization, at most, can be reduced by an
infinitesimal amount, and the SW-synchronized simulation scheme
maintains a nonzero average progress rate. This is favorable
in PDES where global performance requires both finite width and
nonzero utilization. With the SW synchronization scheme, both of
these objectives can be achieved. For example, for $p$$=$$0.1$
$\langle u(\infty) \rangle \simeq 0.242$, while for $p$$=$$1.0$
$\langle u(\infty) \rangle \simeq 0.141$. 
The steady-state utilization as a function of system size
for various values of $p$ can be seen in Fig.~\ref{fig_1d-qrm-util}.

\begin{section}{Two-dimensional Small-World-Connected Synchronization Network}

The de-synchronization (roughening of the virtual time horizon)
again motivates the introduction of the possibly long-range,
quenched random communication links on top of the 2D regular
network. Each node will have exactly one (bi-directional) random
link as illustrated in Fig.~\ref{fig_2d-model}(b). The
actual ``microscopic" rules are analogous to the 1D SW case: with
probability $p$ each node will check the local simulated times of
all of its neighbors, including the random one, and can increment
its local simulated time by an exponentially distributed random
amount only if it is a ``local" minimum (among the four
nearest neighbors and its random neighbor). With probability
$(1$$-$$p)$, only the four regular lattice neighbors are checked for
the local minimum condition.

%%%%%%%%%%%%%%%%%%%%%%%%%%%%%%%%%%%%%%%%%%%%%%%%%%%%%%%%%%%%%%%%%%%%%%%%%%%%%%%%%%%%%%%%%%%%%%%%
\begin{figure}[htb]
\vspace{7cm}
\includegraphics{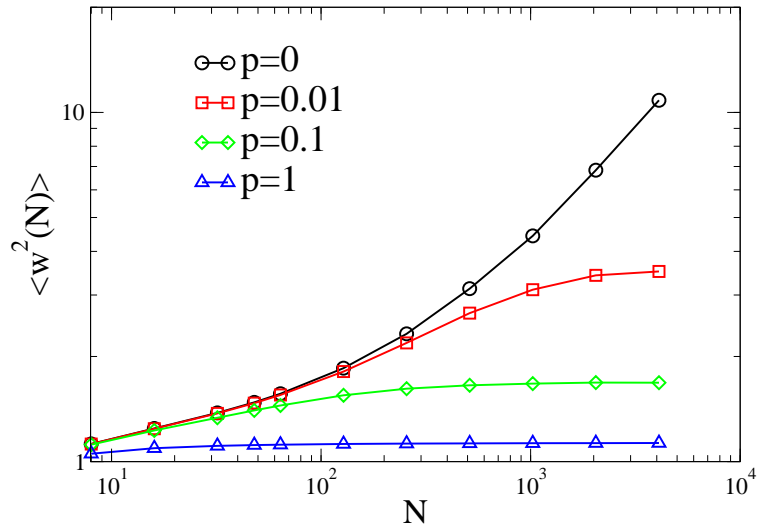}
\vspace{-0.5cm}
\caption[Width scaling for SW network in 2D]{(a) The average steady-state width as a function of linear system size for different
$p$ values in the 2D SW synchronization scheme. The data for $p$=$0$ corresponds to the 2D BCS
scheme on a regular network with only nearest-neighbor connections.}
\vspace{-0.5cm}
\label{fig_2d-qrm-w2}
\end{figure}

The effect of the synchronization through the random links is, 
again, to stop kinetic roughening and to suppress fluctuations in
the synchronization landscapes. Contour plots of the
synchronization landscapes are shown in
Fig.~\ref{fig_2d-surface}(b) and (c) for $p$$=$$0.1$ and $p$$=$$1$,
respectively. Our results indicate that for any nonzero $p$ the
\begin{figure}[htb]
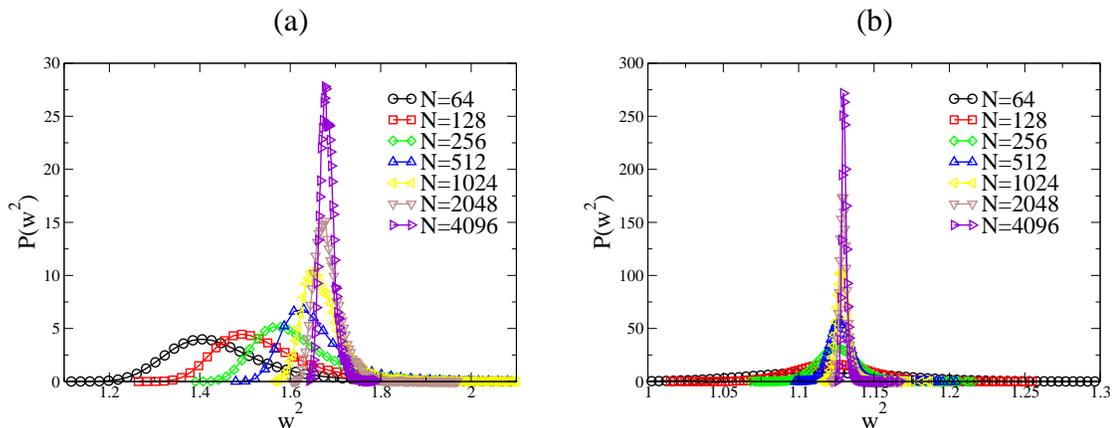

\vspace{6.5cm}
\includegraphics{2D-QRM-P0.10-w2-dist.eps}
\includegraphics{2D-QRM-P1.00-w2-dist.eps}
\vspace{-0.5cm}
\caption[Width distributions for SW network in 2D]{Steady-state width distributions in the 2D SW synchronization scheme
for (a) $p$$=$$0.1$ and for (b) $p$$=$$1.0$ for various system sizes.}
\vspace{-0.5cm}
\label{fig_2d-qrm-w2dist}
\end{figure}
width of the surface approaches a finite value in the limit of
$N$$\to$$\infty$ [Fig.~\ref{fig_2d-qrm-w2}(a)]. At the same time,
the distribution approaches a delta-function in the large
system-size limit as shown in Fig.~\ref{fig_2d-qrm-w2dist}(a) and (b).
The scaled distributions (to zero mean and unit variance) again
show that at least for the finite systems we observed, the shape
of these distribution differs from a Gaussian [Fig.~\ref{fig_2d-qrm-w2dist-scaled}(a) and (b)]. 
The deviation from
the Gaussian around the center of the distribution is stronger for
a larger value of $p$ where the influence of the quenched random
links are stronger. Note that for the 1D SW landscapes as well,
the width distribution only displayed a crossover to Gaussian
behavior for smaller values of $p$ and for very large linear
system sizes ($N>{\cal O}(10^4)$. In the 2D SW case, these linear
system sizes are computationally not achievable, and the
convergence to a Gaussian width distribution remains an open
question.

%%%%%%%%%%%%%%%%%%%%%%%%%%%%%%%%%%%%%%%%%%%%%%%%%%%%%%%%%%%%%%%%%%%%%%%%%%%%%%%%%%%%%%%%%%%%%%%
\begin{figure}[htb]
\vspace{6cm}
\includegraphics{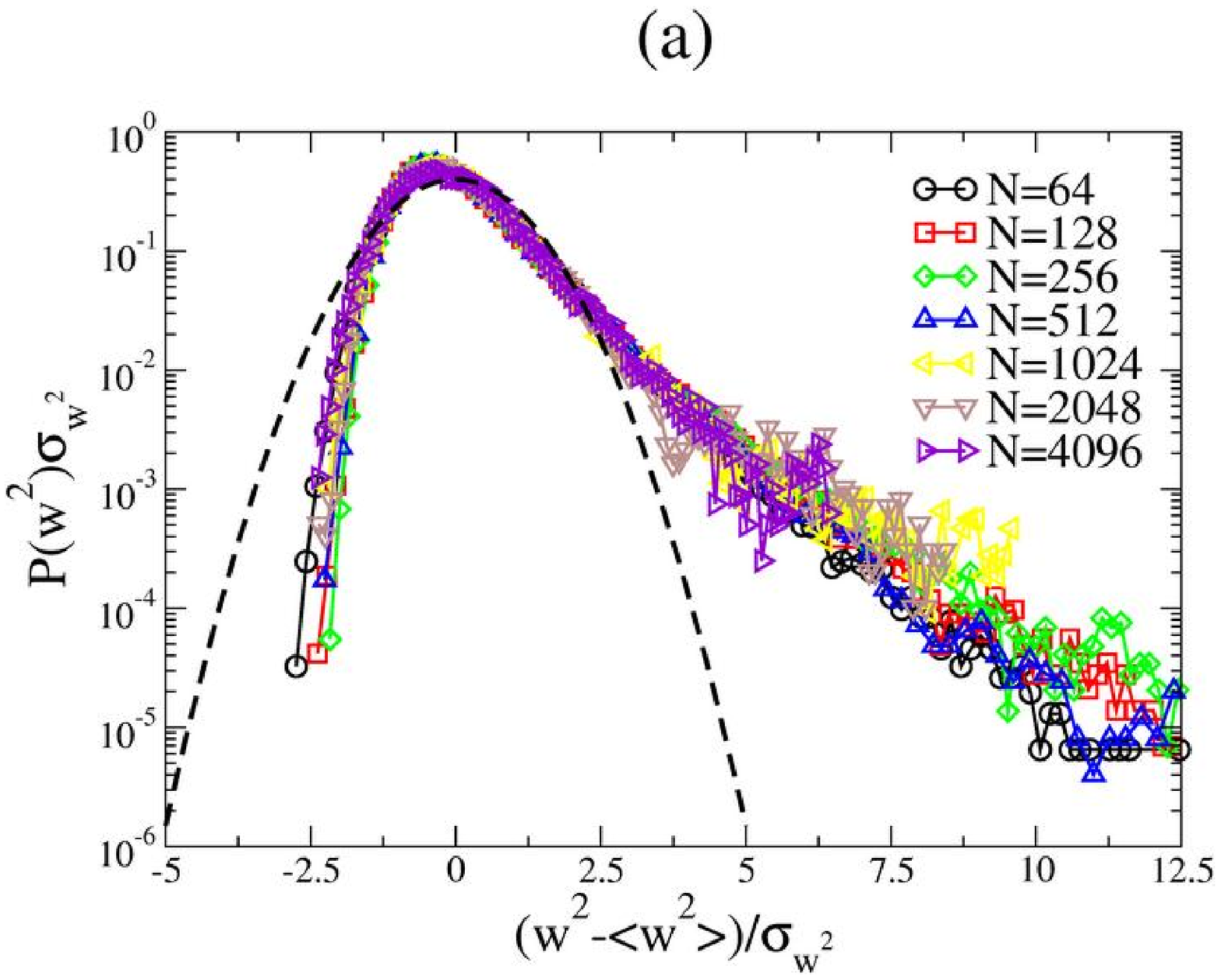}
\includegraphics{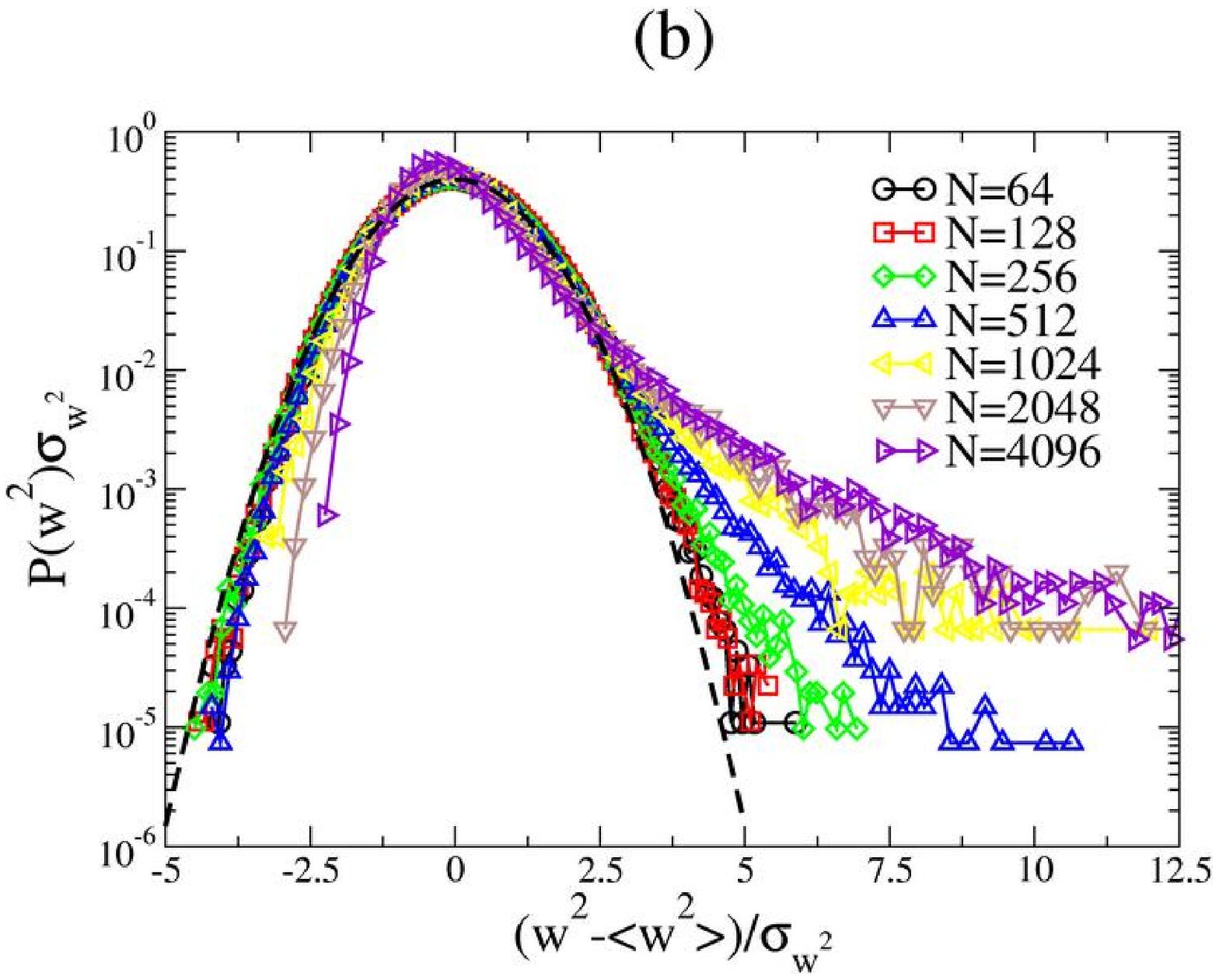}
\vspace{-0.3cm}
\caption[Scaled width distributions for SW network in 2D]{Steady-state width distributions in the 2D SW synchronization scheme
scaled to zero mean and unit variance
for (a) $p$$=$$0.1$ and for (b) $p$$=$$1.0$.
The dashed curves are similarly scaled Gaussians for comparison.}
\vspace{-0.5cm}
\label{fig_2d-qrm-w2dist-scaled}
\end{figure}

The underlying reason for the finite width is again a finite
average correlation length between the nodes. The 2D structure
factor exhibits a massive behavior, i.e., $S(|{\bf k}|)$
approaches a finite value in the limit of $\textbf{k}$$\to$$\textbf{0}$
[Fig.~\ref{fig_2d-sf}(a) and (b)]. For small wave numbers, the
approximate behavior of the structure factor is again
$S(|{\bf k}|) \propto 1/(|{\bf k}|^2+\gamma)$ as can be seen in
the inset of Fig.~\ref{fig_2d-sf}(b), with strong finite-size corrections to $\gamma$.
The relevant feature of the synchronization dynamics on a SW network is the
generation of the effective mass $\gamma$. Nonlinearities can give
rise to a renormalized mass, but the relevant operator is the
Laplacian on the random part of the network.

%%%%%%%%%%%%%%%%%%%%%%%%%%%%%%%%%%%%%%%%%%%%%%%%%%%%%%%%%%%%%%%%%%%%%%%%%%%%%%%%%%%%%%%%%%%%%%%%%
\begin{figure}[htb]
\vspace{6.5cm}
\includegraphics{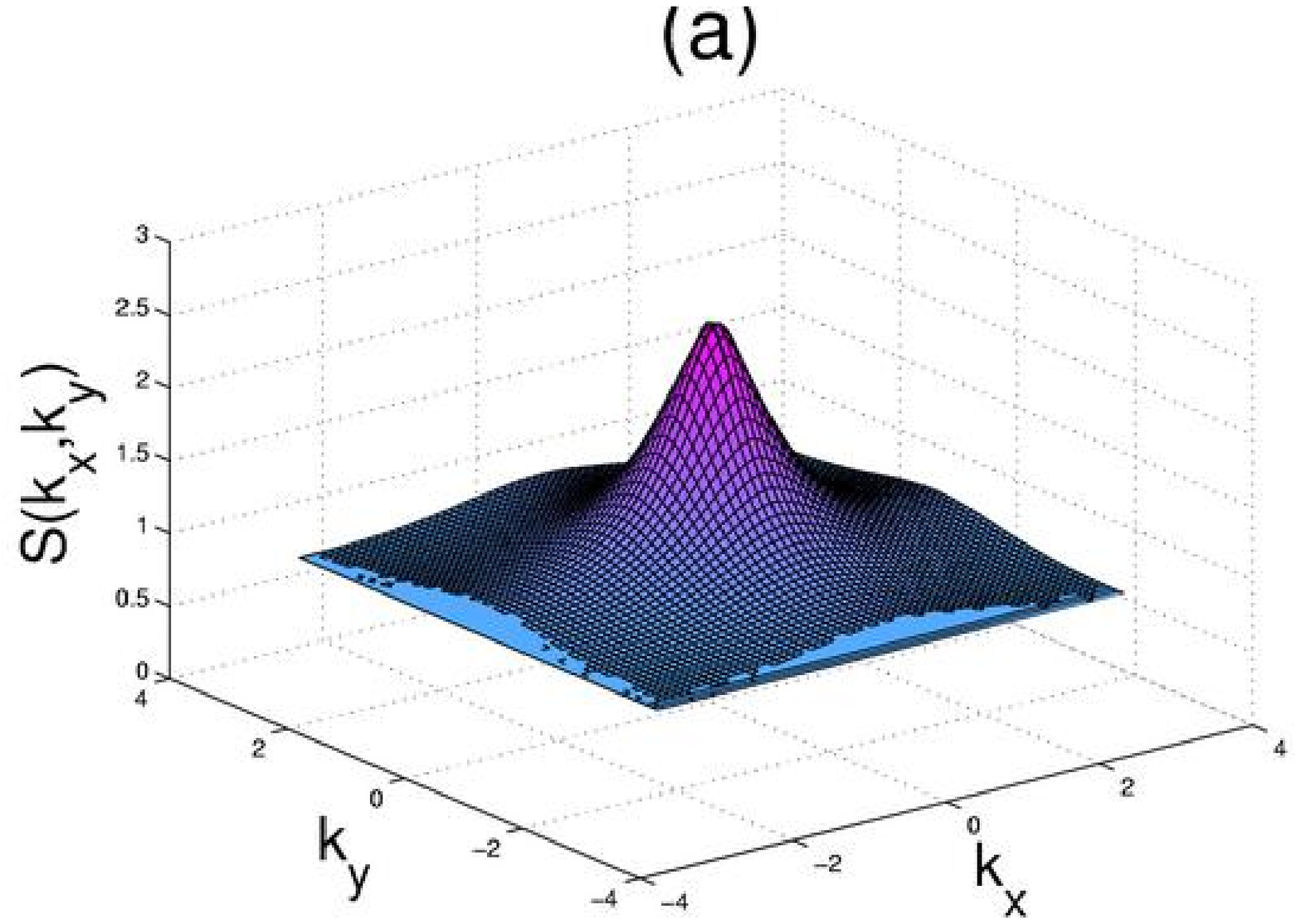}
\includegraphics{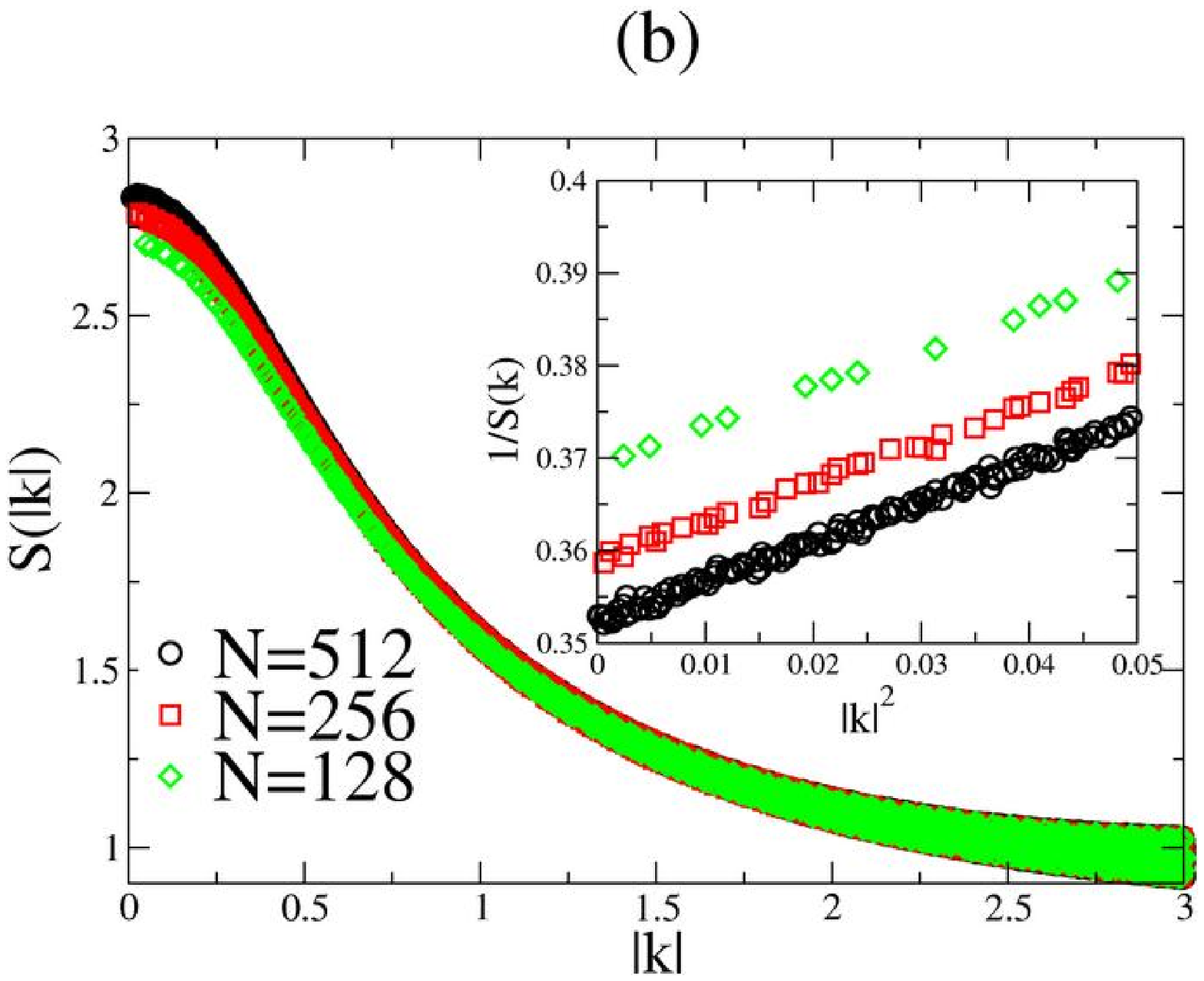}
\vspace{-0.5cm}
\caption[Structure factor for SW network in 2D]{(a) Steady-state structure factor for the 2D SW scheme
as a function of the wave-vector components $k_x$ and $k_y$ for $p$$=$$1.0$.
(b) Structure factor as a function of magnitude of the wave vector, $|\textbf{k}|$, for the 2D SW synchronization scheme
for $p$=$1.0$. The inset shows $1/S(\textbf{k})$ vs. $|\textbf{k}|^2$ for small values of $|\textbf{k}|$.}
\vspace{-0.5cm}
\label{fig_2d-sf}
\end{figure}
%%%%%%%%%%%%%%%%%%%%%%%%%%%%%%%%%%%%%%%%%%%%%%%%%%%%%%%%%%%%%%%%%%%%%%%%%%%%%%%%%%%%%%%%%

In the 2D SW synchronization scheme the steady-state utilization
is smaller than its purely 2D counterpart (BCS in 2D), as a result of the
possible additional checking with the random neighbors. For small
values of $p$, however, it is reduced only by a small amount, and
remains nonzero in the limit of an infinite number of nodes 
[Fig.~\ref{fig_2d-qrm-util}]. For
example, for $p$$=$$0.1$ $\langle u
\rangle_{\infty}$$\simeq$$0.1198$, while for $p$=$1$ $\langle u
\rangle_{\infty}$$\simeq$$0.084$.

%%%%%%%%%%%%%%%%%%%%%%%%%%%%%%%%%%%%%%%%%%%%%%%%%%%%%%%%%%%%%%%%%%%%%%%%%%%%%%%%%%%%%%%%%%%%%%%%%

\begin{figure}[htb]
\vspace{7cm}
\includegraphics{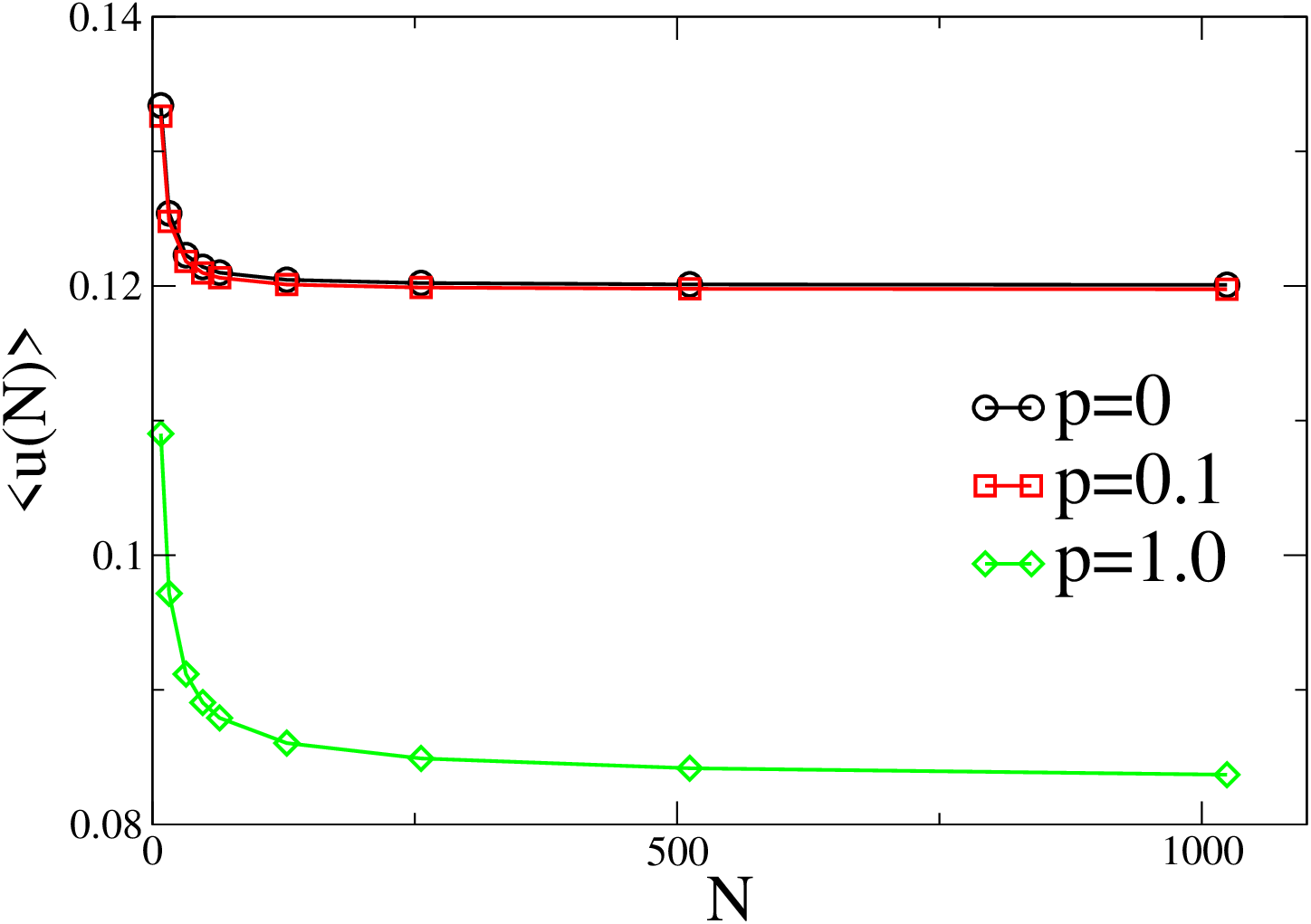}
\vspace{-0.5cm}
\caption[Utilization scaling for SW network in 2D]{Steady-state utilization of SW synchronization network in 2D 
as a function of system-size for three different
values of $p$$=$$0$ (BCS), $0.1$ and $1$.}
\vspace{-0.5cm}
\label{fig_2d-qrm-util}
\end{figure}

\end{section}

\section{Synchronization in scale-free networks} 
\label{sec3}

The Internet is a spontaneously grown collection
of connected computers. The number of (only) webservers
by February 2003 reached
over 35 million \cite{NETCRAFT}. The number of PC-s in use
(Internet users) surpassed 660 million in 2002, and it is
projected to surpass one billion by 2007 \cite{CIA}.
The idea for using it as a giant supercomputer is
rather natural: many computers are in an idle state, running
at best some kind of screen-saver software, and
the ``wasted" computational time is simply immense. Projects such
as SETI@home or the GRID consortium
\cite{GRID} are targeting to harness the power lost in screen-savers.

Most of the problems solved currently with
distributed computation on the Internet are ``embarrassingly parallel"
\cite{KIRKPATRICK03} , i.e., the computed tasks have
little or no connection to each other similar to starting the
same run with a number of different random seeds,
and at the end collecting the data to perform
statistical averages.  However, before 
complex problems can be solved in real time on the Internet
a number of challenges have to be solved, such as
the task allocation problem which is rather complex by itself
\cite{SCHONEVELD99}.

\begin{figure}[htpb]
\vspace{4cm}
\includegraphics{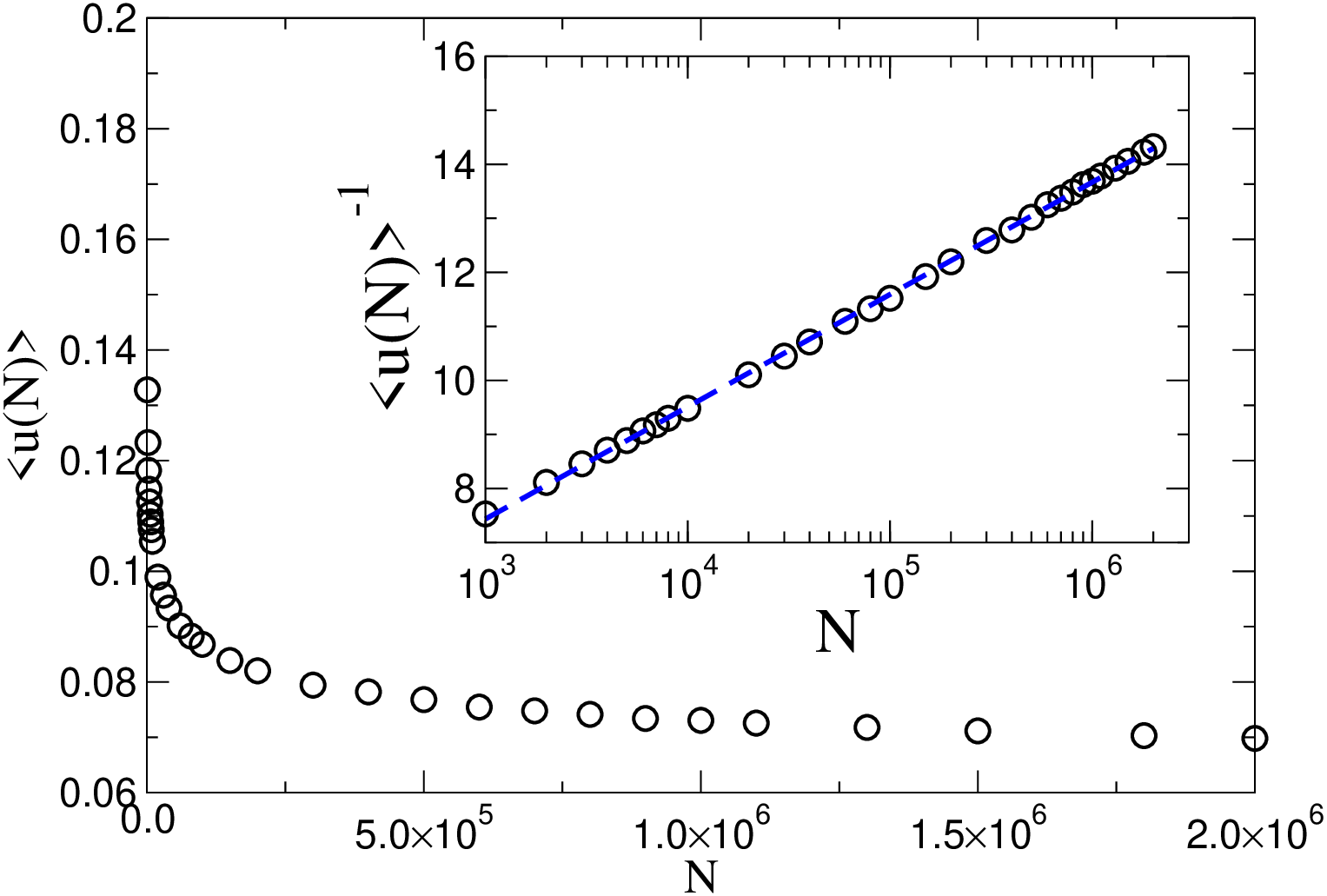}
\vspace{4cm}
\caption[Utilization for the scale-free BA network]{Steady-state utilization for the scale-free BA network}
\vspace{-0.5cm}
\label{uinfg}
\end{figure}

Here one can ask the following question: Given
that task allocation is resolved and the PE communication
topology on the internet is a scale-free network,
what are the scalability properties of a conservative synchronization scheme on
such networks? Here we present numerical results,
for the conservative PDES scheme, as measured on a model of
scale-free networks, namely the Barab\'asi-Albert
model (BA) \cite{BARABASI99,BARABASI99_2}. This network is created through the
stochastic process of preferential attachment: to
the existing network at time $t$ of $N$ nodes,  attaches
the $N$$+$$1^{st}$ node with $m$ links (``stubs") at time
$t$$+$$1$, such that each stub attaches to a node with probability
proportional to the existing  degree of the
node. Here we will only present the $m$$=$$1$ case, when the
network is a scale-free tree. Once we reach a given number of nodes
in the network, we stop the process and use the random network instance
to simulate the synchronization, using the evolution equation
[Eq.~(\ref{evolution})) for the time horizon.
While in case of regular topologies, the degree of a node is constant,
e.g. for $d$-dimensional ``square'' lattices,
$P^{(N^d)}(k)=2d\delta_{k,2d}$ , for the
BA network, it is a power law in the asymptotic ($N$$\to$$\infty$) limit :
$P^{BA}(k)\simeq 2m^2 k^{-3}$. 
The condition for a site to be
updated, i.e., that its virtual time is a local minimum, is a {\em
local} property, and thus we expect that the utilization itself be
correlated with local structural properties of the graph, such as the
degree distribution. 
Figure~\ref{uinfg} shows the steady state ($t$$\to$$\infty$,
on a fixed BA network of $N$ nodes) values of the average utilization
as function of the network size $N$. 
Notice that strictly speaking, the conservative PDES scheme is
computationally {\em non-scalable}. An empirical
fit suggest that
$\langle u(N) \rangle=\langle u(N,t$$=$$\infty)\rangle \simeq
\left[\ln{\left(a N^{b}\right)}\right]^{-1}$ 
with $a\simeq 3.322$ and $b=0.902$,  
i.e., the computation is only logarithmically (or
marginally) non-scalable. For a system of $N$$=$$10^3$ nodes we have found
a steady state utilization (for the worst case scenario) of 
$\langle u \rangle$$=$$0.1328$ ($13.3\%$ efficiency), 
while for a system of a million nodes, $N$$=$$10^6$, the utilization
dropped only to $\langle u \rangle$$=$$0.073$ ($7.3\%$ efficiency), by less than half of its value.
For practical purposes the conservative PDES scheme
can be considered computationally scalable, and this
type of non-scalability we will call {\em logarithmic} (or marginal)
non-scalability. 

\begin{figure}[htbp]
\vspace{8.5cm}
\includegraphics{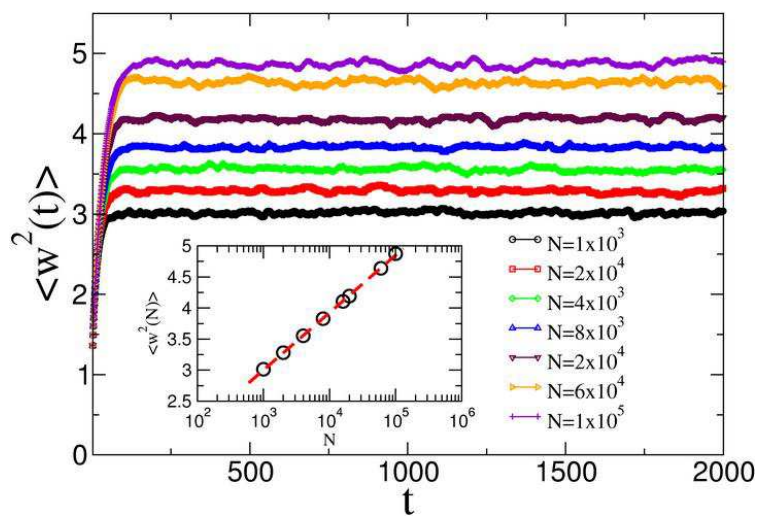}
\vspace{-0.5cm}
\caption[Width for BA network]{Behavior of the time horizon width for the scale-free 
BA network (scale-free tree with $m$$=$$1$).
The inset shows the scaling of the steady-state width as a function
of system size, $N$ (note the log-normal scale on the inset).}
\vspace{-0.5cm}
\label{width1}
\end{figure}

Figure~\ref{width1} shows the scaling of the width of the fluctuations
for the time horizon as function of time, and the scaling of its value
in the steady-state as function of system size (inset). Notice, that
while the steady state width diverges to infinity, it only does so
logarithmically, $\langle w^2(N,t$$=$$\infty) \rangle 
\simeq \left[\ln{\left(cN^{d}\right)}\right]$ 
with $c$$\simeq$$1.25$ and $d$$\simeq$$0.401$. 
Some specific values: $\langle w^2(10^3,t$$=$$\infty) \rangle\simeq 3.01$,
$\langle w^2(10^5,t$$=$$\infty) \rangle$$\simeq$$4.78$. This means that
the measurement phase of the conservative PDES scheme on a scale-free network is
non-scalable either, however, it is so only logarithmically, and for
practical purposes the scheme can be considered scalable. 
Overall, the conservative PDES scheme has logarithmic (or marginal) non-scalability
on scale-free networks.

%% file: rpichap4.tex
%%%%%%%%%%%%%%%%%%%%%%%%%%%%%%%%%%%%%%%%%%%%%%%%%%%%%%%%%%%%%%%%%%% 
%                                                                 %
%                            CHAPTER FOUR                         %
%                                                                 %
%%%%%%%%%%%%%%%%%%%%%%%%%%%%%%%%%%%%%%%%%%%%%%%%%%%%%%%%%%%%%%%%%%% 
 
\chapter{EXTREME FLUCTUATIONS IN SMALL-WORLD NETWORKS}
 
Large fluctuations in networks are to be avoided for many reasons such as
scalability or stability. In the absence of global
intervention or control, this can be a difficult task. Motivated by the results
in Chapter III \cite{KORNISS03} for small-world (SW) \cite{WATTS98}
synchronized autonomous systems in the context of scalable parallel
computing, we investigate the steady-state properties of the extreme
fluctuations in
SW-coupled interacting systems with relaxational dynamics \cite{GUCLU04,GUCLU05}.
Since the introduction of SW networks \cite{WATTS98} it
has been well established that such networks can facilitate autonomous
synchronization \cite{STROGATZ01,BARAHONA02,WATTS99}.
In addition to the average ``load'' in the network, knowing the typical
size and the distribution of the extreme fluctuations
\cite{FISHER28,GUMBEL58,GALAMBOS94} is of great importance from a
system-design viewpoint, since failures and delays are triggered by
extreme events occurring on an individual node.

Relationship between extremal
statistics and universal fluctuations in correlated systems
has been studied intensively 
\cite{BRAMWELL98,RAYCHAUDHURI01,BRAMWELL00,WATKINS02,BRAMWELL02,%
BRAMWELL01,AJI01,ANTAL01,ANTAL02,DAHLSTEDT01,CHAPMAN02,GYORGYI03,BOUCHAUD97,%
BALDASSARRI02,MAJUMDAR04,MAJUMDAR04_2}. 
The focus of a number of
these studies was to find connections, if any, between the probability
distribution of {\em global} observables or order parameters (such as the
width in surface growth problems \cite{BARABASI95} or the magnetization in
magnetic systems \cite{GOLDENFELD92}) and known universal extreme-value limit 
distributions \cite{FISHER28,GUMBEL58,GALAMBOS94}. Recent analytic results
demonstrated \cite{ANTAL02,GYORGYI03} that, in general (except for special
cases \cite{ANTAL01,BERTIN05}), there are no such connections.
Here we discuss to what extent SW couplings
(extending the original dynamics through the random links)
lead to the suppression of the extreme fluctuations of the {\em local} order
parameter or field variable in various noisy environments. We illustrate our
findings on the actual PDES synchronization problem in scalable parallel
computing \cite{KORNISS03}. In Sec.~\ref{sect:ext_EVS} we review the
well-known extreme-value limit distributions for exponential-like and
power-law-tail distributed random
variables. In Sec.~\ref{sect:ext_EW} we discuss the
results \cite{GUCLU04,GUCLU05} on the scaling behavior of the
extreme fluctuations and their distribution.
for the Edwards--Wilkinson model \cite{EDWARDS82} on SW
networks \cite{KOZMA04} with exponential-like noise. 
In Sec.~\ref{sect:ext_EW} we apply these results to
study the extreme load fluctuations in SW-synchronized PDES
schemes \cite{FUJIMOTO90,NICOL94}, applicable to high
performance parallel architectures and large-scale grid-computing
networks. In Sec.~\ref{sect:ext_power} we extend our studies 
\cite{GUCLU04,GUCLU05} and consider the synchronization
problem in the presence of power-law tailed noise.

%%%%%%%%%%%%%%%%%%%%%%%%%%%%%%%%%%%%%%%%%%%%%%%%%%%%%%%%%%%%%
\section{Extreme-Value Distributions for
Independent Random Variables}
\label{sect:ext_EVS}

\subsection{Exponential-like variables}
\label{sect:ext_short_tails}

First, we consider the case when the individual complementary cumulative
distribution $P_{>}(x)$ (the probability that the individual
stochastic variable is greater than $x$) decays faster than any power
law, i.e., exhibits an exponential-like tail in the asymptotic
large-$x$ limit. (Note that in this case the corresponding probability density
function displays the same exponential-like asymptotic tail behavior.)
We will assume $P_{>}(x)\simeq e^{-cx^{\delta}}$
for large $x$ values, where $c$ and $\delta$ are constants.
Then the cumulative distribution $P_{<}^{\max}(x)$
for the largest of the $N$ events (the probability that the maximum
value is less than $x$) can be approximated as \cite{BOUCHAUD97,BOUCHAUD00,
BALDASSARRI00}
\begin{equation}
P_{<}^{\max}(x) = [P_{<}(x)]^N = [1-P_{>}(x)]^N =
e^{N\ln[1-P_{>}(x)]} \simeq e^{-NP_{>}(x)} \;,
\label{extreme_approx}
\end{equation}
%%%%%%%%%%%%%%%%%%%%%%%%%%%%%%%%%%%%%%%%%%%%%%%%%%%%%%%%%%%%%%%%%%%%%%%
where one typically assumes that the dominant contribution to the
statistics of the extremes comes from the tail of the individual
distribution $P_{>}(x)$. With the exponential-like tail in the
individual distribution, this yields
%%%%%%%%%%%%%%%%%%%%%%%%%%%%%%%%%%%%%%%%%%%%%%%%%%%%%%%%%%%%%%%%%%%%%%%
\begin{equation}
P_{<}^{\max}(x) \simeq e^{-e^{-cx^{\delta}+\ln(N)}}.
\label{raw_extreme}
\end{equation}
%%%%%%%%%%%%%%%%%%%%%%%%%%%%%%%%%%%%%%%%%%%%%%%%%%%%%%%%%%%%%%%%%%%%%%%
The extreme-value limit theorem states that there exists a sequence of
scaled variables $\tilde{x}=(x-a_{N})/b_{N}$, such that in the limit of
$N$$\to$$\infty$, the extreme-value probability distribution for $\tilde{x}$
asymptotically approaches the Fisher--Tippett--Gumbel (FTG)
distribution \cite{FISHER28,GUMBEL58}:
%%%%%%%%%%%%%%%%%%%%%%%%%%%%%%%%%%%%%%%%%%%%%%%%%%%%%%%%%%%%%%%%%%%%%%%
\begin{equation}
\tilde{P}_{<}^{\max}(\tilde{x}) \simeq e^{-e^{-\tilde{x}}} \;,
\label{gumbel}
\end{equation}
%%%%%%%%%%%%%%%%%%%%%%%%%%%%%%%%%%%%%%%%%%%%%%%%%%%%%%%%%%%%%%%%%%%%%%%
with mean $\langle\tilde{x}\rangle$$=$$\gamma$$=$$0.577\ldots$ (Euler constant) and variance
$\sigma_{\tilde{x}}^{2}=$ $\langle\tilde{x}^{2}\rangle$$-$$
\langle\tilde{x}\rangle^{2}$ $=\pi^2/6$.
From Eq.~(\ref{raw_extreme}), one can deduce\footnote{Note that for 
$\delta$$\neq$$1$, while the convergence to
Eq.~(\ref{raw_extreme}) is fast, the convergence for the appropriately
scaled variable to the universal FTG distribution Eq. (\ref{gumbel}) is {\em
extremely} slow.} \cite{BALDASSARRI00} that to leading
order the scaling coefficients are
$a_{N} = \left[\frac{\ln(N)}{c}\right]^{1/\delta}$
and
$b_{N} = (\delta c)^{-1}\left[\frac{\ln(N)}{c}\right]^{(1/\delta)-1}$.
The average value of the largest of the $N$ original variables then scales as
%%%%%%%%%%%%%%%%%%%%%%%%%%%%%%%%%%%%%%%%%%%%%%%%%%%%%%%%%%%%%%%%%%%%%%%
\begin{equation}
\langle x_{\max}\rangle = a_N + b_N\gamma \simeq 
\left[\frac{\ln(N)}{c}\right]^{1/\delta}
\label{mean}
\end{equation}
%%%%%%%%%%%%%%%%%%%%%%%%%%%%%%%%%%%%%%%%%%%%%%%%%%%%%%%%%%%%%%%%%%%%%%%
(up to ${\cal O}(\frac{1}{\ln(N)})$ correction) in the asymptotic
large-$N$ limit.
When comparing with experimental or simulation data, instead of
Eq.~(\ref{gumbel}), it is often convenient to use the form of the FTG
distribution which is scaled to zero mean and unit variance, yielding
%%%%%%%%%%%%%%%%%%%%%%%%%%%%%%%%%%%%%%%%%%%%%%%%%%%%%%%%%%%%%%%%%%%%%%%
\begin{equation}
\tilde{P}_{<}^{\max}(y)=e^{-e^{-(ay+\gamma)}}\;,
\label{gumbel2}
\end{equation}
%%%%%%%%%%%%%%%%%%%%%%%%%%%%%%%%%%%%%%%%%%%%%%%%%%%%%%%%%%%%%%%%%%%%%%%
where $a=\pi/\sqrt{6}$ and $\gamma$ is the Euler constant. In
particular, the corresponding FTG density then becomes
%%%%%%%%%%%%%%%%%%%%%%%%%%%%%%%%%%%%%%%%%%%%%%%%%%%%%%%%%%%%%%%%%%%%%%%
\begin{equation}
\tilde{p}^{\max}(y)=ae^{-(ay+\gamma)-e^{-(ay+\gamma)}}\;.
\label{FTG}
\end{equation}
%%%%%%%%%%%%%%%%%%%%%%%%%%%%%%%%%%%%%%%%%%%%%%%%%%%%%%%%%%%%%%%%%%%%%%%

\subsection{Power-law tailed variables}
\label{sect:ext_power_tails}

Now consider independent identically distributed random variables
where the tail of the complementary cumulative distribution decays in
a power law fashion, i.e., $P_{>}(x) \simeq A/x^\mu$ for large values of $x$.
Assuming again that the dominant contribution to the
statistics of the extremes comes from the tail of the individual
distribution \cite{BOUCHAUD97,BOUCHAUD00,
BALDASSARRI00}, Eq.(\ref{extreme_approx}) yields 
%%%%%%%%%%%%%%%%%%%%%%%%%%%%%%%%%%%%%%%%%%%%%%%%%%%%%%%%%%%%%%%%%%%%%%%
\begin{equation}
P_{<}^{\max}(x) \simeq e^{-NP_{>}(x)} 
\simeq e^{-NA/x^\mu} \;.
\label{raw_power}
\end{equation}
%%%%%%%%%%%%%%%%%%%%%%%%%%%%%%%%%%%%%%%%%%%%%%%%%%%%%%%%%%%%%%%%%%%%%%%
Introducing the scaled variable $\tilde{x}=x/b_{N}$, where  
$b_{N}=(AN)^{1/\mu}$, yields the standard form of the so
called Fr\'echet distribution for the extremes in the asymptotic
large-$N$ limit \cite{FISHER28,GALAMBOS94}
%%%%%%%%%%%%%%%%%%%%%%%%%%%%%%%%%%%%%%%%%%%%%%%%%%%%%%%%%%%
\begin{equation}
\tilde{P}_{<}^{\max}(\tilde{x}) = e^{-1/\tilde{x}^\mu}  \;,
\label{frechet}
\end{equation}
%%%%%%%%%%%%%%%%%%%%%%%%%%%%%%%%%%%%%%%%%%%%%%%%%%%%%%%%%%%
and the corresponding probability density
%%%%%%%%%%%%%%%%%%%%%%%%%%%%%%%%%%%%%%%%%%%%%%%%%%%%%%%%%%%
\begin{equation}
\tilde{p}^{\max}(\tilde{x}) = 
\frac{\mu}{\tilde{x}^{\mu+1}}e^{-1/\tilde{x}^\mu} \;.
\label{frechet_density}
\end{equation}
%%%%%%%%%%%%%%%%%%%%%%%%%%%%%%%%%%%%%%%%%%%%%%%%%%%%%%%%%%%
One can note that the tail behavior of the extremes has been
inherited from that of the original individual variables, i.e.,
%%%%%%%%%%%%%%%%%%%%%%%%%%%%%%%%%%%%%%%%%%%%%%%%%%%%%%%%%%%
$\tilde{p}^{\max}(\tilde{x}) \sim 1/\tilde{x}^{\mu +1}$
%%%%%%%%%%%%%%%%%%%%%%%%%%%%%%%%%%%%%%%%%%%%%%%%%%%%%%%%%%%
for large values of $\tilde{x}$. The first moment of the extreme exist if
$\mu>1$ and for the average value of the largest of the $N$ original
power-law  variables one finds
%%%%%%%%%%%%%%%%%%%%%%%%%%%%%%%%%%%%%%%%%%%%%%%%%%%%%%%%%%%
\begin{equation}
\langle x_{\max}\rangle = b_{N}\langle \tilde{x}\rangle \simeq
\Gamma(1-1/\mu)(AN)^{1/\mu} \sim N^{1/\mu} \;
\label{frechet_mean_scale}
\end{equation}
%%%%%%%%%%%%%%%%%%%%%%%%%%%%%%%%%%%%%%%%%%%%%%%%%%%%%%%%%%%
where $\Gamma(z)$ is Euler's gamma function.
For comparison with experimental or simulation data it is often convenient
to use an alternative scaling for the extremes 
$y=x/\langle x_{\max}\rangle$, yielding collapsing ($N$-independent)
probability density functions similar to Eq.(\ref{frechet_density})
%%%%%%%%%%%%%%%%%%%%%%%%%%%%%%%%%%%%%%%%%%%%%%%%%%%%%%%%%%%
\begin{equation}
\tilde{p}^{\max}(y) = 
\frac{\mu}{\Gamma^{\mu}(1-1/\mu)y^{\mu+1}} e^{-1/(\Gamma^{\mu}(1-1/\mu) y^\mu)} \;.
\label{frechet_scaled_density}
\end{equation}
%%%%%%%%%%%%%%%%%%%%%%%%%%%%%%%%%%%%%%%%%%%%%%%%%%%%%%%%%%%

\section{Extreme Fluctuations in 1D BCS Network}

We consider again the simplest stochastic model with linear relaxation
on a SW network used in the previous chapter [Eq.~(\ref{meaf_KPZ})].
In this chapter in addition to the width, we will study
the scaling behavior of the largest fluctuations (e.g., above the
mean) in the steady-state
%%%%%%%%%%%%%%%%%%%%%%%%%%%%%%%%%%%%%%%%%%%%%%%%%%%%%%%%%%%%%%%%%%%%%
\begin{equation}
\langle\Delta_{\max}\rangle \equiv \langle \tau_{\max}-\bar{\tau}\rangle\;. 
\end{equation}

\begin{figure}[ht]
\vspace{6cm}
\includegraphics{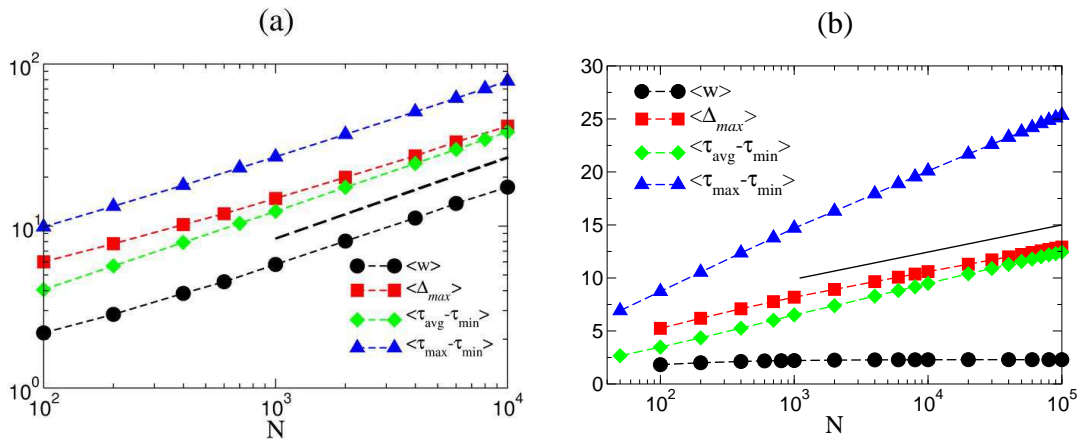}
\includegraphics{G_ext_exp_p0.1_w2L.eps}
\vspace{-0.3cm}
\caption[Width and the extreme fluctuations in regular and SW networks]{(a) Scaling behavior of the average ($w$) and the extreme
(above the mean $\Delta_{\max}$, below the mean $\langle \bar\tau-\tau_{min} \rangle$, 
maximum spread $\langle \tau_{max}-\tau_{min} \rangle$) fluctuations in the virtual time horizon
for the conservative PDES scheme in the steady state.
The processors are connected in a ring-like fashion. Note the log-log
scales. The dashed line represents the theoretical power law with the
roughness exponent $\alpha$$=$$1/2$.
(b) The same quantities as in (a), but the processors are connected
by a small-world topology and the additional synchronization through
the random link is performed with probability $p$$=$$0.10$ at every
parallel step (log-normal scales). The solid straight line indicates the weak
logarithmic increase of the extreme fluctuations with the system size.}
\vspace{-0.5cm}
\label{fig_ext}
\end{figure}

As discussed in Chapter 2 and 3, Eq.~(\ref{meaf_KPZ}) (and its generalization with a
KPZ-like nonlinearity \cite{KARDAR86})
governs the steady-state progress and scalability properties of a
large class of PDES
schemes \cite{KORNISS03,KORNISS00,TOROCZKAI00,KORNISS02,KORNISS01_2}. In this
context, the local height variables $\{\tau_{i}(t)\}_{i=1}^{N}$
correspond to the progress of the individual processors after $t$
parallel steps. The EW/KPZ-type
relaxation at a coarse-grained level originates from the
``microscopic'' (node-to-node) synchronizational rules. 
In the absence of the random links with purely short-range connections,
the corresponding steady-state landscape is rough \cite{BARABASI95}
(de-synchronized state), i.e., it is dominated by large-amplitude
long-wavelength fluctuations. The extreme values of the local
fluctuations emerge through these long-wavelength modes and, 
in one dimension, the
extreme and average fluctuations follow the {\em same} power-law
divergence with the system size \cite{KORNISS02,RAYCHAUDHURI01,MAJUMDAR04,MAJUMDAR04_2,
KORNISS01_2}
%%%%%%%%%%%%%%%%%%%%%%%%%%%%%%%%%%%%%%%%%%%%%%%%%%%%%%%%%%%%%%%%%
\begin{equation}
\langle\Delta_{\max}\rangle \sim w \sim N^{\alpha}\;,
\label{w2_scaling} 
\end{equation}
%%%%%%%%%%%%%%%%%%%%%%%%%%%%%%%%%%%%%%%%%%%%%%%%%%%%%%%%%%%%%%%%%
where $\alpha$ is the roughness exponent \cite{BARABASI95} 
[Fig.~\ref{fig_ext}(a)].

%%%%%%%%%%%%%%%%%%%%%%%%%%%%%%%%%%%%%%%%%%%%%%%%%%%%%%%%%%%%%%%%%%%%%%%%%%
%%%%%%%%%%%%%%%%%%%%%%%%%%%%%%%%%%%%%%%%%%%%%%%%%%%%%%%%%%%%%%%%%%%%%%%%%

The extreme-value limit theorems sketched in the previous section are valid only for independent 
(or short-range correlated)
random variables. Since the heights are strongly correlated in the 1D BCS scheme, the known
extreme-value limit theorems cannot be used. A recent work on this issue
sheds some light on the distribution of the extreme heights in the 1D BCS
\cite{MAJUMDAR04, MAJUMDAR04_2}. Equation~(\ref{w2_scaling}) suggests that the 
normalized probability density function of the maximum relative 
\begin{figure}[htb]
\vspace{8.5cm}
\includegraphics{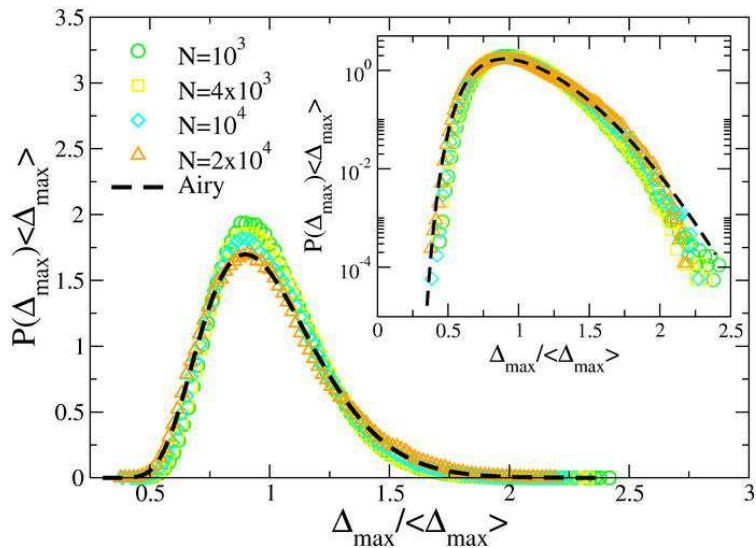}
\vspace{-0.5cm}
\caption[The scaled distribution of the maximum relative height]{The scaled distributions of maximum relative height of the 1D BCS for different
system-sizes. The dashed line
is the Airy distribution function \cite{MAJUMDAR04}. The inset is the log-normal version
of the same data to show the agreement with the theoretical curve in the tail .}
\vspace{-0.5cm}
\label{fig_hmax_airy}
\end{figure}
height $\Delta_{max}$ has a universal scaling form,
$P(\Delta_{max},N)$$\sim$$N^{-\alpha}f(\Delta_{max}/N^\alpha)$. For the 1D EW/KPZ with
periodic boundary conditions ($\alpha$$=$$1/2$), by using
path integral techniques \cite{MAJUMDAR04}
$f(x)$ was found to be the so-called Airy distribution function.
Our simulation results show that the
appropriately scaled maximum relative height distributions are in agreement
with the theoretical distribution from \cite{MAJUMDAR04} [Fig.~\ref{fig_hmax_airy}].

%%%%%%%%%%%%%%%%%%%%%%%%%%%%%%%%%%%%%%%%%%%%%%%%%%%%%%%%%%%%%
\section{Extreme Fluctuations in Small-World-Connected Network}
\label{sect:ext_EW}

The important feature of the EW model on SW networks is the
development of an effective nonzero mass $\gamma(p)$, corresponding to
an actual or pseudo gap in a field theory sense \cite{KOZMA04,MONASSON99,KOZMA04_2}, 
generated by the quenched-random structure 
\cite{KOZMA04}. In turn, both the {\em average} correlation length $\xi\simeq
[\gamma(p)]^{-1/2}$ and the width $w\simeq
(1/\sqrt{2})[\gamma(p)]^{-1/4}$ approach a finite value
(synchronized state) and become self-averaging in the
$N$$\to$$\infty$ limit \cite{GUCLU04_2}. 
Thus, the average correlation length becomes {\em
finite} for an arbitrarily small but nonzero strength of the
random links (one such link per site).
This is the fundamental effect of extending the original dynamics
to a SW network: it decouples the fluctuations of the originally
correlated system. 
Then, the extreme-value limit theorems can be
applied using the number of independent blocks $N/\xi$ in the
system \cite{BOUCHAUD97,BALDASSARRI00}. Further, if the tail of the noise
distribution decays in an exponential-like fashion, the individual
relative height distribution will also do so
\footnote{The exponent $\delta$ for the tail of the local relative height
distribution may differ from that of the noise as a result of the
collective (possibly non-linear) dynamics, but the
exponential-like feature does not change.
}, and
depends on the combination $\Delta_i/w$, where $\Delta_i =
\tau_i$$-$$\bar{\tau}$ is the relative height measured from the mean at
site $i$. Considering, e.g., the fluctuations above the mean for
the individual sites, we will then have
$P_{>}(\Delta_i)\simeq\exp[-c(\Delta_i/w)^{\delta}]$, where
$P_{>}(\Delta_i)$ denotes the ``disorder-averaged'' (averaged over
network realizations) single-site relative height distribution,
which becomes independent of the site $i$ for SW networks. From
the above it follows that the cumulative distribution for the
extreme-height fluctuations relative to the mean
$\Delta_{\max}$$=$$\tau_{\max}$$-$$\bar{\tau}$, if scaled appropriately,
will be given by Eq.~(\ref{gumbel}) [or alternatively by
Eq.~(\ref{gumbel2})] in the 
asymptotic large-$N$ limit (such that $N/\xi$$\gg$$1$). Further,
from Eq.~(\ref{mean}), the average maximum relative height will
scale as
%%%%%%%%%%%%%%%%%%%%%%%%%%%%%%%%%%%%%%%%%%%%%%%%%%%%%%%%%%%%%%%%%%%%%%%
\begin{equation}
\langle \Delta_{\max}\rangle \simeq w
\left[\frac{\ln(N/\xi)}{c}\right]^{1/\delta} \simeq
\frac{w}{c^{1/\delta}}\left[\ln(N)\right]^{1/\delta} \;,
\label{mean_max}
\end{equation}
%%%%%%%%%%%%%%%%%%%%%%%%%%%%%%%%%%%%%%%%%%%%%%%%%%%%%%%%%%%%%%%%%%%%%%%
where we kept only the leading order term in $N$. Note, that both
$w$ and $\xi$ approach their {\em finite} asymptotic
$N$-independent values for SW-coupled systems. Also, the same
logarithmic scaling with $N$ holds for the largest relative
deviations below the mean $\langle\bar{\tau}$$-$$\tau_{\min}\rangle$ and
for the maximum spread $\langle \tau_{\max}$$-$$\tau_{\min}\rangle$.
This weak logarithmic divergence, which one can regard as marginal, ensures
synchronization for practical purposes in SW coupled multi-component
systems with local relaxation in an environment with exponential-like noise.

\begin{figure}[ht]
\vspace{8.5cm}
\includegraphics{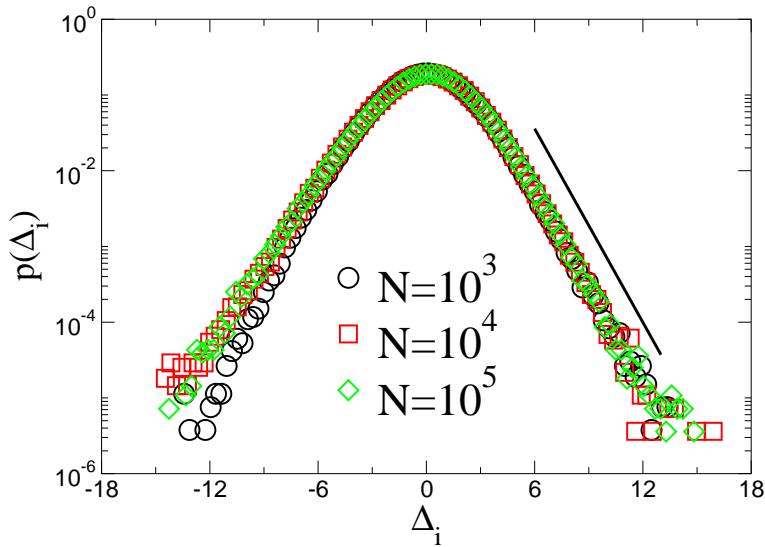}
\vspace{-0.5cm}
\caption[Distribution of the local height fluctuations]{Disorder-averaged probability densities for the local height
fluctuations for the SW-synchronized ($p$$=$$0.10$) landscape for
three system sizes indicated in the figure. Note the log-normal
scales. The solid straight line indicates the exponential tail.}
\vspace{-0.5cm}
\label{fig_P_Delta_i}
\end{figure}

\begin{figure}[ht]
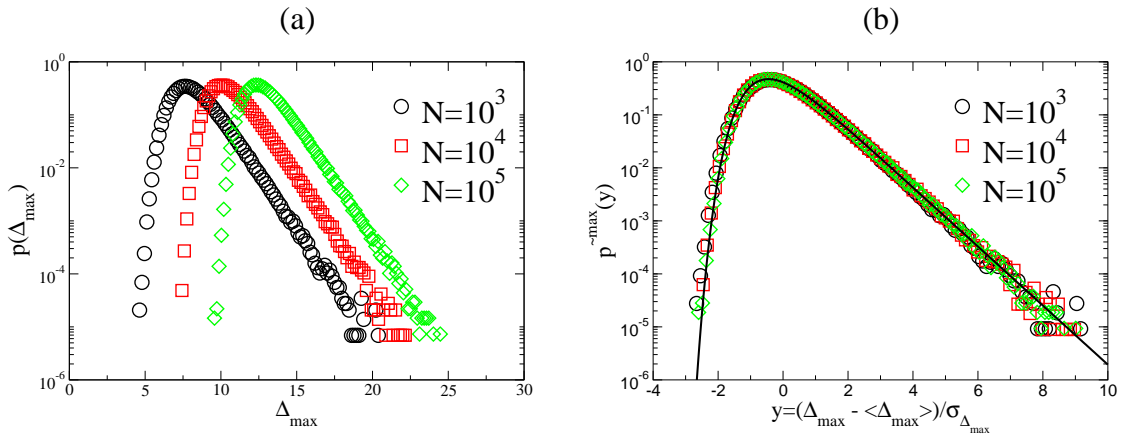

\vspace{5.5cm}
\includegraphics{G_ext_exp_phmax_unscaled.eps}
\includegraphics{G_ext_exp_phmax_scaled.eps}
\vspace{-0.3cm}
\caption[Distribution of the extreme height fluctuations]{(a) Disorder-averaged probability distributions for the extreme height
fluctuations for the SW-synchronized conservative PDES time horizons with
$p$$=$$0.10$ for three system sizes indicated in the figure. Note the
log-normal scales. 
(b) The same as (a), but the probability densities are scaled to zero
mean and unit variance. The solid curve corresponds to the similarly scaled FTG
density Eq.~(\ref{FTG}) for comparison.}
\vspace{-0.5cm}
\label{fig_ftg}
\end{figure}

To study the extreme fluctuations of the SW-synchronized virtual
time horizon, we ``simulated the simulations'', i.e., the
evolution of the local simulated times based on the above exact
algorithmic rules \cite{GUCLU04}. By constructing histograms for $\Delta_i$, we
observed that the tail of the disorder-averaged individual
relative-height distribution decays exponentially ($\delta$$=$$1$)
[Fig.~\ref{fig_P_Delta_i}]. 
%%%%%%%%%%%%%%%%%%%%%%%%%%%%%%%%%%%%%%%%%%%%%%%%%%%%%%%%%%%%%%%%%%%%%%%%%%
%%%%%%%%%%%%%%%%%%%%%%%%%%%%%%%%%%%%%%%%%%%%%%%%%%%%%%%%%%%%%%%%%%%%%%%%%
Then, we constructed histograms for the
extreme-height fluctuations Fig.~\ref{fig_ftg}(a). 
%%%%%%%%%%%%%%%%%%%%%%%%%%%%%%%%%%%%%%%%%%%%%%%%%%%%%%%%%%%%%%%%%%%%%%%%%%
%%%%%%%%%%%%%%%%%%%%%%%%%%%%%%%%%%%%%%%%%%%%%%%%%%%%%%%%%%%%%%%%%%%%%%%%%
The scaled histograms, together with the similarly scaled FTG density
Eq.~(\ref{FTG}), are shown in Fig.~\ref{fig_ftg}(b). 
We also observed that the distribution of the extreme values becomes {\em
self-averaging}, i.e., independent of the network realization.
Figure~\ref{fig_ext}(b) shows that for sufficiently large $N$
(such that $w$ essentially becomes system-size independent) the
average (or typical) size of the extreme-height fluctuations
diverge {\em logarithmically}, according to Eq.~(\ref{mean_max})
with $\delta$$=$$1$. We also found that the largest relative
deviations below the mean $\langle\bar{\tau}$$-$$\tau_{\min}\rangle$,
and the maximum spread $\langle \tau_{\max}$$-$$\tau_{\min}\rangle$
follow the same scaling with the system size $N$ [Fig.~\ref{fig_ext}(b)]. Note, that for
our specific system (PDES time horizon), the ``microscopic''
dynamics is inherently nonlinear, but the effects of the
nonlinearities only give rise to a renormalized mass $\gamma(p)$
(leaving $\gamma(p)$$>$$0$ for all $p$$>$$0$) \cite{KORNISS03}.
Thus, the dynamics is effectively governed by relaxation in a
small world, yielding a finite correlation length and,
consequently, the slow logarithmic increase of the extreme
fluctuations with the system size [Eq.~(\ref{mean_max})]. Also,
for the PDES time horizon, the local height distribution is
asymmetric with respect to the mean, but the average size of the
height fluctuations is, of course, finite for both above and below
the mean. This specific characteristic simply yields different
prefactors for the extreme fluctuations [Eq.~(\ref{mean_max})]
above and below the mean, leaving the logarithmic scaling with $N$
unchanged.

%%%%%%%%%%%%%%%%%%%%%%%%%%%%%%%%%%%%%%%%%%%%%%%%%%%%%%%%%%%%%
\section{Synchronization in the Presence of Power-Law Noise}
\label{sect:ext_power}

Employing SW-like synchronization networks to suppress large
fluctuations was successful in the presence exponential-like ``noise''. We now
investigate the scenario when the noise distribution exhibits a power-law
tail. We consider the synchronization problem from parallel
discrete-event simulations for power-law tailed
asynchrony. The condition for updating the local ``height''
variables in the synchronization landscape (corresponding to the local
virtual times) is {\em unchanged}, i.e., a node is only allowed to increment
its local simulated time $\tau_i$ if it is a local minimum in the virtual
neighborhood (possibly including the random neighbor with probability
$p$). The increment, however, is now drawn from a {\em power-law}
probability density $p(\eta)\sim 1/\eta^{\gamma +1}$.
Since the above local update rule is, essentially, relaxation on
the network, this model also serves as a prototypical model
for relaxation on SW networks in an environment with power-law noise.
The above synchronization rules can be applied to simulating systems 
with non-Poisson asynchrony, relevant to
various transport and transmission phenomena in natural and
artificial systems \cite{MENEZES04,BARABASI04,TOROCZKAI04}.
For example, in Internet or WWW traffic, in part, as a result of
universal power-law tail file-size distributions 
\cite{CROVELLA97,CROVELLA98}, service times exhibit similarly tailed
distributions in the corresponding queuing networks \cite{LELAND94,CSABAI94,PAXSON95}. In turn, when
simulating these systems, the corresponding PDES should use
power-law tail distributed local simulated time increments. 

For a purely one-dimensional connection topology (in the absence of
the random links) we observed kinetic roughening. Since the time to
reach the steady, the relaxation time in the steady state, and the 
surface width all diverge with the number of
nodes, it is difficult to measure the roughness exponent
accurately. It is well documented \cite{HEALY95}, however, that
KPZ-like growth in the presence of power-law noise leads to anomalous
roughening (yielding a roughness exponent greater than $1/2$ in 1D).
%%%%%%%%%%%%%%%%%%%%%%%%%%%%%%%%%%%%%%%%%%%%%%%%%%%%%%%%%%%%%%%%%%%%%%%%%%
\begin{figure}[ht]
\vspace{8.5cm}
\includegraphics{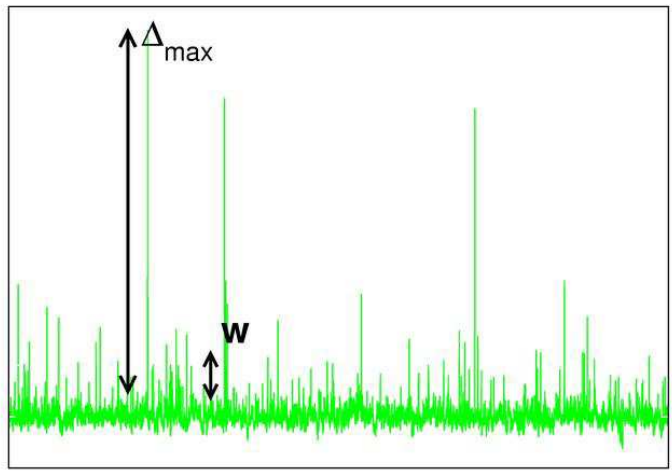}
\vspace{-1cm}
\caption[Virtual time horizon snapshot for SW network with power-law noise]{Snapshot for the SW-synchronized ($p$$=$$0.10$) landscape
in a power-law noise environment ($\gamma$$=$$3$) for $N$$=$$10^4$ nodes.}
\vspace{-0.5cm}
\label{fig_power_height}
\end{figure}
%%%%%%%%%%%%%%%%%%%%%%%%%%%%%%%%%%%%%%%%%%%%%%%%%%%%%%%%%%%%%%%%%%%%%%%%%

%%%%%%%%%%%%%%%%%%%%%%%%%%%%%%%%%%%%%%%%%%%%%%%%%%%%%%%%%%%%%%%%%%%%%%%%%%
\begin{figure}[ht]
\vspace{7cm}
\includegraphics{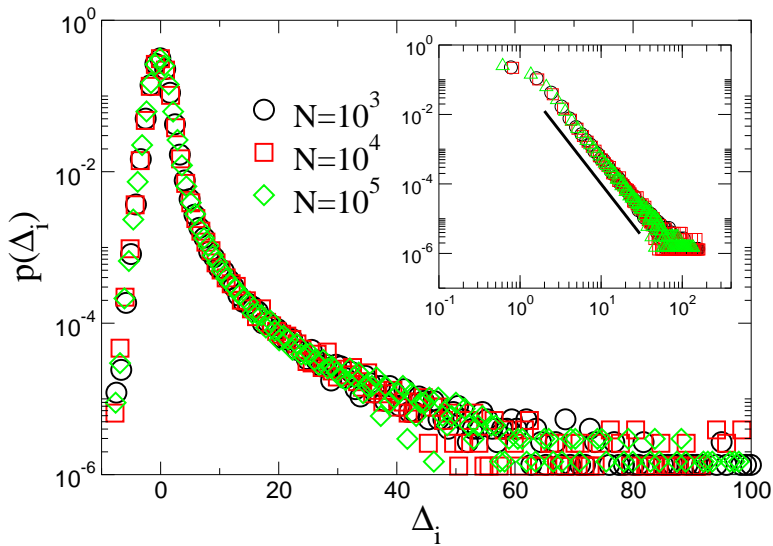}
\vspace{-0.5cm}
\caption[Distribution of the local height fluctuations with power-law noise]{Disorder-averaged probability distributions in a power-law noise 
environment ($\gamma$$=$$3$) with $N$$=$$10^4$ nodes 
for the local height fluctuations for three system sizes indicated in
the figure. Note the log-normal scales. The inset shows the same (for
the positive domain) on log-log scales. The solid line corresponds
to the slope of $\mu+1 \approx 3$.}
\vspace{-0.5cm}
\label{fig_power_height2}
\end{figure}
%%%%%%%%%%%%%%%%%%%%%%%%%%%%%%%%%%%%%%%%%%%%%%%%%%%%%%%%%%%%%%%%%%%%%%%%%

Here we show and discuss results for the power-law noise generated
growth on {\em SW} networks. We have chosen two
values of $\gamma$ governing the tail of the probability density for
the noise: $\gamma$$=$$3$ and $\gamma$$=$$5$. For both of these cases
the noise have a finite mean and variance. 
One can expect a power-law tail (at least for above the mean)
for the probability
density of the individual local height fluctuations $p(\Delta_i)\sim
1/\Delta_{i}^{\mu+1}$, once the noise is ``filtered through'' the
collective dynamics. In Fig.~\ref{fig_power_height} we show a snapshot
for the resulting synchronization landscape, indicating the presence
of some rare but very large fluctuations above the mean.
Since the local update rules lead to nonlinear (KPZ-like) effective
interactions, we could not predict the exponent of the local
height distribution. Instead, we constructed histograms representing
$p(\Delta_i)$. For the above two values of the noise exponent,
$\gamma$$=$$3$ and $\gamma$$=$$5$, we observed power-law tail exponents for 
$p(\Delta_i)\sim 1/\Delta_{i}^{\mu+1}$ as well, but with exponents
clearly differing from that of the noise,
$\mu$$\approx$$2$ and $\mu$$\approx$$4$, respectively. 
Figure~\ref{fig_power_height2} shows $p(\Delta_i)$ for the former.
The figure indicates that for large $\Delta_i$ a power-law tail develops,
while fluctuations below the mean exhibit an exponential-like tail.
This asymmetry is due to the asymmetry in the microscopic update
rules: local minima were {\em incremented} by power-law distributed
random amount, hence anomalously large deviations above the mean can
emerge.

\begin{figure}[htb]
\vspace{8.5cm}
\includegraphics{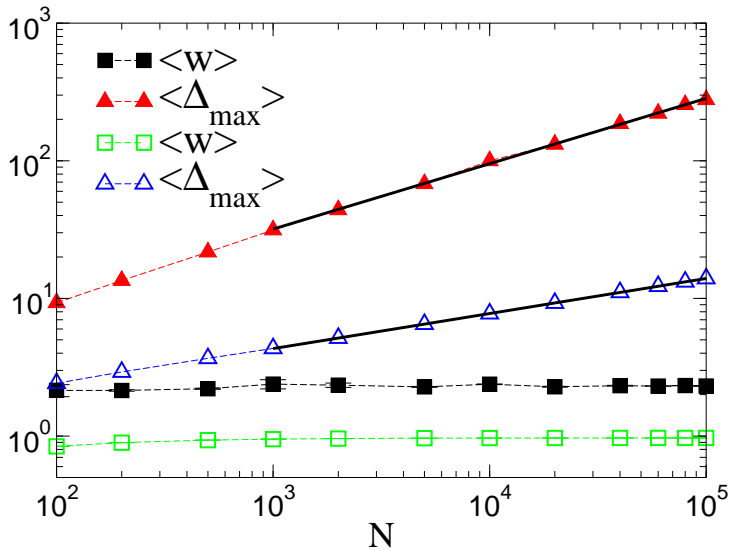}
\vspace{-0.5cm}
\caption[Width and extreme fluctuations in SW network with power-law noise]{Steady-state scaling behavior of the average ($w$) and the extreme
($\Delta_{\max}$) fluctuations in the synchronization landscape
for power-law noise with exponent $\gamma$$=$$3$ (filled symbols) and
$\gamma$$=$$5$ (open symbols) using SW links with $p$$=$$0.10$. The
straight solid-line segments are the best-fit power laws 
($\langle \Delta_{\max}\rangle\sim N^{1/\mu}$ with $1/\mu$$=$$0.47$
and $1/\mu=0.25$, respectively) for the extreme fluctuations 
for system sizes $N \geq 10^3$, according to Eq.~(\ref{frechet_mean_scale}).}
\label{fig_wDelta_scale}
\vspace{-0.5cm}
\end{figure}

\begin{figure}[ht]
\vspace{6cm}
\includegraphics{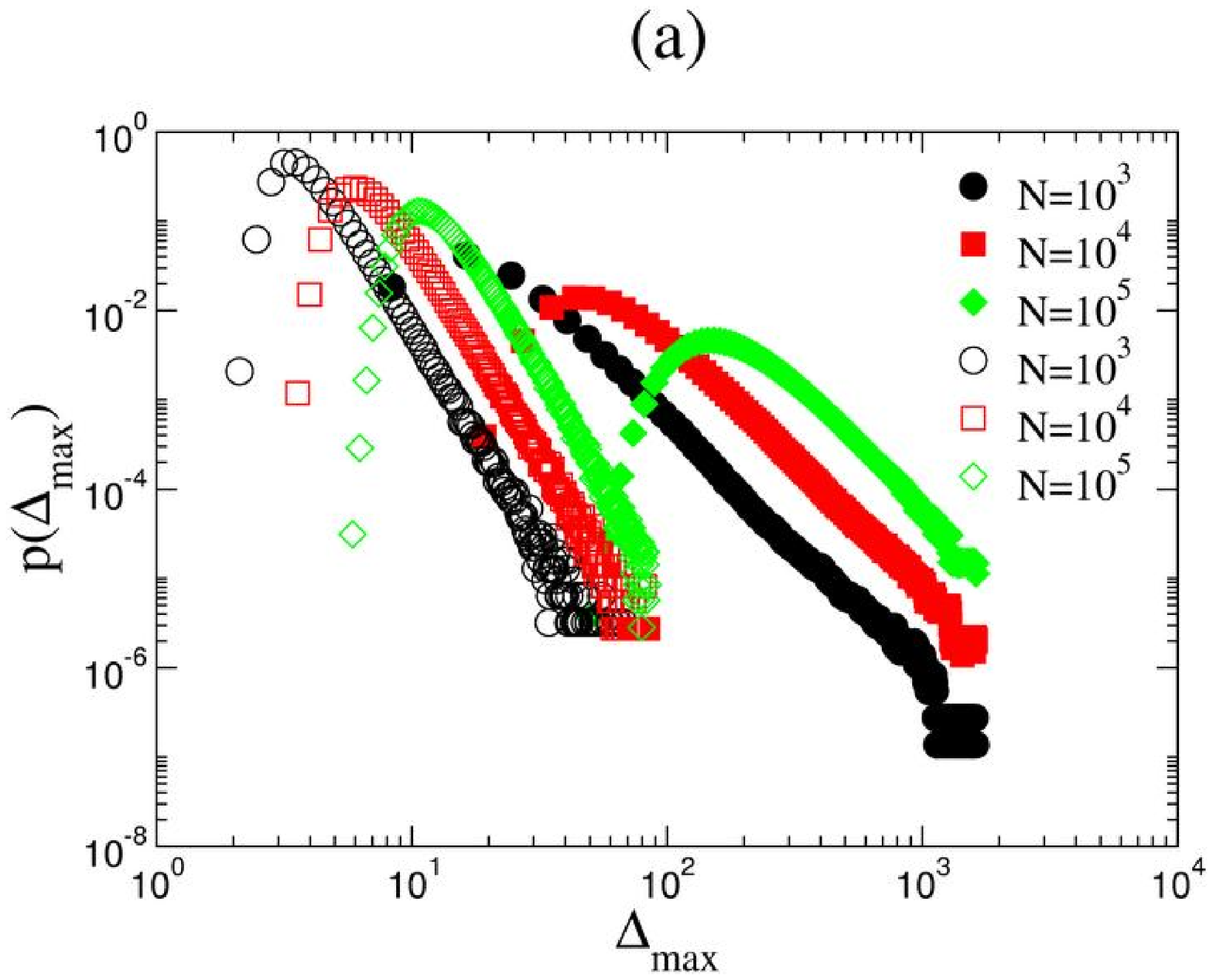}
\includegraphics{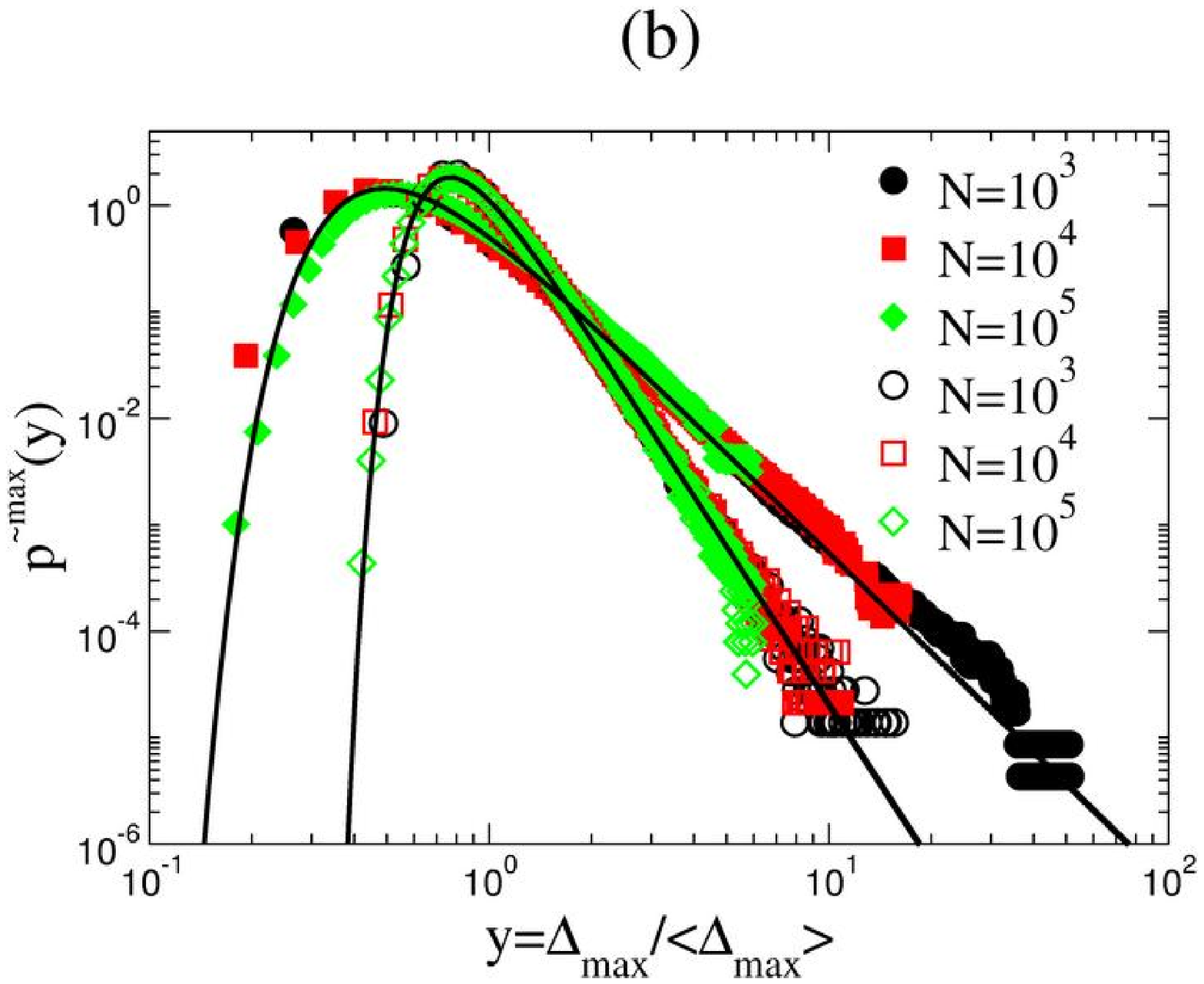}
\vspace{-0.3cm}
\caption[Distribution of the extremes in SW network with power-law noise]{(a) Disorder-averaged probability distributions for the extreme height
fluctuations for the SW-synchronized ($p$$=$$0.10$) landscape
in a power-law noise environment for $\gamma$$=$$3$ (filled symbols)
and $\gamma$$=$$5$ (open symbols) for three system sizes indicated in
the figure. Note the log-log scales.
(b) Scaled form of probability densities in Fig.~\ref{fig_P_Delta_max}. 
The solid curves correspond to the
similarly scaled Fr\'echet density Eq.~(\ref{frechet_scaled_density})
for comparison.} 
\vspace{-0.5cm}
\label{fig_P_Delta_max}
\end{figure}

Then we analyzed the scaling behavior of the average and the
extreme height fluctuations in the associated synchronization landscape.
In the limit of large $N$, $w$ becomes system-size independent, while
the extreme-height fluctuations above the mean diverge in a power-law
fashion according to Eq.~(\ref{frechet_mean_scale})
[Fig.~\ref{fig_wDelta_scale}]. Fitting a power law for 
$N$$\geq$$10^3$ yields $\langle \Delta_{\max}\rangle\sim N^{0.47}$ and 
$\langle \Delta_{\max}\rangle\sim N^{0.25}$ for the two cases analyzed
in Fig.~\ref{fig_wDelta_scale}, for $\gamma$$=$$3$ and $\gamma$$=$$5$,
respectively. 
%%%%%%%%%%%%%%%%%%%%%%%%%%%%%%%%%%%%%%%%%%%%%%%%%%%%%%%%%%%%%%%%%%%%%%%%%%
%%%%%%%%%%%%%%%%%%%%%%%%%%%%%%%%%%%%%%%%%%%%%%%%%%%%%%%%%%%%%%%%%%%%%%%%%
In order to understand the underlying reason for
this divergence, we analyzed the histograms constructed for the
probability density of the extreme height fluctuations $p(\Delta_{\max})$ 
[Fig.~\ref{fig_P_Delta_max}]. The shapes of these histograms suggest
that the limit distribution is of Fr\'echet type. We constructed the
histograms for the scaled variable $y=\Delta_{\max}/\langle
\Delta_{\max}\rangle$. Then using $\mu=1/0.47=2.1$ and
$\mu=1/0.25=4$ as implied by the scaling behavior of
$\langle\Delta_{\max}\rangle$ [Eq.~(\ref{frechet_mean_scale})], 
we plotted the similarly scaled Fr\'echet
density Eq.~(\ref{frechet_scaled_density})
[Fig.~\ref{fig_P_Delta_max}(b)]. 
These results indicate that the effect of the random links in SW
networks is again to decouple the local field variables, an in turn,
the statistics of the extremes are governed by the Fr\'echet
distribution. Consequently, the average size of the extremes diverges
in a power-law fashion $\langle \Delta_{\max}\rangle\sim
N^{1/\mu}$. This power-law divergence is {\em not} the result of a
divergent length scale emerging from the cooperative effects of the
interacting nodes. On the contrary, the local field variables  become
effectively independent using SW synchronization. The tail behavior for them
(power-law with a possibly different exponent), however, is inherited
from the noise. Hence, the statistics of the extremes will be of the
Fr\'echet type, yielding a power-law increase of the average size of the
largest fluctuations above the mean.

%%%%%%%%%%%%%%%%%%%%%%%%%%%%%%%%%%%%%%%%%%%%%%%%%%%%%%%%%%%%%%%%%%%%%%%%%%
%%%%%%%%%%%%%%%%%%%%%%%%%%%%%%%%%%%%%%%%%%%%%%%%%%%%%%%%%%%%%%%%%%%%%%%%%

The above picture is reasonably consistent in that the exponents for the
tail behavior ($\sim$$1/\Delta_{i}^{\mu+1}$) for both the probability density
of the local heights $p(\Delta_i)$ and the extremes
$p(\Delta_{\max})$ were within about $6\%$. Further, the average size
of the extremes increases as $N^{1/\mu}$, in accordance with the
underlying Fr\'echet distribution. 

It is interesting to note that for $\mu$$\approx$$2$, formally, 
$p(\Delta_i)$ does not have a finite variance (associated with the width
$w=\sqrt{\langle (\tau_i-\bar{\tau})^2\rangle}=
\sqrt{\langle (\Delta_i)^2\rangle}$). 
Indeed, in the
simulations we observed large fluctuations in $w$ and error bars
of the order of the width itself. The ``theoretical'' divergence for
$\mu$$=$$2$ is, of course, limited by the logarithm of a large but
finite cutoff in the simulations. This anomalous (formally
divergent) width is not related to a system-size dependent
widening of the individual distributions controlled by a divergent
correlation length. Rather, the individual distributions develop a
heavy-tailed shape independent of the system size. 
 
Examining the largest
fluctuations below the mean reveals that they increase only
logarithmically with the system size. This is simply the result of the
exponential or similar tail of the individual local height fluctuations below the
mean [Fig.~\ref{fig_power_height}], where the governing limit
distribution is of the Gumbel type (Sec.~\ref{sect:ext_short_tails}). 

The above results show, that SW synchronization can be efficient to
control the {\em average} size of the fluctuations, but the largest
fluctuations still diverge in a power-law fashion with the number of
nodes. While the SW-network effectively decouples the fluctuations in
the synchronization landscape, it cannot 
suppress power-law tails already present in local noise
distribution. In fact, the inherited power-law tails for the local
height fluctuations are even ``heavier'' than that of the corresponding noise,
$\mu$$\approx$$2$ for $\gamma$$=$$3$ and $\mu$$\approx$$4$ for
$\gamma$$=$$5$.

%% file: rpichap5.tex
%%%%%%%%%%%%%%%%%%%%%%%%%%%%%%%%%%%%%%%%%%%%%%%%%%%%%%%%%%%%%%%%%%% 
%                                                                 %
%                            CHAPTER FIVE                         %
%                                                                 %
%%%%%%%%%%%%%%%%%%%%%%%%%%%%%%%%%%%%%%%%%%%%%%%%%%%%%%%%%%%%%%%%%%% 
 
\chapter{SUMMARY AND FUTURE WORK}

\section{Summary}

We studied synchronization phenomena in networks in general, 
and considered the scalability problem of the basic conservative synchronization schemes
to test our results. Based on a mapping \cite{KORNISS00} between the evolution of the
virtual time horizon for the basic conservative PDES scheme
\cite{LUBACHEVSKY87,LUBACHEVSKY88} and kinetically grown non-equilibrium surfaces
\cite{BARABASI95}, we constructed a coarse-grained description for the
scalability and performance of such large-scale parallel simulation schemes.
These schemes can be applied to large spatially extended systems with
short-range interactions and asynchronous dynamics.
The one-site-per PE basic PDES was shown to exhibit KPZ-like kinetic roughening.
This scheme is scalable in that the average progress rate of the PEs
approaches a non-zero value. The spread of the virtual time horizon,
however, diverges as the square root of the number of PEs, leading to
``de-synchronization'' and difficulties in data management.

In this work we considered the simplest (and in some regards, the
worst case) scenario, where each nodes carries one site of the
underlying physical system, hence synchronization with nearest
neighbor PEs is required at every step. In actual parallel implementations the
efficiency can be greatly increased by hosting many sites by each
PE \cite{KORNISS99,AMAR_PRB_2005a,AMAR_PRB_2005b}.
That way, communication between PEs is only required when
local variables are to be updated on the boundary region of the
sites hosted by the PEs (within the finite range of the
interactions). While the above procedure clearly increases the
utilization and reduces the actual communication overhead, it
gives rise to an even faster growing early time regime in the
simulated time horizon \cite{KOLA_PRE_2004}. Since the PEs rarely need to
synchronize, up to some crossover time, the evolution of the time
horizon is governed by random deposition \cite{BARABASI95}, a faster
roughening growth, before eventually crossing over to the KPZ
growth and a subsequent saturation.

Our goal here was to achieve synchronization without any
global intervention. We constructed a specific version of the SW
network, where each PE was connected to exactly one other randomly
chosen PE. The extra synchronizational steps through the random links are
merely used to control the width. The virtual time horizon for the
SW-synchronized PDES scheme becomes ``macroscopically'' smooth and
essentially exhibits mean-field like characteristics. The random
links, on top of a regular lattice, generate an effective ``mass'' for
the propagator of the virtual time horizon, corresponding to a nonzero
correlation length. The width becomes finite, for an arbitrary small
rate of synchronization through the random links, while the
utilization remains nonzero, yielding a fully scalable PDES scheme.
The former statement is only marginally weakened by observing that the
extreme fluctuations in the time horizon can exhibit logarithmically
large values as a function of the total number of PEs. The above
predictions of the coarse-grained PDES model were confirmed by actually
``simulating the simulations''.
The generalization when random links are added to a higher-dimensional
underlying regular lattice is clear: since the synchronization landscape of the 1D SW
network is already macroscopically smooth, in higher dimensions it will be even more so
\cite{KOZMA04} (i.e., the critical dimension of the underlying regular substrate
is less than one).

We also studied the scalability properties for a causally
constrained PDES scheme hosted by a network of computers where the
network is {\em scale-free} following a ``preferential attachment''
construction \cite{BARABASI99,BARABASI99_2}. Here the PEs have to satisfy the general
criterion that their simulated time should be smaller than that of all of their links'
simulated times in order to advance their local time. 
Despite some nodes in the network having abnormally large connectivity
(as a result of the scale-free nature of the degree distribution), we
found that the computational phase of the algorithm is only
marginally non-scalable. The utilization exhibited slow
logarithmic decay as a function of the number of PEs. At the same
time, the width of the time horizon diverged logarithmically slowly,
rendering the measurement phase of the simulations marginally
non-scalable as well. The implication of this
finding is that the internet, which is already exploited for
distributed computing for mostly ``embarrassingly parallel'' problems
through existing GRID-based schemes \cite{GRID,KIRKPATRICK03}, may have the
potential to accommodate efficient complex system simulations (such as
asynchronous PDES) where the nodes frequently have to synchronize with
each other. An intriguing question to pursue is how the
logarithmic divergence of the surface fluctuations observed here can be
related to the collective behavior (in particular, the finite-size
effects of the magnetic susceptibility) of Ising ferromagnets on
scale-free networks \cite{ALEKSIEJUK02,LEONE02,DOROGOVTSEV02_2,BIANCONI02} with the same
degree distribution. 

We also considered the extreme-height fluctuations in this prototypical
model with local relaxation, unbounded local variables, and in the
presence of exponential or power-law tailed noise. 
We showed that when the interaction topology
is extended to include random links in a SW fashion, the local height
variables become effectively independent and the
statistics of the extremes is governed by the FTG or the Fr\'echet
distribution, respectively. 
For both types of noise, the average width of the synchronization
landscape becomes independent of the system size. The extreme
fluctuations increase only logarithmically with the number of nodes
for exponential-like noise and in a power-law fashion for the power-law noise.
These findings directly addresses synchronizability in generic
SW-coupled systems where relaxation through the links is the
relevant node-to-node process and effectively governs the
dynamics. We illustrated our results on an actual
synchronizational problem in the context of scalable parallel
simulations.

Our findings are also closely related to critical phenomena and
collective phenomena on networks
\cite{ALBERT02,DOROGOVTSEV02,STROGATZ01,GOLTSEV03}. In particular, in recent
years, a number of prototypical models have been investigated on
SW networks
\cite{WATTS98,SCALETTAR91,
GITTERMAN00,BARRAT00,KIM01,HONG02_2,
KOZMA04,KOZMA05,KOZMA05b,MONASSON99,BLUMEN_2000a,BLUMEN_2000b,
ALMAAS_2002,HASTINGS04,HONG02,WATTS99,NEWMAN00,NEWMAN99,PEKALSKI01,HONG02_3,HERRERO02,JEONG03,HASTINGS03}.
Of these, the ones most closely related to our work are the XY model
\cite{KIM01}, the EW model \cite{KOZMA04,KOZMA05,KOZMA05b}, and
diffusion \cite{KOZMA04,KOZMA05,KOZMA05b,MONASSON99,BLUMEN_2000a,BLUMEN_2000b,ALMAAS_2002,HASTINGS04} on SW networks.
The findings suggest that systems without inherent frustration
exhibit (strict or anomalous) \cite{KOZMA04,KOZMA05,KOZMA05b,HASTINGS04,HASTINGS03} mean-field-like
behavior when the original short-range interaction topology is
modified to a SW network.
In essence, the SW couplings, although sparse, induce an effective
relaxation to the mean of the respective local field variables, and in
turn, the system exhibits a mean-field-like behavior \cite{HASTINGS03}.
This effect is qualitatively similar to those observed in models with
``annealed'' long-range random couplings \cite{BLUMEN_2000b,DROZ90,BERGERSEN91},
but on (quenched) SW networks, the scaling properties can differ from those with annealed interactions
\cite{KOZMA04,KOZMA05,KOZMA05b,HASTINGS04,HASTINGS03}.

\begin{section}{Future Work}

In many real networks the cost of the random and arbitrarily long-range links could
be unaffordable. Recent works show that in cortical \cite{LAUGHLIN03} and on-chip 
logic networks \cite{DAVIS98}
power-law-suppressed link-length distributions are observed because these
are spatially embedded networks with wiring-costs \cite{PETERMANN05}. To study these
networks one can consider a special model of SW network with distance-dependent power-law
probability distribution of long-range links, where the probability of two
nodes being connected is $r^{-\alpha}$, with a varying exponent $\alpha$ and the distance $r$
between the nodes \cite{KOZMA05,KOZMA05b}. By varying the exponent $\alpha$, one can
control the distribution of the random links. Changing $\alpha$ changes the topology of the network
from the ``plain'' SW in which there is no wiring cost as we discussed in Chapter 3 ($\alpha$$=$$0$), 
to a short-range network in which only nearest neighbor links are present ($\alpha$$=$$\infty$).

The synchronization problem described in the previous chapters can also be studied on this network.
The roughness of the time-surface as a function of the exponent $\alpha$ 
can give valuable information about the conditions on the scalability of the PDES.
One can follow the same way of constructing the network; one random link per PE (chosen with
distance-dependent probability) in addition to the nearest neighbors.
The width of the linear EW model obtained through adjacency matrix diagonalization
exhibits a smooth transition from system-size
independent width (plain SW) to the KPZ width dependence ($\langle w^2 \rangle$$\sim$$N$) [Fig.~\ref{fig_r_alpha}(a)].
%%%%%%%%%%%%%%%%%%%%%%%%%%%%%%%%%%%%%%%%%%%%%%%%%%%%%%%%%%%%%%%%%%%%%%%%%%%%%%%%%%%%%%%%
\begin{figure}[htb]
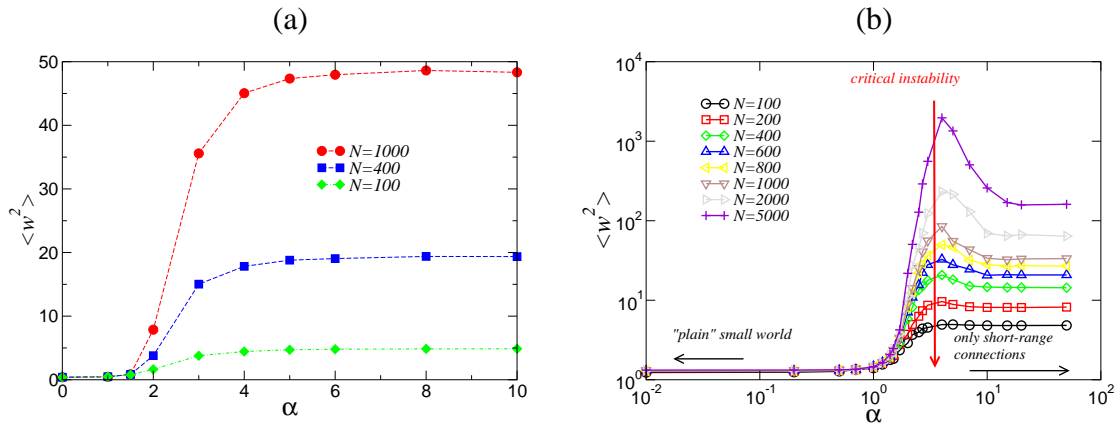

\vspace{6cm}
\includegraphics{r_alpha_EW.eps}
\includegraphics{r_alpha_SW.eps}
\vspace{-0.5cm}
\caption[Width for the power-law SW network]{(a) Width as a function of $\alpha$ for the EW model on a $1/r^\alpha$ SW network. 
(b) Width as a function of $\alpha$ for
the conservative synchronization dynamics on a $1/r^\alpha$ SW network.}
\vspace{-0.5cm}
\label{fig_r_alpha}
\end{figure}
%%%%%%%%%%%%%%%%%%%%%%%%%%%%%%%%%%%%%%%%%%%%%%%%%%%%%%%%%%%%%%%%%%%%%%%%%%%%%%%%%%%%%%%%
Simulating the ``microscopic" dynamics in the actual PDES (incrementing the ``local'' minima) can give rise to
nonlinear effects, typical for the KPZ surface, as can be seen in Fig.~\ref{fig_r_alpha}(b).
We see the transition from plain SW to KPZ again but through critical instability. By
analyzing the Fig.~\ref{fig_r_alpha}(b) we can argue that for some critical range of $\alpha$,
the scalability of the PDES is worse than the BCS scheme (short-range network). Another
surface growth model, single-step model (KPZ class), has also the same effect suggesting that
as long as there is nonlinearity in the growth, the similar peaks in the width will be seen.

Another possible future work might be to study the scale-free Barab\'asi-Albert network further.
In the preferential attachment process to construct the BA network at each time step we added
only one node to the network ($m$$=$$1$). This led to logarithmic (marginal) scalability of the
PDES scheme. It might help to increase $m$ in suppressing the virtual time fluctuations and thus
making the PDES scheme fully scalable in this network as well.

\end{section}

%% file: rpibib.tex
%%%%%%%%%%%%%%%%%%%%%%%%%%%%%%%%%%%%%%%%%%%%%%%%%%%%%%%%%%%%%%%%%%% 
%                                                                 %
%                           BIBLIOGRAPHY                          %
%                                                                 %
%%%%%%%%%%%%%%%%%%%%%%%%%%%%%%%%%%%%%%%%%%%%%%%%%%%%%%%%%%%%%%%%%%% 
 
%This method produces a numbered bibliography where the numbers
%correspond to the \cite commands in the text. See the LaTeX manual.

\specialhead{LITERATURE CITED}

\begin{singlespace}

\end{singlespace}
   
%This is an alternative method.  It's an unnumbered bibliography
%with hanging indentation.
%\specialhead{BIBLIOGRAPHY}
%\begin{singlespace}
%\bibentry This is the first item in the Bibliography.
%Let's make it very long so it takes more than one line.
%Let's make it very long so it takes more than one line.
%Let's make it very long so it takes more than one line.
%Let's make it very long so it takes more than one line.
%Let's make it very long so it takes more than one line. 
%\bibentry The second item in the Bibliography.
%\bibentry Another item in the Bibliography.
  
%\end{singlespace} 

%% file: rpiapp.tex
%%%%%%%%%%%%%%%%%%%%%%%%%%%%%%%%%%%%%%%%%%%%%%%%%%%%%%%%%%%%%%%%%%%
%                                                                 %
%                            APPENDICES                           %
%                                                                 %
%%%%%%%%%%%%%%%%%%%%%%%%%%%%%%%%%%%%%%%%%%%%%%%%%%%%%%%%%%%%%%%%%%%
 
\appendix    % This command is used only once!
%\addcontentsline{toc}{chapter}{APPENDICES}             %toc entry  or:
\addtocontents{toc}{\parindent0pt\vskip12pt APPENDICES} %toc entry, no page #

\chapter{Steady-State Structure Factor in Linear Growth Models}

The time evolution of linearly interacting local field variables
on a network can be written in the form of a Langevin equation as
\begin{equation}
\partial_t\tau_i(t) = -\sum_j\Gamma_{ij}\tau_j(t)+\eta_i(t)
\;,
\label{eq1}
\end{equation}
where $\Gamma_{ij}$ is the coupling matrix containing the topology of the network,
and $\eta_i(t)$ is the noise delta-correlated in space/time with
zero average as expressed in the following equations
\begin{equation}
\langle \eta_i(t) \rangle = 0
\label{eq2}
\end{equation}
and
\begin{equation}
\langle \eta_i(t)\eta_j(t^\prime) \rangle = 2D\delta_{i,j}\delta(t-t^\prime)
\;.
\label{eq3}
\end{equation}
For the nearest-neighbor network $\Gamma_{ij}$ is the discrete Laplacian
\begin{equation}
\Gamma_{ij}=\Gamma_{ij}^0 = 2\delta_{i,j}-\delta_{i-1,j}-\delta_{i+1,j}
\;.
\label{eq4}
\end{equation}
For the maximal-distance network
\begin{equation}
\Gamma_{ij}=\Gamma_{ij}^0 + \gamma(\delta_{i,j}-\delta_{i-N/2,j})
\;.
\label{eq5}
\end{equation}
For the fully-connected network
\begin{equation}
\Gamma_{ij}=\Gamma_{ij}^0 + \gamma(\delta_{i,j}-\frac{1}{N})
\;.
\label{eq6}
\end{equation}

We consider cases where $\Gamma_{ij}$ is translationally invariant, i.e., it does not depend on
$i$ and $j$ explicitly but depends only on the distance $l$$=$$i-j$ between them,
\begin{equation}
\Gamma_{ij} = \Gamma(i-j) = \Gamma(l)
\;.
\label{eq6_2}
\end{equation}

The structure factor [as defined through Eq.~(\ref{str_factor})] contains all
the physics we need to describe the evolution of the network. One needs
the Fourier transforms of the local field variables defined as

\begin{equation}
\tilde\tau_k(t) = \sum_{j=1}^N e^{-ikj}[\tau_j(t)-\bar\tau(t)]
\;.
\label{eq7}
\end{equation}

Taking the Fourier transformation of Eq.~(\ref{eq1}) one finds
\begin{equation}
\partial_t\tilde\tau_k(t) = -\tilde\Gamma(k)\tilde\tau_k(t)+\tilde\eta_k(t)
\;,
\label{eq8}
\end{equation}
where $\tilde\Gamma(k)$ and $\tilde\eta_k(t)$ are the Fourier transforms
of $\Gamma(l)$ and $\eta_i(t)$, respectively. The wave number $k$
goes from $1$ to $N$$-$$1$ since we exclude the zero-mode contribution,
$\tilde\tau_0(t)$$\equiv $$0$ for all $t$. We can see that the evolution
decouples for different $k$ values in Fourier space. The second moment of the
Fourier transform of the noise is
\begin{equation}
\langle \eta_k(t)\eta_{k^\prime}(t^\prime) \rangle = 
2DN\delta_{k+k^\prime,0}\delta(t-t^\prime)
\;.
\label{eq9}
\end{equation}

Integrating Eq.~(\ref{eq8}) we obtain
\begin{equation}
\tilde\tau_k(t) = e^{-\tilde\Gamma(k)t}\int_0^t dt^\prime e^{\tilde\Gamma(k)t^\prime}\tilde\eta_k(t^\prime)
\;.
\label{eq10}
\end{equation}

By using Eq.~(\ref{eq10}) one can write the equal-time 
correlations for the local field variables as
\begin{equation}
\langle \tilde\tau_k(t) \tilde\tau_{k^\prime}(t) \rangle = 
e^{-[\tilde\Gamma(k)+\tilde\Gamma(k^\prime)]t}\int dt^\prime \int dt^{\prime\prime}
e^{\tilde\Gamma(k)t^\prime+\tilde\Gamma(k^\prime)t^{\prime\prime}} 
\langle \tilde\eta_k(t^\prime) \tilde\eta_{k^\prime}(t^{\prime\prime})\rangle
\;.
\label{eq11}
\end{equation}
By substituting the second moment of the Fourier transform of the noise
[Eq.~(\ref{eq9})] into Eq.~(\ref{eq11}), and by using the basic
property of the delta function $\delta(t^\prime-t^{\prime\prime})$ in the integral, one obtains
\begin{equation}
\langle \tilde\tau_k(t) \tilde\tau_{k^\prime}(t) \rangle = 
2DN\delta_{k+k^\prime,0}e^{-[\tilde\Gamma(k)+\tilde\Gamma(k^\prime)]t} \int_0^t dt^\prime
e^{[\tilde\Gamma(k)+\tilde\Gamma(k^\prime)]t^\prime}
\;.
\label{eq12}
\end{equation}
The integral in the equation above can be evaluated easily and one obtains
\begin{equation}
\langle \tilde\tau_k(t) \tilde\tau_{k^\prime}(t) \rangle = 
2DN\delta_{k+k^\prime,0}e^{-[\tilde\Gamma(k)+\tilde\Gamma(k^\prime)]t} 
\frac{e^{[\tilde\Gamma(k)+\tilde\Gamma(k^\prime)]t}-1}{\tilde\Gamma(k)+\tilde\Gamma(k^\prime)}
\;.
\label{eq13}
\end{equation}
After rearranging the terms and rewriting the delta function as 
$\delta_{k+k^\prime,0}$$=$$\delta_{k^\prime,-k}$, one obtains
\begin{equation}
\langle \tilde\tau_k(t) \tilde\tau_{k^\prime}(t) \rangle = 
\frac{2DN\delta_{k^\prime,-k}}{\tilde\Gamma(k)+\tilde\Gamma(-k)}  
\{1-e^{-[\tilde\Gamma(k)+\tilde\Gamma(-k)]t}\}
\;.
\label{eq14}
\end{equation}
From the present form of Eq.~(\ref{eq14}) and by using Eq.~(\ref{str_factor})
one can deduce the general time-dependent structure factor as
\begin{equation}
S(k,t) = \frac{2D}{\tilde\Gamma(k)+\tilde\Gamma(-k)}\{1-e^{-[\tilde\Gamma(k)+\tilde\Gamma(-k)]t}\}
\;.
\label{eq15}
\end{equation}
In the steady-state ($t$$\to$$\infty$) the structure factor
becomes
\begin{equation}
S(k) = \lim_{t\to\infty} S(k,t) = \frac{2D}{\tilde\Gamma(k)+\tilde\Gamma(-k)}
\;.
\label{eq16}
\end{equation}

The steady-state structure factor can be calculated easily
once the Fourier transform of the coupling function 
$\tilde\Gamma(k)$ is known. Now we calculate
the structure factors for a few simple interaction topologies.

\section{Nearest-neighbor network}

The coupling matrix $\Gamma_{ij}$ for the nearest-neighbor network
is a Laplacian,
\begin{equation}
\Gamma_{ij}=\Gamma_{ij}^0 = 2\delta_{i,j}-\delta_{i-1,j}-\delta_{i+1,j}
\;,
\label{eq17}
\end{equation}
and one can rewrite the equation above by using the distance-dependent coupling function
\begin{equation}
\Gamma^0(l) = 2\delta_{l,0}-\delta_{l,1}-\delta_{l,-1}
\;,
\label{eq17_2}
\end{equation}
then the Fourier transform of $\Gamma^0(l)$ becomes
\begin{equation}
\tilde\Gamma^0(k)=(2-e^{ik}-e^{-ik})=2[1-\cos(k)]
\;.
\label{eq18}
\end{equation}
So the structure factor is, as in Eq.~(\ref{sf_toro}),
\begin{equation}
S(k) = \frac{D}{2[1-\cos(k)]}
\;.
\label{eq19}
\end{equation}

\section{Maximal-distance network}

Starting with the coupling matrix for the
maximal-distance network, one obtains the distance-dependent coupling function as
\begin{equation}
\Gamma(l) = \Gamma^0(l)+\gamma(\delta_{l,0}-\delta_{l,\frac{N}{2}})
\label{eq20}
\;.
\end{equation}
Then the Fourier transform $\tilde\Gamma(k)$ becomes, 
\begin{equation}
\tilde\Gamma(k) = \tilde\Gamma^0(k)+\gamma(1-e^{\frac{ikN}{2}})
=2[1-\cos(k)]+\gamma(1-e^\frac{ikN}{2}).
\label{eq20_2}
\;.
\end{equation}
By using Eq.~(\ref{eq16}) we obtain
\begin{equation}
S(k) = \frac{D}{2[1-\cos(k)]+\gamma[1-\cos(\frac{kN}{2})]}
\label{eq21}
\;.
\end{equation}

\section{Fully-connected network}

In this network, as we mentioned in Chapter 3, each node is
connected to all other nodes with strength $\gamma/N$ and the 
coupling function is
\begin{equation}
\Gamma(l) = \Gamma^0(l)+\gamma(\delta_{l,0}-\frac{1}{N})
\label{eq22}
\;.
\end{equation}
Its Fourier transform becomes
\begin{equation}
\tilde\Gamma(k) = \tilde\Gamma^0(k)+\gamma=2[1-\cos(k)]+\gamma
\label{eq22_2}
\;.
\end{equation}
Thus one can obtain the structure factor as
\begin{equation}
S(k) = \frac{D}{2[1-\cos(k)]+\gamma}
\label{eq23}
\;.
\end{equation}

%% file: rpithes.bbl
\begin{thebibliography}{99}

\bibitem{BARYAM03}
Y. Bar-Yam,
\textit{Dynamics of Complex Systems (Studies in Nonlinearity)}
(Westview Press, Boulder, CO, 1997). [\texttt{http://necsi.org/publications/dcs/index.html}]

\bibitem{ALBERT02}
R. Albert and A.-L. Barab\'asi,
``Statistical mechanics of complex networks'',
\textit{Rev. Mod. Phys.} \textbf{74}, pp. 47--97 (2002).
[cond-mat/0106096] \footnote{Where available the references to the
preprint server, arxiv.org are given. Papers may be downloaded from 
\texttt{http://arxiv.org/abs/*}, where \texttt{*} is the reference. 
For example; the link for the paper cond-mat/0311575 is
\texttt{http://arxiv.org/cond-mat/0311575}. For preprints this is a 
primary reference; for published papers there may be differences 
between the published versions and the preprint.}

\bibitem{DOROGOVTSEV02}
S.N. Dorogovtsev and J.F.F. Mendes,
``Evolution of networks'',
\textit{Adv. Phys.} \textbf{51}, pp. 1079--1187 (2002).
[cond-mat/0106144] 

\bibitem{NEWMAN03}
M.E.J. Newman,
``The structure and function of complex networks'',
\textit{SIAM Review} \textbf{45}, pp. 167--256 (2002).
[cond-mat/0303516]

\bibitem{BARABASI99}
A.-L. Barab\'asi and R. Albert,
``Emergence of scaling in random networks'',
\textit{Science} \textbf{286}, pp. 509--512 (1999).
[cond-mat/9910332]

\bibitem{FALOUTSOS99}
M. Faloutsos, P. Faloutsos, and C. Faloutsos,
``On power-law relationships of the Internet topology'',
\textit{Proc. ACM SIGCOMM, Comput. Commun. Rev.} \textbf{29},  
pp. 251--262 (1999).

\bibitem{LERNER03}
E.J. Lerner,
``What's wrong with the electric grid?'',
\textit{The Industrial Physicist}, Oct./Nov., pp. 8--13 (2003).

\bibitem{KORNISS03}
G. Korniss, M.A. Novotny, H. Guclu, Z. Toroczkai, and P.A. Rikvold, 
``Suppressing roughness of virtual times in parallel discrete-event simulations'',
\textit{Science} \textbf{299}, pp. 677--679 (2003).
[cond-mat/0302050]

\bibitem{RABANI98}
Y. Rabani, A. Sinclair, and R. Wanka,
``Local divergence of Markov chains and the analysis of iterative
load-balancing schemes'', in 
\textit{Proceedings of the 39th Annual Symposium on Foundations of
Computer Science}, pp. 694--703 (IEEE Comput. Soc., Los Alamitos, CA, 1998).

\bibitem{PETERSON00}
L.L. Peterson and B.S. Davie, 
\textit{Computer Networks, A Systems Approach},
2nd edn (Morgan Kaufmann, San Francisco, CA, 2000).

\bibitem{ALBERT99}
R. Albert, H. Jeong, and A.-L. Barab\'asi,
``Internet: Diameter of the World-Wide Web'',
\textit{Nature} \textbf{401}, pp. 130--131 (1999).

\bibitem{JEONG00}
H. Jeong, B. Tombor, R. Albert, Z.N. Oltvai, and A.-L. Barab\'asi,
``The large-scale organization of metabolic networks'',
\textit{Nature} \textbf{407}, pp. 651--654 (2000).
[cond-mat/0010278]

\bibitem{EUBANK04}
S. Eubank, H. Guclu, V.S.A. Kumar, M.V. Marathe, A. Srinivasan, Z. Toroczkai, and N. Wang,
``Modelling disease outbreaks in realistic urban social networks'',
\textit{Nature} \textbf{429}, pp. 180--184 (2004).

\bibitem{SENTURK05}
R. Senturk,
``Narrative social structure'' (Stanford University Press, Stanford, CA, 2005).

\bibitem{TOP500_IBM}
See, e.g., \texttt{www.top500.org} and \texttt{www.research.ibm.com/bluegene/}.

\bibitem{KIRKPATRICK03}
S. Kirkpatrick, 
``Rough times ahead'',
\textit{Science} \textbf{299}, pp. 668--669 (2003).

\bibitem{GRID}
See, e.g., \texttt{www.gridforum.org} and \texttt{setiathome.ssl.berkeley.edu}.

\bibitem{MILGRAM67}
S. Milgram,
``The small-world problem'',
\textit{Psychology Today} \textbf{1}, pp. 61--68 (1967).

\bibitem{WATTS98}
D.J. Watts and S.H. Strogatz,
``Collective dynamics of small-world networks'',
\textit{Nature} \textbf{393}, pp. 440--442 (1998).

\bibitem{ERDOS60}
P. Erd\H{o}s and A. R\'enyi, 
``On the evolution of random graphs'', 
\textit{Publ. Math. Inst. Hung. Acad. Sci.} \textbf{5}, pp. 17--61 (1960).

\bibitem{SCALETTAR91}
R.T. Scalettar,
``Critical properties of an Ising model with dilute long range interactions'',
\textit{Physica A} \textbf{170}, pp. 282--290 (1991).

\bibitem{GITTERMAN00}
M. Gitterman,
``Small-world phenomena in physics: the Ising model'',
\textit{J. Phys. A} \textbf{33}, pp. 8373--8381 (2000).

\bibitem{BARRAT00}
A. Barrat and M. Weigt,
``On the properties of small-world network models'',
\textit{Eur. Phys. J. B} \textbf{13}, pp. 547--560 (2000).
[cond-mat/9903411]

\bibitem{NOVOTNY04}
M.A. Novotny and S.M. Wheeler,
``On the possibility of quasi small-world nanomaterials'',
\textit{Braz. J. Phys.} \textbf{34}, pp. 395--400 (2004).
[cond-mat/0308602]

\bibitem{KIM01}
B.J. Kim, H. Hong, P. Holme, G.S. Jeon, P. Minnhagen, and M.Y. Choi,
``\textit{XY} model in small-world networks'',
\textit{Phys. Rev. E} \textbf{64}, 056135 [5 pages] (2001).
[cond-mat/0108392]

\bibitem{HONG02_2}
H. Hong, M.Y. Choi, and B.J. Kim,
``Phase ordering on small-world networks with nearest-neighbor edges'',
\textit{Phys. Rev. E} \textbf{65}, 047104 [4 pages] (2002).
[cond-mat/0203177]

\bibitem{KOZMA04}
B. Kozma, M.B. Hastings, and G. Korniss,
``Roughness scaling for Edwards-Wilkinson relaxation in small-world
networks'',
\textit{Phys. Rev. Lett.} \textbf{92}, 108701 [4 pages] (2004).
[cond-mat/0309196]

\bibitem{KOZMA05}
B. Kozma, M.B. Hastings, and G. Korniss,
``Processes on annealed and quenched power-law small-world networks'',
in \textit{Noise in Complex Systems and Stochastic Dynamics III},
edited by L.B. Kish, K. Lindenberg, Z. Gingl,
Proceedings of SPIE \textbf{5845}, pp. 130--138 (SPIE, Bellingham, WA, 2005) .

\bibitem{KOZMA05b}
B. Kozma, M.B. Hastings, and G. Korniss,
``Diffusion Processes on Power-Law Small-World Networks'',
\textit{Phys. Rev. Lett.} (in press, 2005).
[cond-mat/0501509]

\bibitem{MONASSON99}
R. Monasson,
``Diffusion, localization and dispersion relations on small-world lattices'',
\textit{Eur. Phys. J. B} \textbf{12}, pp. 555--567 (1999).
[cond-mat/9903347]

\bibitem{BLUMEN_2000a}
S. Jespersen, I.M. Sokolov, and A. Blumen,
``Relaxation properties of small-world networks'',
\textit{Phys. Rev. E} {\bf 62}, pp. 4405--4408 (2002).
[cond-mat/0004214]

\bibitem{BLUMEN_2000b}
S. Jespersen and A. Blumen,
``Small-world networks: Links with long-tailed distributions'',
\textit{Phys. Rev. E} {\bf 62}, pp. 6270--6274 (2002).
[cond-mat/0009082]

\bibitem{ALMAAS_2002}
E. Almaas, R.V. Kulkarni, and D. Stroud,
``Characterizing the structure of small-world networks'',
\textit{Phys. Rev. Lett.} {\bf 88}, 098101 [4 pages] (2002).
[cond-mat/0109227]

\bibitem{HASTINGS04}
M.B. Hastings,
``An $\epsilon$-expansion for small-world networks'',
\textit{Eur. Phys. J. B} {\bf 42}, pp. 297--301 (2004).
[cond-mat/0407374]

\bibitem{WIESENFELD96}
K. Wiesenfeld, P. Colet, and S.H. Strogatz,
``Synchronization transitions in a disordered josephson series array'',
\textit{Phys. Rev. Lett.} \textbf{76}, pp. 404--407 (1996).

\bibitem{STROGATZ01}
S.H. Strogatz,
``Exploring complex networks'',
\textit{Nature} \textbf{410}, pp. 268--276 (2001).

\bibitem{BARAHONA02}
M. Barahona and L.M. Pecora,
``Synchronization in small-world systems,
\textit{Phys. Rev. Lett.} \textbf{89}, 054101 [4 pages] (2002).
[nlin.CD/0112023]

\bibitem{HONG02}
H. Hong, B.J. Kim, and M.Y. Choi,
``Comment on `Ising model on a small-world network' '',
\textit{Phys. Rev. E} \textbf{66}, 018101 [2 pages] (2002).
[cond-mat/0204357]

\bibitem{KORNISS00}
G. Korniss, Z. Toroczkai, M.A. Novotny, and P.A. Rikvold,
``From massively parallel algorithms and fluctuating time horizons to 
non-equilibrium surface growth''
\textit{Phys. Rev. Lett.} \textbf{84}, pp. 1351--1354 (2000).
[cond-mat/9909114]

\bibitem{FUJIMOTO90}
R. Fujimoto, 
``Parallel discrete event simulation'',
\textit{Commun. of the ACM} \textbf{33}, pp. 30--53 (1990).

\bibitem{NICOL94}
D.M. Nicol and R.M. Fujimoto,
``Parallel simulation today'',
\textit{Ann. Oper. Res.} \textbf{53}, pp. 249--285 (1994).

\bibitem{LUBACHEVSKY00}
B.D. Lubachevsky, 
``Fast simulation of multicomponent dynamic systems'',
\textit{Bell Labs Tech. J.} \textbf{5} (April-June), pp. 134--156 (2000).
[cs.DS/0405077]

\bibitem{KOLA_REV_2005}
A. Kolakowska, M. A. Novotny, in 
{\it Progress in Computer Science Research} 
(Nova Science Publishers, in press) (2004).
[cs.DC/0409032]

\bibitem{HWANG93}
K. Hwang, 
\textit{Advanced Computer Architecture: Parallelism,
Scalability, and Programmability} (McGraw--Hill, New York, NY, 1993).

\bibitem{BINDER98}
K. Binder and D.W. Heermann, 
\textit{Monte Carlo Simulation in Statistical Physics : An Introduction}
(Springer, Berlin, 1998).

\bibitem{GREENBERG94}
A.G. Greenberg, B.D. Lubachevsky, D.M. Nicol, and P.E. Wright,
``Efficient Massively parallel simulation of dynamic channel
assignment schemes for wireless cellular communications'',
\textit{Proc. 8th Workshop on Parallel and Distributed Simulation (PADS'94)},
pp. 187--194 (SCS, San Diego, CA, 1994).

\bibitem{KORNISS99}
G. Korniss, M.A. Novotny, and P.A. Rikvold,
``Parallelization of a dynamic Monte Carlo algorithm: A partially
rejection-free conservative approach'',
\textit{J. Comput. Phys.} \textbf{153}, pp. 488--508 (1999).
[cond-mat/9812344]

\bibitem{KORNISS01_3}
G. Korniss, C.J. White, P.A. Rikvold, and M.A. Novotny,
``Dynamic phase transition, universality, and finite-size scaling in the
two-dimensional kinetic Ising  model in an oscillating field'',
\textit{Phys. Rev. E} \textbf{63}, 016120 [15 pages] (2001).
[cond-mat/0008155]

\bibitem{DEELMAN96}
E. Deelman, B.K. Szymanski, and T. Caraco,
``Simulating lyme disease using parallel discrete-event simulations'',
in \textit{Proc. 28th Winter Simulation Conference}, pp. 1191--1198
(ACM, New York, NY, 1996).

\bibitem{AMAR_PRB_2005a}
Y. Shim and J.G. Amar,
``Rigorous synchronous relaxation algorithm for parallel 
kinetic Monte Carlo simulations of thin film growth'', 
\textit{Phys. Rev. B} {\bf 71}, 115436 [12 pages] (2005).
[cond-mat/0406540]

\bibitem{AMAR_PRB_2005b}
Y. Shim and J.G. Amar,
``Semirigorous synchronous sublattice algorithm for parallel
kinetic Monte Carlo simulations of thin film growth'', 
\textit{Phys. Rev. B} {\bf 71}, 125432 [13 pages] (2005).
[cond-mat/0406379]

\bibitem{NICOL87}
D.M. Nicol,
``Performance issues for distributed battlefield simulations'',
\textit{Proc. $19^{th}$ Conf. Winter. Simul.} ed. by A. Thesen, H. Grant,
and W.D. Kelton, pp. 624--628 (1987).

\bibitem{COWIE99}
J. Cowie, D.M. Nicol, and A. Ogielski,
``Modeling the global Internet'',
\textit{IEEE Computing in Science and Engineering} \textbf{1}, pp. 42--50 (1999).

\bibitem{GLAUBER63}
R.J. Glauber,
``Time-dependent statistics of the Ising model'',
\textit{J. Math. Phys.} \textbf{4}, pp. 294--307 (1963).

\bibitem{LUBACHEVSKY87}
B.D. Lubachevsky,
``Efficient parallel simulations of asynchronous cellular arrays'',
\textit{Complex Systems} \textbf{1}, pp. 1099--1123 (1987).
[cs.DC/0502039]

\bibitem{LUBACHEVSKY88}
B.D. Lubachevsky,
``Efficient parallel simulations of dynamic Ising spin systems'',
\textit{J. Comput. Phys.} \textbf{75}, pp. 103--122 (1988).

\bibitem{JEFFERSON85}
D.R. Jefferson,
``Virtual time'', \textit{Assoc. Comput. Mach. Trans.
Programming Languages and Systems} \textbf{7}, pp. 404--425 (1985).

\bibitem{CHANDY79}
K.M. Chandy and J. Misra,
``Distributed simulation: A case study in design and verification
of distributed programs'',
\textit{IEEE Trans. Softw. Eng.} \textbf{SE-5}, pp. 440--452 (1979).

\bibitem{CHANDY81}
K.M. Chandy, and J. Misra,
``Asynchronous distributed simulation via a sequence of parallel
computations'', 
\textit{Commun. ACM} \textbf{24}, pp. 198--205 (1981).

\bibitem{DEELMAN97}
E. Deelman and B.K. Szymanski,
``Breadth-first rollback in spatially explicit simulations'',
\textit{Proc. of the 11$^{th}$ Workshop on Parallel and Distributed
Simulation, PADS '97}, pp. 124--131
(IEEE Computer Society, Los Alamitos, CA, 1997).

\bibitem{CAROTHERS99}
C.D. Carothers, K.S. Perumalla, and R.M. Fujimoto,
``Efficient optimistic parallel simulations using reverse computation'',
\textit{ACM Transactions on Modeling and Computer Simulation (TOMACS)} \textbf{9}, 
pp. 224--253 (1999).

\bibitem{CHEN02}
G. Chen and B.K. Szymanski,
``Lookback: a new way of exploiting parallelism in discrete event simulation'',
\textit{Proc. of the 16$^{th}$ Workshop on Parallel and Distributed
Simulation, PADS '02}, pp. 153--162 (IEEE Computer Society, Washington, DC, 2002).

\bibitem{SHCHUR04}
L. N. Shchur and M. A. Novotny,
``Evolution of time horizons in parallel and grid simulations'',
\textit{Phys. Rev. E} \textbf{70}, 026703 [9 pages] (2004).
[cond-mat/0401229]

\bibitem{GREENBERG96}
A.G. Greenberg, S. Shenker, and A.L. Stolyar,
``Asynchronous updates in large parallel systems'',
\textit{Performance Eval. Rev.} \textbf{24}, pp. 91--103 (1996).

\bibitem{SLOOT01}
P.M.A. Sloot, B.J. Overeinder, and A. Schoneveld, 
``Self-organized criticality in simulated correlated systems'', 
\textit{Comput. Phys. Commun.} \textbf{142}, pp. 76--81 (2001).

\bibitem{SCHONEVELD99}
A. Schoneveld,
``Parallel complex systems simulation'', 
Ph.D. Thesis, Universiteit van Amsterdam (1999).

\bibitem{BAK87}
P. Bak, C. Tang, and K. Wiesenfeld, 
``Self-organized criticality: An explanation of $1/f$ noise'', 
\textit{Phys. Rev. Lett.} \textbf{59}, pp. 381--384 (1987).

\bibitem{BAK88}
P. Bak, C. Tang, and K. Wiesenfeld, 
``Self-organized criticality'',
\textit{Phys. Rev. A} \textbf{38}, pp. 364--374 (1988).

\bibitem{COMPLEXITY}
\textit{Frontiers in Problem Solving: Phase Transitions and
Complexity}, ed. by T. Hogg, B.A. Huberman, and C. Williams,
Artif. Intell. \textbf{81}, issue 1-2 (1996).

\bibitem{MONASSON99_2}
R. Monasson, R. Zecchina, S. Kirkpatrick, B. Selman, and L. Troyansky, 
``Determining computational complexity from characteristic phase transitions''
\textit{Nature} \textbf{400}, pp. 133--137 (1999).

\bibitem{NICOL91}
D.M. Nicol, 
``Performance bounds on parallel self-initiating discrete-event simulations'',
\textit{ACM Trans. Model. Comput. Simul.} \textbf{1}, pp. 24--50 (1991).

\bibitem{FELDERMAN91}
R.E. Felderman and L. Kleinrock,
``Bounds and approximations for self-initiating distributed simulation
without lookahead'',
\textit{ACM Trans. Model. Comput. Simul.} \textbf{1}, 386 (1991).

\bibitem{BARABASI95}
A.-L. Barab\'asi and H.E. Stanley,
\textit{Fractal Concepts in Surface Growth},
(Cambridge Univ. Press, Cambridge, MA, 1995).

\bibitem{HEALY95}
T. Halpin-Healy and Y.-C. Zhang,
``Kinetic roughening phenomena, stochastic growth, directed polymers,
and all that'',
\textit{Physics Reports} \textbf{254}, pp. 215--414 (1995).

\bibitem{KRUG97}
J. Krug,
``Origins of scale invariance in growth processes'',
\textit{Adv. Phys.} \textbf{46}, pp. 139--282 (1997).

\bibitem{FAMILY85}
F. Family and T. Vicsek,
``Scaling of the active zone in the Eden process on percolation networks
and the ballistic deposition model'',
\textit{J. Phys. A} \textbf {18}, pp. L75--L81 (1985).

\bibitem{TOROCZKAI00}
Z. Toroczkai, G. Korniss, S. Das Sarma, and R.K.P. Zia,
``Extremal point densities of interface fluctuations'',
\textit{Phys. Rev. E} \textbf{62}, pp. 276--294 (2000).
[cond-mat/0002143]

\bibitem{KRUG90}
J. Krug and P. Meakin,
``Universal finite-size effects in the rate of growth process'',
\textit{J. Phys. A} \textbf{23}, pp. L987--L994 (1990).

\bibitem{KORNISS02}
G. Korniss, M.A. Novotny, A.K. Kolakowska, and H. Guclu,
``Statistical properties of the simulated time horizon in conservative
parallel discrete-event simulations'',
SAC 2002, \textit{Proceedings of the 2002 ACM Symposium on Applied Computing},
pp. 132--138 (2002).

\bibitem{KOLAKOWSKA03}
A. Kolakowska, M.A. Novotny, and G. Korniss,
``Algorithmic scalability in globally constrained conservative parallel
discrete-event simulations of asynchronous systems'',
\textit{Phys. Rev. E} \textbf{67}, 046703 [13 pages] (2003).
[cs.DC/0211013]

\bibitem{KOLAKOWSKA03_2}
A. Kolakowska, M.A. Novotny, and P.A. Rikvold,
``Update statistics in conservative parallel discrete event simulations 
of asynchronous systems'',
\textit{Phys. Rev. E} \textbf{68}, 046705 [14 pages] (2003).
[cond-mat/0306222]

\bibitem{KOLAKOWSKA04}
A. Kolakowska and M.A. Novotny,
``Discrete-event analytic technique for surface growth problems'',
\textit{Phys. Rev. B} \textbf{69}, 075407 [5 pages] (2004).
[cond-mat/0311015]

\bibitem{PLISCHKE87}
M. Plischke, Z. R\'acz, and D. Liu,
``Time-reversal invariance and universality of two-dimensional growth models'',
\textit{Phys. Rev. B} \textbf{35}, pp. 3485--3495 (1987).

\bibitem{KARDAR86}
M. Kardar, G. Parisi, and Y.-C. Zhang,
``Scaling of growing interfaces'',
\textit{Phys. Rev. Lett.} \textbf{56}, pp. 889--892 (1986).

\bibitem{BURGERS74}
M. Burgers,
\textit{The Nonlinear Diffusion Equation} (Riedel, Boston, MA, 1974).

\bibitem{EDWARDS82}
S.F. Edwards and D.R. Wilkinson,
``The surface statistics of a granular aggregate'',
\textit{Proc. R. Soc. London, Ser. A} \textbf{381}, pp. 17--31 (1982).

\bibitem{TOROCZKAI03}
Z. Toroczkai, G. Korniss, M. A. Novotny, and H. Guclu,
``Virtual time horizon control via communication network design'',
in \textit{Computational Complexity and Statistical Physics}, 
edited by A. Percus, G. Istrate, and C. Moore, 
Santa Fe Institute Studies in the Sciences of Complexity Series (Oxford University Press, 2004). 
[cond-mat/0304617]

\bibitem{FOLTIN94} 
G. Foltin, K. Oerding, Z. R\'acz, R.L. Workman, and R.K.P. Zia, 
``Width distribution for random-walk interfaces'',
\textit{Phys. Rev. E} \textbf{50}, pp. R639--R642 (1994). 

\bibitem{PLISCHKE94}
M. Plischke, Z. R\'acz, and R.K.P. Zia,
``Width distribution of curvature-driven interfaces: A study of universality'',
\textit{Phys. Rev. E} \textbf{50}, pp. 3589--3593 (1994).

\bibitem{RACZ94}
Z. R\'acz and M. Plischke,
``Width distribution for $(2$$+$$1)$-dimensional growth and deposition processes'',
\textit{Phys. Rev. E} \textbf{50}, pp. 3530--3537 (1994).

\bibitem{ANTAL96}
T. Antal and Z. R\'acz,
``Dynamic scaling of width distribution in Edwards--Wilkinson type 
models of interface dynamics'',
\textit{Phys. Rev. E} \textbf{54}, pp. 2256--2260 (1996).
[cond-mat/9510170]

\bibitem{KORNISS01}
G. Korniss, M.A. Novotny, Z. Toroczkai, and P.A. Rikvold,
``Nonequilibrium surface growth and scalability of parallel algorithms
for large asynchronous systems'',
\textit{Computer Simulation Studies in Condensed Matter
Physics XIII}, Springer Proceedings in Physics, Vol. 86, editors D.P. Landau, S.P.
Lewis, and H.-B. Sch{\"u}ttler (Springer-Verlag, Berlin, Heidelberg), 
pp. 183--188 (2001).
[cond-mat/0002469]

\bibitem{MARINARI00}
E. Marinari, A. Pagnani, and G. Parisi,
``Critical exponents of the KPZ equation via multi-surface 
coding numerical simulations'',
\textit{J. Phys. A: Math. Gen.} \textbf{33}, pp. 8181--8192 (2000).
[cond-mat/0005105]

\bibitem{KIM89}
J.M. Kim and J.M. Kosterlitz,
``Growth in a restricted solid-on-solid model'',
\textit{Phys. Rev. Lett.} \textbf{62}, pp. 2289--2292 (1989).

\bibitem{MARINARI02}
E. Marinari, A. Pagnani, G. Parisi, and Z. R\'acz,
``Width distributions and the upper critical dimension of KPZ interfaces''
\textit{Phys. Rev. E} \textbf{65}, 026136 [4 pages] (2002).
[cond-mat/0105158]

\bibitem{BOLLOBAS01}
B. Bollob\'as, 
\textit{Random Graphs} (Cambridge University Press,
Cambridge, UK, 2001).

\bibitem{FELLER68}
W. Feller, 
\textit{An Introduction to Probability Theory and Its Applications}, 
Vol. 1, 3rd ed. (Wiley, New York, NY, 1968). 

\bibitem{FELLER71}
W. Feller, 
\textit{An Introduction to Probability Theory and Its Applications}, 
Vol. 2, 3rd ed. (Wiley, New York, NY, 1971). 

\bibitem{GUCLU04_2}
H. Guclu, G. Korniss, Z.  Toroczkai, and  M.A. Novotny,
``Small-world synchronized computing networks for scalable parallel 
discrete-event simulations'', in \textit{Complex Networks}, 
edited by E. Ben-Naim, H. Frauenfelder, and Z. Toroczkai,
Lecture Notes in Physics \textbf{650}, pp. 255--275 (Springer-Verlag, Berlin, 2004).

\bibitem{NETCRAFT}
\texttt{http://www.netcraft.com/survey/}

\bibitem{CIA}
\texttt{http://www.c-i-a.com}

\bibitem{BARABASI99_2}
A.-L. Barab\'asi, R. Albert, and H. Jeong,
``Mean-field theory for scale-free random networks'',
\textit{Physica A} \textbf{272}, pp. 173--187 (1999).
[cond-mat/9907068]

\bibitem{GUCLU04}
H. Guclu and G. Korniss,
``Extreme fluctuations in small-world networks with relaxational dynamics'',
\textit{Phys. Rev. E} \textbf{69}, 065104(R) [4 pages] (2004).
[cond-mat/0311575]

\bibitem{GUCLU05}
H. Guclu and G. Korniss,
``Extreme fluctuations in small-world-coupled autonomous systems with relaxational dynamics'',
\textit{Fluctuations and Noise Letters} \textbf{5}, pp. L43-L62 (2005).

\bibitem{WATTS99}
D.J. Watts,
\textit{Small Worlds} 
(Princeton Univ. Press, Princeton, NJ, 1999).

\bibitem{FISHER28}
R.A. Fisher and L.H.C. Tippett,
``The frequency distribution of the largest or smallest member of
a sample'',
\textit{Proc. Camb. Philos. Soc.} \textbf{24}, pp. 180--191 (1928).

\bibitem{GUMBEL58}
E.J. Gumbel, 
\textit{Statistics of Extremes} 
(Columbia University Press, New York, NY, 1958).

\bibitem{GALAMBOS94}
\textit{Extreme Value Theory and Applications},
edited by J. Galambos, J. Lechner, and E. Simin (Kluwer, Dordrecht, 1994).

\bibitem{BRAMWELL98}
S.T. Bramwell, P.C.W. Holdsworth, and J.-F. Plinton,
``Universality of rare fluctuations in turbulance and critical phenomena'',
\textit{Nature} \textbf{396}, pp. 552--554 (1998).

\bibitem{RAYCHAUDHURI01}
S. Raychaudhuri, M. Cranston, C. Przybyla, and Y. Shapir,
``Maximal height scaling of kinetically growing surfaces''
\textit{Phys. Rev. Lett.} \textbf{87}, 136101 [4 pages] (2001).
[cond-mat/0105176]

\bibitem{BRAMWELL00}
S.T. Bramwell, K. Christensen, J.-Y. Fortin, P.C.W. Holdsworth,
H.J. Jensen, S. Lise, J.M. L\'opez, M. Nicodemi, J.-F. Pinton, and M. Sellitto,
``Universal fluctuations in correlated system'',
\textit{Phys. Rev. Lett.} \textbf{84}, pp. 3744--3747 (2000).
[cond-mat/9912255]

\bibitem{WATKINS02}
N.W. Watkins, S.C. Chapman, and G. Rowlands,
``Comment on `universal fluctuations in correlated systems' '',
\textit{Phys. Rev. Lett.} \textbf{89}, 208901 [1 page] (2002).
[cond-mat/0209398]

\bibitem{BRAMWELL02}
S.T. Bramwell, K. Christensen, J.-Y. Fortin, P.C.W. Holdsworth,
H.J. Jensen, S. Lise, J.M. L\'opez, M. Nicodemi, J.-F. Pinton, and M. Sellitto,
``Reply to `comment on `universal fluctuations in correlated systems' ' '',
\textit{Phys. Rev. Lett.} \textbf{89}, 208902 [1 page] (2002).
[cond-mat/0209416]

\bibitem{BRAMWELL01}
S.T. Bramwell, J.-Y. Fortin, P.C.W. Holdsworth, S. Peysson,
J.-F. Pinton, B. Portelli, and M. Sellitto,
``Magnetic fluctuations in the classical \textit{XY} model: The origin of
an exponential tail in complex systems'',
\textit{Phys. Rev. E} \textbf{63}, 041106 [22 pages] (2001).
[cond-mat/0008093]

\bibitem{AJI01}
V. Aji and N. Goldenfeld,
``Fluctuations in finite critical and turbulent systems'',
\textit{Phys. Rev. Lett.} \textbf{86}, pp. 1007--1010 (2001).
[cond-mat/0008243]

\bibitem{ANTAL01}
T. Antal, M. Droz, G. Gy\"orgyi, and Z. R\'acz,
``$1/f$ Noise and extreme-value statistics'',
\textit{Phys. Rev. Lett.} \textbf{87}, 240601 [4 pages] (2001).
[cond-mat/0105599]

\bibitem{BERTIN05}
E. Bertin,
``Global fluctuations and Gumbel statistics'' (2005).
[cond-mat/0506166]

\bibitem{ANTAL02}
T. Antal, M. Droz, G. Gy\"orgyi, and Z. R\'acz,
``Roughness distributions for $1/f^{\alpha}$ signals'',
\textit{Phys. Rev. E} \textbf{65}, 046140 [12 pages] (2002).
[cond-mat/0112277]

\bibitem{DAHLSTEDT01}
K. Dahlstedt and H.J. Jensen,
``Universal fluctuations and extreme value statistics'',
\textit{J. Phys. A} \textbf{34}, pp. 11193--11200 (2001).
[cond-mat/0108007]

\bibitem{CHAPMAN02}
S. C. Chapman, G. Rowlands, and N. W. Watkins,
``Extremum statistics: A framework for data analysis'',
\textit{Nonlin. Proc. Geophys.} \textbf{9}, pp. 409--418 (2002).
[cond-mat/0106015]

\bibitem{GYORGYI03}
G. Gy\"orgyi, P.C.W. Holdsworth, B. Portelli, and Z. R\'acz,
``Statistics of extremal intensities for Gaussian interfaces'',
\textit{Phys. Rev. E} \textbf{68}, 056116 [14 pages] (2003).
[cond-mat/0307645]

\bibitem{BOUCHAUD97}
J.-P. Bouchaud and M. M\'ezard,
``Universality classes for extreme-value statistics'',
\textit{J. Phys. A} \textbf{30}, pp. 7997--8015 (1997).
[cond-mat/9707047]

\bibitem{BALDASSARRI02}
A. Baldassarri, A. Gabrielli, and B. Sapoval,
``Chemical fracture statistics and universal distribution of extreme values'',
\textit{Europhys. Lett.} \textbf{59}, pp. 232--238 (2002). 
[cond-mat/0205130]

\bibitem{MAJUMDAR04}
S.N. Majumdar and A. Comtet,
``Exact maximal height distribution of fluctuating interfaces'',
\textit{Phys. Rev. Lett.} \textbf{92}, 225501 [4 pages] (2004).

\bibitem{MAJUMDAR04_2}
S.N. Majumdar and A. Comtet,
``Airy distribution function: From the area under a Brownian excursion to the
maximal height of fluctuating interfaces'' (2004).
[cond-mat/0409566] 

\bibitem{GOLDENFELD92}
N. Goldenfeld,
\textit{Lectures on Phase Transition and the Renormalization Group},
(Addison-Wesley, Reading, MA, 1992).

\bibitem{BOUCHAUD00}
J.-P. Bouchaud and M. Potters,
\textit{Theory of Financial Risk},
(Cambridge Univ. Press, Cambridge, UK, 2000)

\bibitem{BALDASSARRI00}
A. Baldassarri, 
\textit{Statistics of Persistent Extreme Events},
Ph.D. Thesis, De l\'{}Universit\'e Paris XI Orsay (2000)
\texttt{http://axtnt3.phys.uniroma1.it/\~{}andreab/these.html}

\bibitem{KORNISS01_2}
G. Korniss, M.A. Novotny, P.A. Rikvold, H. Guclu, and Z. Toroczkai,
``Going through rough times: From non-equilibrium surface growth to
algorithmic scalability'',
\textit{Materials Research Society Symposium Proceedings
Series} \textbf{700}, pp. 297--308, Fall Meeting, Boston, (2001).
[cond-mat/0112103]

\bibitem{KOZMA04_2}
B. Kozma and G. Korniss,
``Stochastic growth in a small world'',
in \textit{Computer Simulation Studies in Condensed Matter Physics XVI}, edited
by D.P. Landau, S.P. Lewis, and H.-B. Sch\"uttler, Springer
Proceedings in Physics \textbf{95}, pp. 29--33 (Springer-Verlag, Berlin, 2004).
[cond-mat/0305025]

\bibitem{MENEZES04}
M. Argollo de Menezes and A.-L. Barab\'asi
``Fluctuations in network dynamics'',
\textit{Phys. Rev. Lett.} \textbf{92}, 028701 [4 pages] (2004).
[cond-mat/0306304]

\bibitem{BARABASI04}
A.-L. Barab\'asi, M. Argollo de Menezes, S. Balensiefer, and J. Brockman,
``Hot spots and universality in network dynamics'',
\textit{Eur. Phys. J. B} \textbf{38}, pp. 169--175 (2004).

\bibitem{TOROCZKAI04}
Z. Toroczkai and K.E. Bassler,
``Jamming is limited in scale-free systems'',
\textit{Nature} \textbf{428}, p. 716 [1 page] (2004).

\bibitem{CROVELLA97}
M.E. Crovella and A. Bestavros,
``Self-similarity in world wide web traffic: evidence and possible causes'',
\textit{IEEE/ACM Trans. on Networking} \textbf{5}, pp. 835--846 (1997).

\bibitem{CROVELLA98}
M.E. Crovella, M.S. Taqqu, and A. Bestavros,
``Heavy-tailed probability distributions in the world wide web'',
in \textit{A Practical guide to heavy-tails: Statistical techniques and
applications}, ed. by R.J. Adler, R.E. Feldman, and M.S. Taqqu, pp 3--25
(Birkh\"auser, Boston, MA, 1998).

\bibitem{LELAND94}
W.E. Leland, M.S. Taqqu, W. Willinger, and D.V. Wilson,
``On the self-similar nature of ethernet traffic'',
\textit{IEEE/ACM Trans. on Networking} \textbf{2}, pp. 1--15 (1994).

\bibitem{CSABAI94}
I. Csabai,
``$1/f$ noise in computer network traffic'',
\textit{J. Phys. A} \textbf{27}, pp. L417--L421 (1994).

\bibitem{PAXSON95}
V. Paxson and S. Floyd,
``Wide area traffic: The failure of poisson modeling'',
\textit{IEEE/ACM Trans. on Networking} \textbf{3}, pp. 226--244 (1995).

\bibitem{KOLA_PRE_2004}
A. Kolakowska, M.A. Novotny, and P.S. Verma,
``Roughening of the interfaces in (1+1)-dimensional 
two-component surface growth with an admixture of random deposition'',
\textit{Phys. Rev. E} {\bf 70}, 051602 [16 pages] (2004).
[cond-mat/0403341]

\bibitem{ALEKSIEJUK02}
A. Aleksiejuk, J.A. Holyst, and D. Stauffer,
``Ferromagnetic phase transition in Barab\'asi-Albert networks'',
\textit{Physica A} \textbf{310}, pp. 260--266 (2002).
[cond-mat/0112312]

\bibitem{LEONE02}
M. Leone, A. V\'azquez, A. Vespignani, and R. Zecchina,
``Ferromagnetic ordering in graphs with arbitrary degree distribution'',
\textit{Eur. Phys. J. B} \textbf{28}, pp. 191--197 (2002).
[cond-mat/0203416]

\bibitem{DOROGOVTSEV02_2}
S.N. Dorogovtsev, A.V. Goltsev, and J.F.F. Mendes,
``Ising model on networks with an arbitrary distribution of connections'',
\textit{Phys. Rev. E} \textbf{66}, 016104 [5 pages] (2002).
[cond-mat/0203227]

\bibitem{BIANCONI02}
G. Bianconi,
``Mean-field solution of the Ising model on a Barab\'asi-Albert network'',
\textit{Phys. Lett. A} \textbf{303}, pp. 166--169 (2002).
[cond-mat/0204455]

\bibitem{GOLTSEV03}
A.V. Goltsev, S.N. Dorogovtsev, and J.F.F. Mendes,
``Critical phenomena in networks'',
\textit{Phys. Rev. E} \textbf{67}, 026123 [5 pages] (2003).
[cond-mat/0204596]

\bibitem{NEWMAN00}
M.E.J. Newman,
``Models of a small sorld'',
\textit{J. Stat. Phys.} \textbf{101}, pp. 819--841 (2000).
[cond-mat/0001118]

\bibitem{NEWMAN99}
M.E.J. Newman and D.J. Watts, 
``Renormalization group analysis of the Small-World network model'',
\textit{Phys. Lett. A} \textbf{263}, pp. 341--346 (1999).
[cond-mat/9903357]

\bibitem{PEKALSKI01}
A. Pekalski,
``Ising model on a small-world network'',
\textit{Phys. Rev. E} \textbf{64}, 057104 [4 pages] (2001).

\bibitem{HONG02_3}
H. Hong, M.Y. Choi, and B.J. Kim,
``Synchronization on small-world networks'',
\textit{Phys. Rev. E} \textbf{65}, 026139 [5 pages] (2002).
[cond-mat/0110359]

\bibitem{HERRERO02}
C.P. Herrero,
``Ising model in small-world networks'',
\textit{Phys. Rev. E} \textbf{65}, 066110 [6 pages] (2002).
[cond-mat/0206079]

\bibitem{JEONG03}
D. Jeong, H. Hong, B.J. Kim, and M.Y. Choi,
``Phase transition in the Ising model on a small-world network with
distance-dependent interactions'',
\textit{Phys. Rev. E} \textbf{68}, 027101 [4 pages] (2003).
[cond-mat/0306017]

\bibitem{HASTINGS03}
M.B. Hastings,
``Mean-field and anomalous behavior on a small-world network'',
\textit{Phys. Rev. Lett.\/} \textbf{91}, 098701 [4 pages] (2003).
[cond-mat/0304530]

\bibitem{DROZ90}
M. Droz, Z. R\'acz, and P. Tartaglia,
``One-dimensional kinetic Ising model with competing spin-flip and
spin-exchange dynamics: Ordering in the case of long-range exchanges'',
\textit{Phys. Rev. A} \textbf{41}, pp. 6621--6624 (1990).

\bibitem{BERGERSEN91}
B. Bergersen and Z. R\'acz,
``Dynamical generation of long-range interactions: random Levy flights
in the kinetic Ising and spherical models'',
\textit{Phys. Rev. Lett.} \textbf{67}, pp. 3047--3050 (1991).

\bibitem{LAUGHLIN03}
S.B. Laughlin and T.J. Sejnowski,
``Communication in neuronal networks'',
\textit{Science} \textbf{301}, pp. 1870--1874 (2003).

\bibitem{DAVIS98}
J.A. Davis, V.K. De, and J.D. Meindl,
``A stochastic wire length distribution for gigascale integration (GSI) 
Part I: Derivation and validation''
\textit{IEEE Trans. Elec. Dev.} \textbf{45}, pp. 580--589 (1998).

\bibitem{PETERMANN05}
T. Petermann and P. de los Rios,
``Spatial small-world networks: A wiring-cost perspective''  (2005).
[cond-mat/0501420]




























%
%\bibitem{NEWMAN01}
%M.E.J. Newman,
%``The structure of scientific collaboration networks'',
%\textit{Proc. Natl. Acad. Sci. USA} \textbf{98}, pp. 404--409 (2001).
%[cond-mat/0007214]
%
%\bibitem{RAVASZ02}
%E. Ravasz, A.L. Somera, D.A. Mongru, Z.N. Oltvai, and A.-L. Barab\'asi,
%``Hierarchical organization of modularity in metabolic networks'',
%\textit{Science} \textbf{297}, pp. 1551--1555 (2002).
%[cond-mat/0209244]
%
%\bibitem{STEINMAN93}
%J.S. Steinman,
%``Breathing time warp'',
%\textit{Proceedings of the 7th Workshop on Parallel and Distributed
%Simulation (PADS'93)} edited by R. Bagrodia and D. Jefferson
%, Vol. 23, No. 1, pp. 109--118 (Soc. Comput. Simul., San Diego, CA, 1993).
%
%\bibitem{STEINMAN94}
%J.S. Steinman,
%``Discrete event simulation and the event horizon'',
%\textit{Proceedings of the 8th Workshop on Parallel and Distributed 
%Simulation (PADS'94)} edited by D.K. Arvind, R. Bagrodia, and
%J.Y.-B. Lin, pp. 39--49 (Soc. Comput. Simul., San Diego, CA, 1994).
%
%\bibitem{STEINMAN96}
%J.S. Steinman,
%``Discrete-event simulation and the event horizon II: event list management'',
%\textit{Proceedings of the 9th Workshop on Parallel and Distributed Simulation
%(PADS'96)}, pp. 170--178, (IEEE Comput. Soc. Press, Los Alamitos, CA, 1996).
%
%\bibitem{MIDDLETON92}
%A.A. Middleton,
%``Asymptotic uniqueness of the sliding state for charge-density waves'',
%\textit{Phys. Rev. Lett.} \textbf{68}, pp. 670--673 (1992).
%
%\bibitem{KLEINBERG00}
%J. Kleinberg,
%``Navigation in a small world'',
%\textit{Nature} \textbf{406}, p. 845 [1 page], (2000).
%
%
%
%\bibitem{MENEZES00}
%M. Argollo de Menezes, C.F. Moukarzel, and T.J. Penna,
%``First-order transition in small-world networks'',
%\textit{Europhys. Lett.} \textbf{50}, pp. 574--579 (2000).                         
%[cond-mat/9903426]
%
%
%\bibitem{BORST97}
%S.C. Borst, S.A. Grandhi, C.L. Kahn, K. Kumaran, B.D. Lubachevsky, and D.M. Sand,
%``Wireless simulation and self-organizing spectrum management'',
%\textit{Bell Labs Tech. J.} \textbf{2}, pp. 81--98 (1997).
%
%\bibitem{KORNISS02_2}
%G. Korniss, P.A. Rikvold, and M.A. Novotny,
%``Absence of first-order transition and tricritical point in the
%dynamic phase diagram of a spatially extended bistable system in an
%oscillating field'', 
%\textit{Phys. Rev. E} \textbf{66}, 056127 [12 pages] (2002).
%[cond-mat/0207275]
%
%
%\bibitem{OVEREINDER00}
%B.J. Overeinder, 
%``Distributed event-driven simulation: scheduling strategies and resource management'', 
%Ph.D. thesis, Universiteit van Amsterdam (2000).
%
%\bibitem{OVEREINDER01}
%B.J. Overeinder, A. Schoneveld, and P.M.A. Sloot,
%``Spatio-temporal correlations and rollback distributions in optimistic simulations'',
%\textit{Proceedings of the 15th Workshop on Parallel and Distributed Simulation},
%pp. 145--152, (IEEE Comput. Soc., Lake Arrowhead, CA, 2001).
%
%\bibitem{BREZIN94}
%E. Br\'ezin and A. Zee,
%``Correlation functions in disordered systems'',
%\textit{Phys. Rev. E} \textbf{49}, pp. 2588--2596, 1994.
%
%\bibitem{SCALETTAR00}
%R.T. Scalettar,
%``Critical properties of an Ising model with dilute long-range
%Interactions'', \textit{Physica A} \textbf{170}, pp. 282--290 (1991).
%[cond-mat/0409208]
%
%\bibitem{IGLOI02}
%F. Igloi and L. Turban
%``First- and second-order phase transitions in scale-free networks'',
%\textit{Phys. Rev. E} \textbf{66}, 036140 [4 pages] (2002).
%[cond-mat/0206522]
%
%\bibitem{DOROGOVTSEV03}
%S.N. Dorogovtsev, A.V. Goltsev, and J.F.F. Mendes,
%``Potts model on complex networks'',
%\textit{Eur. Phys. J. B} \textbf{38}, pp. 177--182 (2004).
%[cond-mat/0310693]

\end{thebibliography}
